\documentclass[fleqn,usenatbib]{mnras}

\usepackage{newtxtext,newtxmath}

\usepackage[T1]{fontenc}

\DeclareRobustCommand{\VAN}[3]{#2}
\let\VANthebibliography\thebibliography
\def\thebibliography{\DeclareRobustCommand{\VAN}[3]{##3}\VANthebibliography}

\usepackage{graphicx}	
\usepackage{amsmath}	
\usepackage{comment}
\usepackage{subcaption}
\usepackage{newtxtext,newtxmath}
\usepackage{comment}
\usepackage{multirow}
\usepackage{subcaption}
\usepackage{longtable}
\usepackage{supertabular}
\usepackage[T1]{fontenc}
\usepackage{booktabs}
\usepackage{siunitx}
\usepackage{graphicx}	
\usepackage{amsbsy}

\newcommand{\eazy}{{\tt{EAZY}}}
\newcommand{\bagpipes}{{\tt{Bagpipes}}}

\newcommand{\galfit}{{\tt{GALFIT}}}

\newcommand\mkms{\rm km \, s^{-1}}
\newcommand\sfive{$\Sigma_5$}
\newcommand\PDFMz{$\mathrm{PDF}_{i}(M_{\star}, z)$}
\newcommand{\jaguar}{{\tt{JAGUAR}}}
\newcommand{\oiii}{[O{\sc iii}]}

\usepackage{xcolor}

\title[Galaxy environment at high-$z$]{EPOCHS Paper X: Environmental effects on Galaxy Formation and Protocluster Galaxy candidates at $4.5<z<10$ from JWST observations}

\author[Li et al.]{
Qiong Li\thanks{ qiong.li@manchester.ac.uk}$^{1}$,
Christopher J. Conselice$^{1}$, 
Florian Sarron\thanks{florian.sarron@irit.fr}$^{2}$ $^{3}$,
Tom Harvey$^{1}$,
Duncan Austin$^{1}$,
Nathan Adams$^{1}$,
\newauthor
James A.\@ A.\@ Trussler$^{1}$,
Qiao Duan$^{1}$,
Leonardo Ferreira$^{4}$,
Lewi Westcott$^{1}$,
Honor Harris$^{1}$,
Herv\'e Dole$^{5}$,
\newauthor
Norman A. Grogin$^{6}$,
Brenda Frye$^{7}$,
Anton M. Koekemoer$^{6}$,
Clayton Robertson$^{8}$, 
Rogier A. Windhorst$^9$, 
\newauthor
Maria del Carmen Polletta$^{10}$, 
Nimish P. Hathi$^{6}$
\\
$^{1}$ Jodrell Bank Centre for Astrophysics, University of Manchester, Oxford Road, Manchester UK \\
$^{2}$ IRAP, Institut de Recherche en Astrophysique et Planétologie, Université de Toulouse, UPS-OMP, CNRS, CNES, 14 avenue E. Belin, F-31400 Toulouse,
France \\
$^{3}$ Institut de Recherche en Informatique de Toulouse (IRIT) et Centre de Biologie Intégrative (CBI), Laboratoire MCD, CNRS, Université de Toulouse, France \\
$^{4}$ School of Physics and Astronomy, University of Victoria, Victoria, BC, Canada \\
$^{5}$Universit\'e Paris-Saclay, CNRS, Institut d’Astrophysique
Spatiale, 91405, Orsay, France\\
$^{6}$Space Telescope Science Institute, 3700 San Martin Drive, Baltimore, MD 21218, USA\\
$^{7}$Department of Astronomy/Steward Observatory, University of Arizona, 933 N Cherry Ave,
Tucson, AZ, 85721-0009, USA\\
$^{8}$Department of Physics, University of Louisville, Natural Science Building 102, Louisville KY 40292, USA\\
$^9$ School of Earth and Space Exploration, Arizona State University, Tempe, AZ 85287-6004, USA\\
$^{10}$ INAF – Istituto di Astrofisica Spaziale e Fisica Cosmica Milano,  Via A. Corti 12, I-20133 Milano, Italy
}

\date{Accepted SSS. Received YYY; in original form ZZZ}

\pubyear{2015}

\begin{document}
\label{firstpage}
\pagerange{\pageref{firstpage}--\pageref{lastpage}}
\maketitle

\begin{abstract}
In this paper we describe our search for galaxy protocluster candidates at $4.5< z < 10$ and explore the environmental and physical properties of their member galaxies identified through JWST wide-field surveys within the CEERS, JADES, and PEARLS NEP-TDF fields. Combining with HST data, we identify 2948 robust $z>4.5$ candidates within an area of 185.4 arcmin$^2$. We determine nearest neighbour statistics and galaxy environments. We find that high-$z$ galaxies in overdense environments exhibit higher star formation activity compared to those in underdense regions. Galaxies in dense environments have a slightly increased SFR at a given mass compared with galaxies in the lower density environments. At the high mass end we also find a gradual flattening of the $M_{\star}$-SFR slope. We find that galaxies in high-density regions often have redder UV slopes than those in low-density regions, suggesting more dust extinction, weaker Lyman-alpha emission and/or a higher damped Lyman-alpha absorption. We also find that the mass–size relation remains consistent and statistically similar across all environments. Furthermore, we quantitatively assess the probability of a galaxy belonging to a protocluster candidate. In total, we identified 26 overdensities at $z=5-7$ and estimate their dark matter halo masses. We find that all proto-cluster candidates could evolve into clusters with $M_{\rm halo} > 10^{14}M_{\odot}$ at $z = 0$, thereby supporting the theoretical and simulation predictions of cluster formation. Notably, this marks an early search for protocluster candidates in JWST wide field based on photometric data, providing valuable candidates to study cosmic structure formation at the early stages.
\end{abstract}






\section{Introduction}

Galaxy clusters, along with their high-redshift progenitor structures, have played a prominent role throughout the history of cosmology. Many observational and theoretical studies show out these objects have been central to discussions on diverse cosmological models, structure formation scenarios, and the nature of dark matter (e.g. \citealt{Overzier2016,Chiang2017,Alberts2022}). As unique laboratories, they provide valuable insights into cosmic mass assembly, the intricate interplay between galaxy and structure formation, and the complex interactions between galaxies and the intergalactic medium. 

However, studying cluster formation using standard galaxy surveys is challenging due to their rarity and the need for extensive surveys to find their progenitors. To identify overdense regions in the early universe, researchers have employed a variety of strategies. In addition to surveys at X-ray and sub-mm wavelengths through the the Sunyaev-Zeldovich (S-Z) effect, the optical and NIR remain a powerful way to find clusters through the distribution of their galaxies. This includes conducting spectroscopic and photometric redshift surveys centered around rare and luminous objects believed to be associated with overdense structures. Numerous successful discoveries of overdense regions have been made using this approach. Notable examples include the studies around Quasi-Stellar Objects \citep[QSOs; e.g.][]{Morselli2014,QiongLi2023}, High-redshift radio galaxies (HzRGs; e.g., \citealt{Miley2008,Shimakawa2018,Wylezalek2013}), and sub-millimeter galaxies (SMGs, e.g. \citealt{Oteo2018,Harikane2019,Chapman2009,LiQiong2021,Umehata2015,Polletta2021}). Overall, these studies have provided insights into the formation and evolution of galaxy clusters and their precursor structures, significantly advancing our understanding of structure formation.

Protoclusters have been discovered in certain fields using techniques such as rest-frame UV spectroscopy and through the Lyman-alpha emitters/Lyman break galaxies (LAEs/LBGs) selection (e.g. \citealt{Lee2014,Ouchi2005,Shimakawa2018,Brinch2024,Toshikawa2016,Toshikawa2018,Steidel2005,Harikane2019,Dey2016}). To accelerate this process, pre-selection of overdense regions from photometric surveys has been employed, followed by spectroscopic confirmation of the most significant overdensities (e.g., \citealt{Chiang2015,Diener2015,WangTao2016,Cucciati2014,Lemaux2014}). Examples of successful protocluster surveys include the Subaru Deep Field ($z = 4.86$ and $z = 6.01$, \citealt{Shimasaku2003,Toshikawa2012}), the Subaru/XMM-Newton Deep Field ($z = 5.7$, \citealt{Ouchi2005}), and the Canada–France–Hawaii Telescope Legacy Survey \citep{Toshikawa2016}. 

The James Webb Space Telescope (JWST) stands as a transformative telescope for addressing this problem in new ways, offering a broad wavelength range, high sensitivity, and advanced imaging and spectroscopy \citep{Gardner2006,Rieke2023} particularly adept for finding distant cluster/groups and proto-clusters. JWST/NIRCam, with two neighboring 2$\times$2 arcmin detectors, despite its small size can likely efficiently identify substantial samples of high-redshift galaxy clusters and their progenitors, facilitating large-scale statistical studies of galaxy clustering in the early universe. Furthermore, JWST's deep observations extend the exploration of galaxies to unprecedented redshifts of perhaps $z\sim15$ \citep{Harikane2023}. Its high spatial resolution of $\sim$0.03" enables us to also resolve individual galaxies within dense environments, supplying detailed information about their morphology and spatial distribution, including discerning merger events and galaxy interactions.

Recently, there have been some notable successes in the search for galaxy protocluster candidates at $z\sim6$ utilizing the powerful capabilities of JWST. For instance, the deep imaging and wide-field slitless spectroscopy (WFSS) provided by JWST/NIRCam has already uncovered extensive galaxy overdensities around ultraluminous QSOs such as J0100+2802 at $z = 6.3$ \citep{Kashino2023} and J0305–3150 at $z = 6.6$ \citep{wangfg2023}. Additionally, noteworthy discoveries include the  identification of an overdensity around a well-known dusty star-forming galaxy (DSFG), HDF850.1, at $z = 5.2$ \citep{Sun2023,Herard-Demanche2023}, and in a blank region of the sky near the Hubble Ultra Deep Field (HUDF) at $z = 5.4$ \citep{Helton2023}. Furthermore, at higher redshifts, researchers have utilized the Near-Infrared Spectrograph (NIRSpec; \citealt{ferruit2022near}) to reveal the highest redshift spectroscopically confirmed protocluster to date at $z = 7.9$ \citep{Morishita2023}, a lensed protocluster candidate at $z = 7.66$ \citep{Laporte2022} and the identification of a candidate protocluster core surrounding GN-z11 at $z = 10.6$ \citep{Scholtz2023}.

In this paper, we use recent photometric surveys conducted by the JWST to probe the impact of environment on galaxy evolution at high redshifts. Here, we have selected regions with relatively contiguous observations by JWST, including CEERS, NEP, and JADES. The total surveyed area we study spans 185 arcmin$^2$. The main emphasis of this investigation lies in the discrimination between high-density and low-density environments. Furthermore, we investigate the physical properties of galaxies in different environments, including mass, morphology, and star formation activity. Additionally, our analysis includes the search for potential protocluster candidates and the quantitative assessment of the probability that a galaxy belongs to a specific galaxy cluster. We then discuss how the overdense areas relate to these proto-clusters.

The organizational structure of this paper is as follows: In \S\ref{sec:obs}, we provide an overview of the JWST data used in this study and the relevant HST data acquisition. \S\ref{sec:photo_measure} delves into the establishment of our high-redshift sample, discussing the reliability of photometric redshifts and discussing the completeness of our sample. In \S\ref{sec:over_under}, we categorize the sample into dense and under dense regions, comparing galaxy mass and star forming activity. We explore the impact of galaxy environments on the physical properties of galaxies. In \S\ref{sec:Group galaxy membership probabilities} we detail our search for galaxy clusters in early-universe using JWST and HST photometric data. We employ Bayesian methods to quantitatively assign group galaxy membership probabilities within these clusters. In \S\ref{sec: discussion}, we compare our findings with relevant spectroscopic studies at these redshifts, and discuss the process of cluster assembly by integrating simulations and previous observational data at lower redshifts. Finally, in \S\ref{sec:conclusions}, we summarize our findings and outline the challenges faced in our exploration of high-redshift galaxy protocluster candidates. Additionally, we discuss future plans and directions for further research in this field.

Throughout this paper, we assume a flat cosmological model with $\Omega_{\Lambda} = 0.7, \Omega_{m} = 0.3$ and $H_0 = 70 \mkms \, \rm Mpc^{-1}$. All magnitudes used in this paper are in the AB system \citep{Oke1983}.

\begin{figure*}
\centering
 \includegraphics[width=0.95\textwidth]{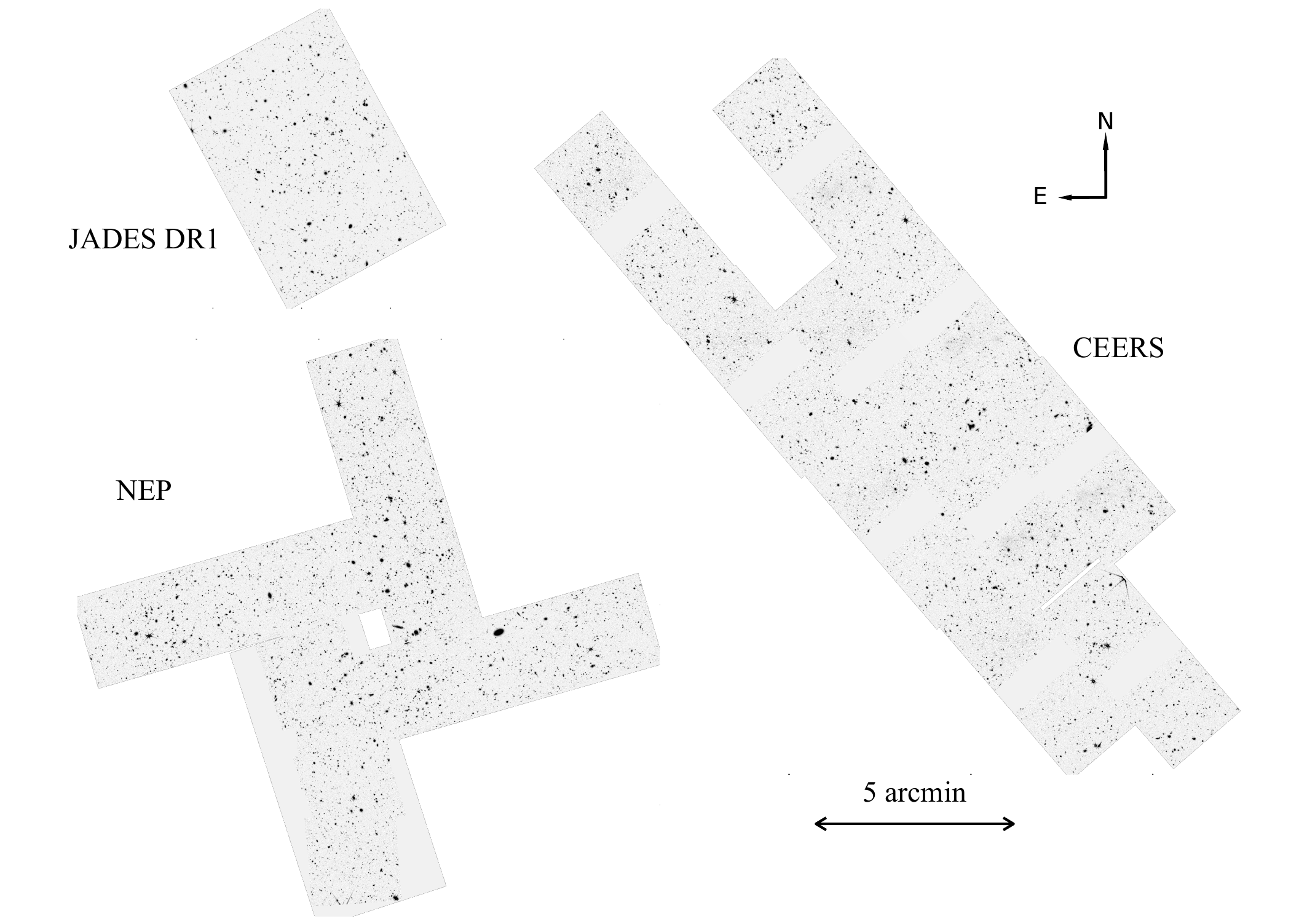} 
    \caption{NIRCam/F444W images of the three fields (CEERS, JADES, and NEP) analyzed in this work. We have created masks to avoid diffraction spikes from bright stars, image artifacts, and field edges. These masks are used throughout the analysis in this article. CEERS field has 10 NIRCam pointings and covers $\sim$94 arcmin$^2$ in total; JADES GOODS-S DR1 has 4 pointings, $\sim$27 arcmin$^2$ and NEP has 8 pointings, $\sim$64 arcmin$^2$.
    }
    \label{fig:plot_fields}
\end{figure*}


\section{Observations and Data Reductions}\label{sec:obs}

In this study, we selected data from continuous deep fields imaged by the NIRCam instrument of three JWST observation programs: JADES \citep{Rieke2023}, CEERS \citep{Bagley2023}, and PEARLS/NEP-TDF \citep{Windhorst2023}. These fields offer a contiguous area, enabling a more comprehensive investigation of the galactic environment at the Mpc scale compared to other existing JWST observations. Additionally, to study objects at lower redshifts ($z \sim 4.5 - 6$), we incorporated datasets from Hubble Space Telescope (HST) imaging. Specifically, the HST CANDELS survey \citep{Grogin2011,Koekemoer2011} includes the fields: GOODS-S \citep{2013ApJS..207...24G}, GOODS-N \citep{2023yCat..22430022B}, COSMOS \citep{2017ApJS..228....7N}, and EGS \citep{2017ApJS..229...32S}. The combined area of these fields reaches 185.4 arcmin$^2$. This section provides an overview of the observational details for each program.


\subsection{CEERS JWST NIRCam and HST Imaging}

We use data from JWST/NIRCam observing runs of the Cosmic Evolution Early Release Science Survey (CEERS; \citealt{Bagley2023}; ID: 1345, PI: S. Finkelstein) conducted in July 2022 and December 2022. This survey involves 10 NIRCam pointings that were mosaiced over the Extended Groth Strip field (EGS) area covered by CANDELS \citep{Grogin2011,Koekemoer2011}. JWST/NIRCam observations in this field span seven photometric bands: F115W, F150W, F200W, F277W, F356W, F410M, and F444W. In our JWST observations, CEERS field provided the largest single area, about 94.32 square arcminutes. 

We processed the CEERS JWST data utilizing the STScI JWST Pipeline\footnote{\url{https://github.com/spacetelescope/jwst}} v1.9.6  \citep{Bushouse2022} and CRDS\footnote{\url{https://jwst-crds.stsci.edu/}} using the calibration files v1084 independently ourselves with a bespoke reduction. The final resolution of the drizzled images is 0.03\arcsec/pixel. The average photometric depth is 28.6 AB magnitudes using 0.32 arcsec circular apertures. More detailed calibration process and data products can be found in our EPOCHS paper II \citep{Adams2023}. 

To complement our photometry for lower redshift systems, we also collect Hubble Space Telescope (HST) Advanced Camera for Surveys (ACS) images from the Cosmic Assembly Near-Infrared Deep Extragalactic Legacy Survey (CANDELS; \citealt{Grogin2011, Koekemoer2011}). The HST ACS observations we used include the F606W and F814W filters and were part of the CANDELS survey. Notably, the CEERS team has refined the astrometric alignment of these HST ACS observations to the Gaia Early Data Release 3 (EDR3). Detailed information regarding this can be found on the CEERS team's website \footnote{\url{https://ceers.github.io/hdr1.html}}.
The average 5$\sigma$ local depth is about 28.8-28.3 mag. The pixel scale of the processed data was 0.03".

The inclusion of these HST F606W and F814W observations in our dataset serves as a valuable complement, particularly in compensating for the absence of observations in the F090W filter within the CEERS survey.  This data also expands our sample's lower redshift range to $z\sim4.5$. However, it is important to note that redshift measurements for sources falling within the ranges of $7 < z < 7.5$ and $9 < z < 10$ have larger uncertainties due to these redshifts coinciding with the large transmission gaps between two NIRCam filters in CEERS.

\subsection{JADES JWST NIRCam and HST Imaging}
The JWST Advanced Deep Extragalactic Survey (JADES) NIRCam imaging observation \citep{Rieke2023} covers 25 arcmin$^2$ in and around the Hubble Ultra Deep Field (HUDF), and both the GOODS-S and GOODS-N fields. Here we use the JADES DR1 Release data of the GOODS-S field (PI: Eisenstein, N. Lützgendorf, ID:1180, 1210). There are data in nine filters in JADES: F090W, F115W, F150W, F200W, F277W, F335M, F356W, F410M, and F444W for these fields.

We process the images with the JWST Calibration Pipeline (version 1.9.6, CRDS v1084) with our custom steps, as described in EPOCHS II \citep{Adams2023}. We have considered various issues, including flat-field inaccuracies, cosmic ray, dark, distortion, bad pixel masks. This results in $5\sigma$ depths could deep to 29.8-29.6 mag in the aperture sizes of 0.32 arcsec. The final image resolution for this data is 0.03\arcsec/pixel. 

To to extend the wavelength coverage, we use imaging data from the CANDELS \citep{Grogin2011, Koekemoer2011}, which covers a broader area in GOODS-S. 
We use the Hubble Legacy Fields reduction\footnote{https://archive.stsci.edu/prepds/hlf/} for the ACS images in F435W, F606W, F775W, F814W, and F850LP, as well as WFC3 images in F105W, F125W, and F160W. It is noteworthy that we typically combine JWST NIRCam data with HST ACS data in our analyses. These images, derived from the CANDELS program, provide blue photometry for the JADES sources. 

\subsection{NEP-TDF NIRCam Imaging}
\label{sec: NEP NIRCam}
 The North Ecliptic Pole Time-Domain Fields (NEP-TDF) \citep{Windhorst2023} is part of the JWST PEARLS observational program. PEARLS, the Prime Extragalactic Areas for Reionization and Lensing Science project \citep{Diego2023,Frye2023,Windhorst2023}, is a JWST Guaranteed Time Observation (GTO) program (PID 1176, 2738, PI: Rogier Windhorst). The NEP-TDF observations involve imaging in eight filter bands: F090W, F115W, F150W, F200W, F277W, F356W, and F444W, along with one medium band: F410M. The total area for the four TDF epochs is 64 arcmin$^2$. We process the images with the JWST Calibration Pipeline (version 1.9.6, CRDS v1084) independently , as described in \citep{Adams2023}. The $5\sigma$ depths reach 28.7-28.5 mag in aperture sizes of 0.32 arcsec in diameter. The final image resolution is 0.03\arcsec/pixel. 

We obtained deep HST observations with WFC3/UVIS F275W and ACS/WFC F435W and F606W through GO programs 15278 (PI: R. Jansen) and 16252+16793 (PI: R. Jansen \& N. Grogin) from October 2017 to October 2022. Details of these observations will be presented in R. Jansen et al. (2024, in preparation).



\subsection{Source Extraction}
We carry out source identification and extraction using the code {\tt SExtractor} \citep{Bertin1996}. We run this in dual-image mode with the weighted stack of the F277W+F356W+F444W bands used for object selection, from which we carry out forced aperture photometry for multi-band measurements. This photometry is calculated within 0.32 arcsecond diameter circular apertures, and we include an aperture correction derived from simulated \texttt{WebbPSF} point spread functions for each band used \citep{Perrin2012,Perrin2014}. We use $5\sigma$ local depth as the error of the photometry to avoid the known underestimate of the photometric errors. We calculate local depths by placing apertures across the 'empty' regions of our images. We use the normalised mean absolute deviation of the nearest 200 apertures for each source. The details can be found in our EPOCHS paper I (Conselice in preparation) and paper II \citep{Adams2023}.

\section{Photometric redshifts and SED fitting}\label{sec:photo_measure}
In this section, we outline our methodology for fitting galaxy SEDs using \eazy\, and \bagpipes\, and selecting a high-redshift galaxy sample. We also discuss the differences between our photometric redshifts and the measured spectroscopic redshifts to ensure the accuracy of our photometric data. Additionally, we conduct a statistical analysis of number counts in each field and calculate both detection and selection completeness to assess the reliability of our dataset.

\subsection{SED Fitting}\label{sec: SED Fitting}

To obtain our photometric redshift measurements, we first use the EAZY-py \citep{brammer2008eazy} photometric redshift code \citep{brammer2008eazy}. We employ the blue templates described in \citet{larson2022spectral}, with a Chabrier initial mass function for our analyses, as detailed in \citet{Chabrier2003}. This is well suited for modelling galaxies up to $z>4$, as they are the same as templates and parameters as used in previous papers of the EPOCHS series (e.g. \citealt{Adams2023}). We used ten characteristic timescales ranging from $0.01<\tau<13$~Gyr, extending beyond the default template sets to include galaxies with bluer colors and stronger emission lines. We apply an intergalactic medium (IGM) attenuation derived from \citet{1995ApJ...441...18M}. These enhance the accuracy of redshift measurements for high-redshift samples. We consider the dust effects using the prescription of \citet{calzetti2000}. We allow for $E(B-V)$ values up to 3.5 to account for potential very dusty galaxies at these high redshifts \citep{Smail2023,Polletta2024} and to assess likely errors arising from low redshift contamination. We set a minimum error of 10\% to our measured photometry to take into account potential zero point offsets and inaccuracies in our theoretical templates.


To compute the probability density functions for redshift and stellar mass, we employ the photometric redshifts obtained with \eazy\ as input values for each galaxy candidate. These redshifts contain uncertainties, modeled using a Gaussian redshift prior, and input into the Bayesian SED fitting code Bayesian Analysis of Galaxies for Physical Inference and Parameter EStimation {\tt{Bagpipes}} \citep{carnall2018inferring}. {\tt{Bagpipes}} leverages a Bayesian approach and SED templates to infer the most likely properties of galaxies, including their star formation history, stellar mass, stellar population age, metallicities, among other properties. 

We utilize BC03 stellar population models \citep{Bruzual2003} with a Kroupa IMF \citep{kroupa2001MNRAS.322..231K} in our fits using {\tt{Bagpipes}}. The fitting process of \bagpipes\ we use the lognormal star formation history (SFH) parametrization, a well-established model used in various other studies \citep[e.g.][]{2023Natur.619..716C,2023MNRAS.519.5859W}. Additionally, we include the emission lines and nebular continuum based on {\tt CLOUDY} models \citep{2013RMxAA..49..137F}. We apply Log10 priors for dust, metallicity, and age, considering the expectation of young, less dusty, and lower metallicity galaxies at high redshifts. The fitting process incorporates the \cite{calzetti2000} dust attenuation model. We set prior ranges for metallicity in the range of $[10^{-6}, 10] \, \text{Z}_{\odot}$, dust prior in the range of $[0.0001, 10]$ in $A_\text{V}$, ionization parameter $\mathrm{Log}_{10}(\mathrm{U})$ in range of $[-3, -1]$, the time assumed for star formation to start at 0.001 Gyr, the time assumed for star formation to stop at $\tau_\text{U}$, with $\tau_\text{U}$ denoting the age of the Universe.

\subsection{A robust high-z galaxy sample}

In this section we list our selection criteria for locating $z>4.5$ galaxies. Our sample closely align with those utilized in other papers of the EPOCH series, including a detailed breakdown in previous papers \citep[e.g.,][]{Adams2024, Harvey2024}. In summary these criteria are: 
\begin{enumerate}
    \item The source must not be detected $\geq 3 \sigma$ in band(s) blueward of the Lyman break; it also must have $\geq 5\sigma$ detections in the 2 bands directly redward of the Lyman break, and $\geq 2\sigma$ detections in all other redward bands. This insures a strong detection of the Lyman-break for these systems, as well as a strong detection for measuring accurate redshfits and other galaxy properties based on the spectrum redward of the break. 
    \item We require that the integration of the photometric redshift place the galaxy with high certainity at high redshift.  The mathematical express for this is $\int^{1.10\times z_{\rm phot}}_{0.90\times z_{\rm phot}} \ P(z) \ dz \ \geq \ 0.6 $, which ensures the majority of the redshift PDF is located within the primary peak. Here $z_{\rm phot}$ is the maximum likelihood redshift from the PDF.
    \item We required a good fit such that $\chi^2_\text{reduced} < 3$ for a best-fitting SED to be classed as robust.
    \item We also require that the secondary photometric redshift solution $P(z_{\rm sec}) < 0.5 \times P(z_{\rm phot})$ to ensure the low probability of a secondary peak, if it exists, is less than 50\% of the high-z solution.
    \item We remove contamination sources by checking the morphology and SED fitting by eye, e.g. hot pixels, artefacts, blended features. 98\% are kept in our final sample.    
\end{enumerate}

High redshift galaxies with strong emission lines, such as $\rm H_{\beta}$+$\rm O_{III}$, may lead to additional flux in the corresponding photometric band at those wavelengths, resulting in elevated $\chi^2$ values. To mitigate this, we retain these galaxies in our sample if they have higher $\chi^2$ values ($<6$) but satisfy the other criteria. However, it is important to note that the fitting results for these galaxies require more careful manual checking. The median value of the photometric redshift uncertainty from SED fitting is $\Delta z \sim 0.14$.

By incorporating HST bands F606W and F814W using the Lyman break method, we can identify galaxies with $z>4.5$. However, due to the incomplete wavelength coverage across the full optical spectrum, our high-redshift sample may not be complete in the redshift direction. We will discuss the implications of this incompleteness for our work in \S\ref{sec: Detection and Selection Completeness}.

\subsection{Comparison between photometric and spectroscopic redshifts}

\begin{figure*}
\centering
 \includegraphics[width=0.95\textwidth]{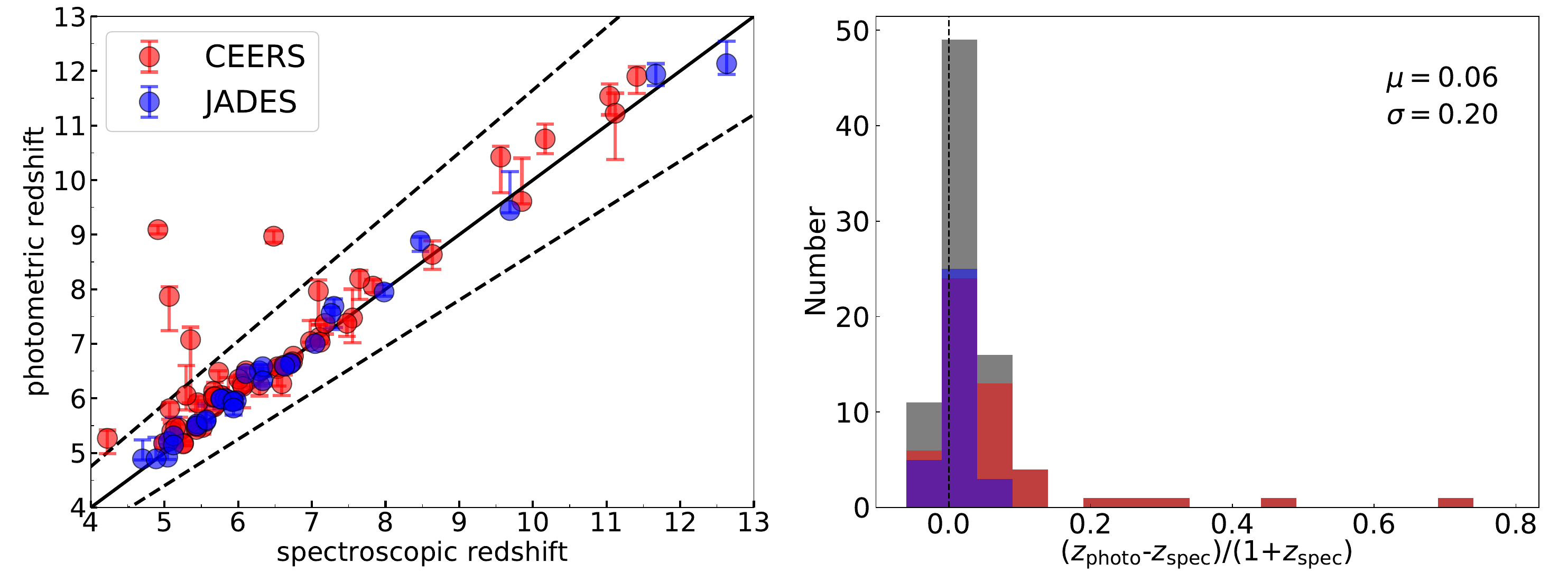} 
    \caption{Left: Diagnostic plot showing the comparison of spectroscopic redshifts with photometric redshifts for all the sample at $z>4$. The spectroscopic redshifts are from the observations of Subaru, VLT/MUSE, JWST/NIRISS and JWST/NIRSpec \citep{carnall2023surprising,Noirot2023}. The black solid line shows the one-to-one relation; the dashed lines show 15\,percent offsets in ($1+z$).  
    Right: the histogram of the relative difference between the photometric redshifts from our \bagpipes\, fits and the spectroscopic redshift in $(1+z_{\rm spec})$. The labelled scatter indicates the median of the relative difference, respectively. The error bars show the range of the 16th-84th percentiles.
    }
    \label{fig:spec_photo_z}
\end{figure*}

\begin{table}
\centering
\caption{Accuracy of our photometric redshift measurements for the JADES, CEERS, and combined data at $z>4.5$. The outlier fraction, $\eta$ quantifies the proportion of outliers in the redshift comparison. The NMAD, serves as a measure of the scatter in the redshift differences, adjusted for scale. The low values observed for both $\eta$ and NMAD across the datasets affirm the precision and accuracy of our photometric redshift measurements. $\mu$ and $\sigma$ represent the median and standard deviation, respectively, of the relative difference between the photometric redshifts obtained from our \eazy\ fits and the spectroscopic redshifts in $(1+z_{\rm spec})$.}

\label{tab:eta_NMAD}
\begin{tabular}{cccc}
\hline
Parameters   & JADES  & CEERS   & Combined  \\
\hline
$\eta$       & 2.9 \%   & 9.6 \%& 7.0 \%\\
NMAD         & 0.031  & 0.044   & 0.038  \\
$\mu$        &0.056   &0.064    &0.061 \\
$\sigma$     &0.273   &0.127    &0.198 \\
\hline
\end{tabular}
\end{table}

To investigate how well the photometric redshifts we have determined compare to the spectroscopic redshifts, we conducted a matching process involving all galaxies within our sample. There are 86 galaxies at $z>4.5$ with a reliable spectroscopic redshift measurement matched within the projected distance of $\Delta<0.3$\arcsec, 34 from JADES and 52 from CEERS. As described above in \S\ref{sec: SED Fitting}, the photometric redshifts were obtained via \eazy\,. For spectroscopic redshifts, within CEERS, we combine catalogues obtained from NIRSpec prism data of the publicly DAWN JWST Archive\footnote{https://dawn-cph.github.io/dja/}\citep{Heintz2023}, the DEEP2 survey \citep{Newman2013}, 3D-HST \citep{Momcheva2016} and MOSDEF/MOSFIRE surveys \citep{Kriek2015} of the Extended Goth Strip (EGS). Within JADES, we use JADES DR1 spectroscopy PRISM catalogue released from the JADES collaboration \citep{bunker2023jades}.

The results of our redshift comparison of 86 matched galaxies are represented in Figure~\ref{fig:spec_photo_z}. The photometric redshift data predominantly fall within 15\% of the spectroscopic redshifts. We assess photo-$z$ quality using two statistical measures, the outlier fraction \(\eta\), and the Normalized Median Absolute Deviation (NMAD). The fraction of outliers \(\eta\), defined as the fraction of the photometric redshifts differing from the spectroscopic redshifts by more than 15\% in $(1+z)$, ($|\Delta\,z|$/(1+spec-$z$)$>$0.15), is given by:
\begin{equation}
\eta = \frac{N_{115} + N_{85}}{N_{\text{total}}},
\label{eta}
\end{equation}
where \(N_{115}\) and \(N_{85}\) represent the counts of points lying above the line \(z_{\text{phot}} = 1.15 \times (z_{\text{spec}} + 1)\) and below the line \(z_{\text{phot}} = 0.85 \times (z_{\text{spec}} + 1)\), respectively.

The NMAD is the quantification of the dispersion in the redshift variances, normalized for scale, defined by the following expression:
\begin{equation}
\text{NMAD} = 1.48 \times \text{med}\left|\frac{z_{\text{spec}} - z_{\text{phot}}}{1 + z_{\text{spec}}}\right|.
\label{NMAD}
\end{equation}
All the statistical measurement values for these parameters are detailed in \autoref{tab:eta_NMAD}. The diagnostic of photometric versus spectroscopic redshift (Figure~\ref{fig:spec_photo_z}) shows that the fraction of outliers \(\eta\) is 7.0\% for the combined high-$z$ galaxies, and the normalized median absolute deviation of the whole sample is 0.038. This suggests a good agreement between our photometric and spectroscopic redshift measurements. The values of $\mu$ and $\sigma$ denote the average and standard deviation, respectively, of the relative difference distribution between the photometric redshifts derived from our \bagpipes\ fits and the spectroscopic redshifts in $(1+z_{\rm spec})$.  The proximity of $\mu$ values to 0, along with their low standard deviation, further underscores the reliability and accuracy of the photometric redshift measurements from our data. 

Our photometric redshift determination demonstrates their reliability. It is crucial to note that the spectroscopic observations of high-redshift objects are insufficient for mapping galaxy environments on scales of several Mpc, e.g. these detect only <$10$ kpc scale protocluster cores around GN-z11 \citep{Scholtz2023}. In contrast, \citet{Tacchella2023} previously identified a photometric candidate overdensity surrounding GN-z11 within a radius of $\sim5$ Mpc, corresponding to $\sim$430 pkpc at $z=10.6$. Our analysis, based on precise photometric redshift data, provides an effective approach for identifying protocluster galaxy candidates while minimizing selection bias. Subsequent redshift facilities, such as JWST or ALMA spectroscopic observations, will further validate whether these galaxies indeed reside within an interacting proto-cluster or a larger-scale structure.

\subsection{Number counts}\label{sec: number counts}

\begin{figure}
    \includegraphics[width=8.5cm]{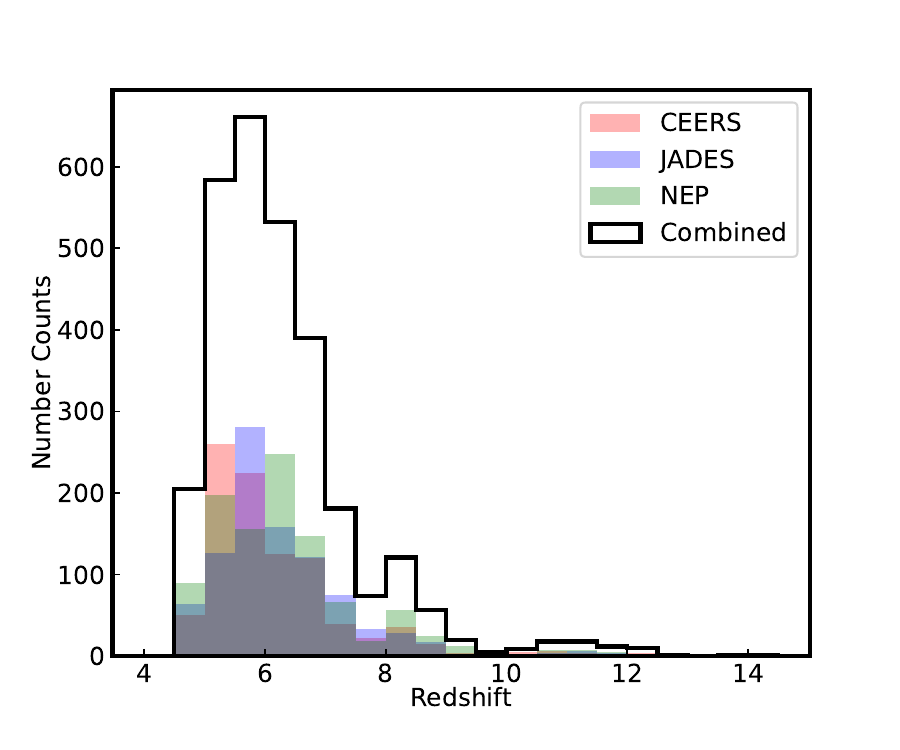}
    \caption{Redshift distribution of the high-$z$ galaxies in the CEERS, JADES, NEP fields, in bins of $z = 0.5$. The black histogram combines all the data we used in this paper. These samples are those galaxies with a high probability of existing at high redshifts \citep[][]{Adams2024, Harvey2024}.}   \label{fig:high-z sample}
\end{figure}

\begin{table}
\centering
\caption{The high-$z$ samples used in this work, with their main properties: the unmasked region area, the 5$\sigma$ depth in rest frame UV bands, the number counts at $z>4.5$. The unmasked region area refers to the area remaining after applying a mask, where we have excluded regions at the edges of observations, gaps, and portions of the data with noticeably poor quality. $C_{p\,90\%}$ is the 90\% selection completeness; $N$ is the total galaxy number at $z>4.5$ in each unmasked fields.}
\label{tab:feilds_counts}
\begin{tabular}{c|cccc}
\hline
Fields    & Sky area   & 5$\sigma$ depth $_{\rm UV_{rest}}$  &$C_{p\,90\%}$ & N  \\
    & arcmin$^2$ & mag & mag &  \\
\hline
CEERS      &94.32 &28.8-28.3 & 28.9 & 926\\
JADES      &26.88 &29.8-29.6 & 29.8 & 932\\
NEP        &64.21 &28.7-28.5 & 28.9 & 1090\\
Combined &185.43  &        &     &   2948\\
\hline
\end{tabular}
\end{table}

In Table~\ref{tab:feilds_counts}, we provide a summary of the galaxy catalogues in each field mentioned above, covering most of the protocluster footprint compared to the typical scale we expect protoclusters to subtend at these high redshifts. Figure~\ref{fig:high-z sample} displays the redshift distribution N$(z)$ for each catalogue, spanning a redshift range $z= 4.5-14.0$. However, due to issues related to the gaps in the photometry bands, the galaxy density is underestimated at certain redshifts (e.g., $z\sim9.5$). This results in a potential underestimation of density for high-density regions already identified. To address this, we implement a simple redshift mask as follows:
\begin{equation}
    \mathcal{M}^z(z) = 
    \begin{cases}
    0 & \text{if } z[z_{\rm i, lower,68\%}, z_{\rm i, upper,68\%}] \in z_\text{masked} \\ 
    1 & \text{otherwise}
    \end{cases}.
\end{equation}
where $z_{\rm i, lower,16\%}$, $z_{\rm i, upper,68\%}$ are the lower and upper limit of 68\% confidence interval on PDF($z$) respectively. $z_\text{masked}$ is the redshift masked region. It is important to note that this effect is attributed to the limitations arising from the discontinuous redshift of photometric data.

We limit the sample to $z < 9.5$, by masking galaxies whose 68\% integral of the $P(z)$ falls at $z>9.5$, totalling 128 galaxies (3.9\%). These galaxies are distributed over a wide redshift range from $9.5 < z < 16$. Due to limitations in observational depth, the number of galaxies in these redshift bins is too small to conduct a meaningful analysis of the galaxy environment. In the future, observations from even deeper JWST fields can extend this work to higher redshifts. Here we set $z_\text{masked}$ as [9.5, $\infty$].

\subsection{Detection and Selection Completeness}\label{sec: Detection and Selection Completeness}

\begin{figure*}
    \centering
    \includegraphics[width=0.95\textwidth]{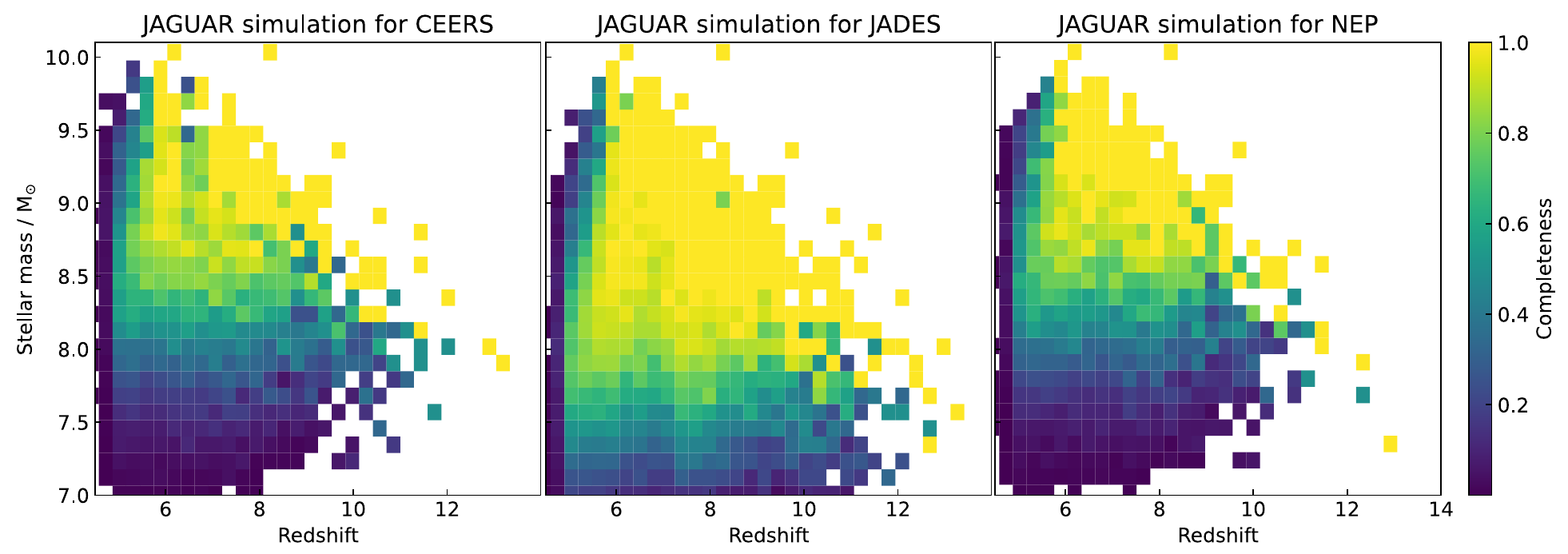}
    \caption{The stellar masses and redshifts of simulated \jaguar\, galaxies in in all three fields (CEERS, JADES, NEP), shown with their completeness values. We include only the brighter galaxies with completeness $\mathcal{CP}(z, M_*)>20\%$ in our calculations. White region show where no galaxies exist in the simulation.}
    \label{fig:figure_completeness}
\end{figure*}

\begin{table}
\centering
\begin{tabular}{c|ccc}
\hline
\textbf{redshift bins} & \textbf{>25\%} & \textbf{>50\%} & \textbf{>75\%}\\
\hline
 4-5  & -18.87 & -18.50  & -18.21 \\
 5-6  & -19.26 & -18.73 & -18.33 \\
 6-7  & -19.50  & -18.95 & -18.54 \\
 7-8  & -19.63 & -19.12 & -18.75 \\
 8-9  & -19.53 & -19.09 & -18.55 \\
 9-10 & -19.78 & -19.31 & -18.85 \\
\hline
\end{tabular}
\caption{An example of detection completeness in CEERS. We list the corresponding $M_{\rm UV}$ values in different redshift bins for completeness of 25\%, 50\% and 75\%. The completeness for the other fields are described in more detail in \citet{Adams2024} and \citet{Harvey2024}.}
\label{tab:completeness}
\end{table}

In this section, we briefly describe how we calculate the completeness and selection completeness measures and their impact on our results. We use the same methods with the EPOCHS II and IV papers, for more detailed methods, please refer to these papers (\citealt{Adams2023,Harvey2024}). 

To assess JWST data detection completeness, we simulated galaxies with absolute magnitudes randomly ranging from -16 to -24 mag in the F444W band. Using Monte Carlo methods, we randomly placed 1000 mock points on source-masked images to simulate the presence of detected sources. We applied this progress to all fields in our study. Applying SExtractor to these images with the same detected parameters and selection criteria, our detection completeness surpasses 90\% across redshift and magnitude ranges for our high-$z$ sample. The overall completeness is primarily influenced by our selection criteria. The results indicate that the detected sources are generally reliable.

For the selection completeness, we used \jaguar\,\footnote{https://fenrir.as.arizona.edu/jwstmock/} version 1.2 \citep{2018ApJS..236...33W}, a synthetic galaxy catalog at $z \leq 10$ \citep{2014ApJ...783...85T, Bouwens2015, stefanon2017b} based on observational constraints. We match the $10^6$ mock sources to the sources which are detectable given the depths of our observations. Realistic observations were simulated by perturbing catalog photometry based on 1$\sigma$ local depth errors. Our same selection procedure is then applied to assess redshift recovery. Using \bagpipes, we tested the reconstruction of stellar masses for \jaguar\, galaxies. Completeness and contamination are estimated for each redshift and mass bin, revealing a strong dependence on galaxy stellar mass. For masses above $\approx 10^8 M_{\odot}$, we achieved $\geq 50\%$ completeness in most areas. Figure~\ref{fig:figure_completeness} shows an example of the selection completeness results in CEERS.

\section{Galaxy local environment}\label{sec:over_under}
To study the local environments of high-$z$ galaxies, we utilized the k-dimensional tree (KDTree) data structure to search for nearest neighbors. We investigate the properties of galaxies as a function of their local environment (number of neighbours / local density), and compare this to the physical parameters exhibited by galaxies residing in overdense and underdense regions. Through this analysis, we aim to not only discern the existing disparities but also to shed light on the potential redshift evolution characterizing the galaxy environment. By examining these aspects, we can gain insights into the complex interplay and evolution between the galaxy and its surroundings across cosmic time.

\subsection{Galaxies in underdense and overdense environments}\label{sec: Galaxy density estimates}

To ensure a robust evaluation of the galaxy environment, accounting for variations in the selected surrounding area, we employ the nearest neighbor method to measure the local environment of the galaxy, and select possible galaxy group members \citep{Lopes2016}. 

Specifically, we utilize a k-dimensional tree (KDTree) data structure to facilitate nearest neighbor queries. For each galaxy in our dataset, we calculate its projected distance, denoted as $d_n$, to the $n$th nearest neighbouring galaxy, considering a maximum velocity offset relative to the velocity of the galaxy under investigation. The velocity offset mask is defined as:

\begin{equation}
    \mathcal{M}^v(z) = 
    \begin{cases}
    1 & \text{if } {\Delta z\,<\,0.1\ }  \\ 
    0 & \text{otherwise}
    \end{cases}.
\end{equation}
Here, we chose galaxies within a $\Delta z<0.1$ redshift cube, in order to better match our photometric redshift uncertainty constraints. Additionally, according to the predictions by \citet{Chiang2013} and \citet{Muldrew2015}, the effective diameter of protoclusters, defined as 2 $\times$ the effective radius (2Re), is about 20 cMpc at $z = 5$. Hence, selecting galaxies with $\Delta z<0.1$ conveniently covers the entire overdensity region.

The local galaxy density, $\Sigma_n (z)$ is denoted as defined as: 
\begin{equation}
\Sigma_n (z) = n/\pi d_n^2
\end{equation}
where $d_n$ is the projected distance in Mpc of the $n$th neighbour, $\Sigma_n$ is expressed in units of galaxies/Mpc$^2$. The choice of the parameter $n$, representing the rank of the density-defining neighbor, is crucial. It is imperative to ensure that $n$ is not greater than the number of galaxies within the halo, as this could lead to a loss of sensitivity in our environmental analysis. Consequently, we adopt the \sfive\,, local galaxy density estimator with $n=5$. This choice is often preferred in the literature, as it tends to be smaller than the number of galaxies typically found in a cluster, ensuring a robust and commonly used estimate (e.g., \citealt{Lopes2016,Santos2014}).

\begin{figure}
    \includegraphics[width=8.5cm]{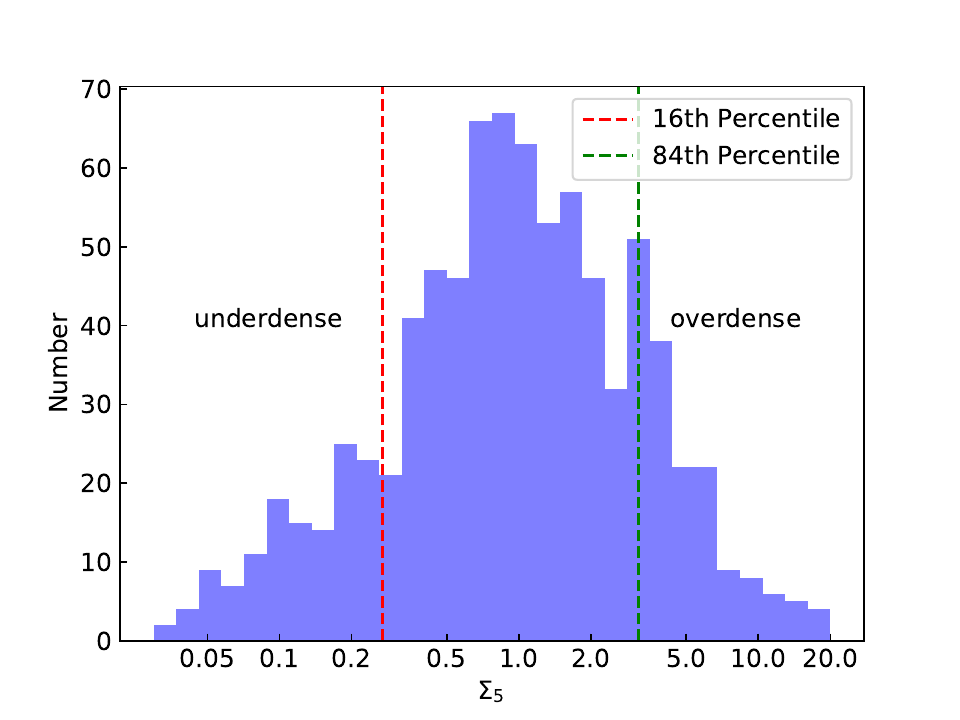}
    \caption{Selection of overdense and underdense galaxies using \sfive\,. The green and red dashed lines are the 84\% and 16\% confidence interval on \sfive\, at given redshift bin respectively. Within each redshift bin, we classify galaxies in the distribution of \sfive\, as `overdense galaxies' if \sfive\, fall within the > 84\% and as `underdense galaxies' if they fall within the < 16\%, refer to Section~\ref{sec: over under galaxies}. This plot has not been corrected for any observational selection effects or incompleteness corrections.}
    \label{fig:under_vs_dense}
\end{figure}


\subsubsection{Completeness Due to Image Borders}\label{sec: Completeness Due to Image Borders}
For galaxies located in the proximity to image boundaries or masked regions (e.g., bright stars), there is a tendency to underestimate their $\Sigma_5$. While this impact may not significantly alter the classification of galaxies in dense regions, it can have a substantial effect on galaxies situated in underdense areas, necessitating careful consideration. 

The simplest approach is to exclude galaxies significantly affected by the boundaries. For each galaxy, we calculate its distance to the boundary, denoted as $d_{\text{edge}}$. We define a mask for galaxies affected by the boundary as: 
\begin{equation}
    \mathcal{M}_\text{edge} = 
    \begin{cases}
    0 & \text{if } \Sigma_{5} \leq \Sigma_{5, 16\%} \text{ and } d_\text{edge}<1{\rm cMpc}  \\ 
    1 & \text{otherwise}
    \end{cases}.
\end{equation}
where $d_5$ is the distance to the fifth-nearest galaxy from the central galaxy. For galaxies already situated within dense regions, adjusting for boundary effects would only further accentuate their overdensity, thus having no impact on the classification of overdense galaxies. Therefore, we are solely addressing boundary effects for underdense galaxies. We apply the mask to galaxies in the 16 percentile lower density regions, that is the galaxies in the most underdense regions. We do not apply this correction for the highest density systems given that these are very compact and would only increase the over densities of these regions.


\subsubsection{Completeness Cut from JAGUAR simulation}\label{sec: Completeness Cut from JAGUAR simulation}

The galaxy number counts in our study are flux-limited and affected by the depths of the observations. Thus, density estimations are primarily derived for galaxies falling within a specific luminosity range. In this section, we employ a set of simulated \jaguar\, galaxies to assess the completeness of CEERS, JADES and NEP, and subsequently remove faint galaxies from our density estimates. \jaguar\ is a simulation of the spectral energy distributions of galaxies as observed with JWST. These mock catalogs can then be used to determine how well we are able to retrieve galaxies of various types and at various redshifts. 

The photometric selection completeness, denoted as $\mathcal{CP}(z, M_*)$, is determined as the selection proportion of JAGUAR simulated galaxies, which is a function of redshift and stellar mass as shown in Figure~\ref{fig:figure_completeness}. In this mock selection process, we introduce photometric noise to the measurements of each JAGUAR galaxy to mimic our observations. The noise levels set to match the actual median CEERS/NIRCam F444W depths, which are 29.8-28.3 magnitudes deep based on observations in Table~\ref{tab:feilds_counts}. However, we also apply the depths of the similar fields to determine how these findings might change. 

Our analysis reveals that for galaxies with apparent rest-UV magnitudes of $M_{\rm UV}$ = -18.50, -18.95, -19.31 at $z=$ 4, 6, and 9, over 50 percent of them are accurately selected to fall within the correct photometric redshift range. 
The \jaguar\, simulation data corroborates that by restricting our analysis to galaxies brighter than these specified magnitudes, we effectively control the contamination fraction for low-redshift galaxies, maintaining it at reasonably low values ($<$15 percent). As a result, we choose to include only the brightest galaxies with completeness exceeding 20 percent in our density calculations. We thus define a selection completeness mask as:
\begin{equation}
    \mathcal{M}_\text{CP} = 
    \begin{cases}
    0 & \text{if } \mathcal{CP}(z, M_*)<20\%  \\ 
    1 & \text{otherwise}
    \end{cases}.
\end{equation}
The completeness mask we selected here is a compromise for subsequent analysis. We aimed to correct for incompleteness while ensuring sufficient data for statistical analysis. We also experimented with a $\mathcal{CP}(z, M_*)<$50\% mask, finding that the overall conclusions remained unchanged. However, the differences are diluted due to increased statistical uncertainty.

\subsubsection{Selection of overdense and underdense galaxies}\label{sec: over under galaxies}
The measurements of \sfive\, are influenced by variations in galaxy counts at different redshifts. When galaxies are scarce, it leads to a lower estimate of \sfive\,. The number of galaxies is limited by the flux detection limits and selection biases. At highest redshifts, typically, only the bright end of the luminosity function is sampled, resulting in the detection of only a small number of galaxies. Another limitation arises from the presence of transmission gaps between the filters, which leads to the omission of galaxies with redshift tracers (e.g., strong emission lines) falling into the gaps. 

To mitigate the influence of redshift on the galaxy environment, we have divided galaxies into different redshift bins from $z=[4.5,10.0]$ with the bins having a width of 0.5 that are chosen large enough to compensate against biases in certain redshifts.
Within each redshift bin, we classify galaxies in the distribution of \sfive\, as `overdense galaxies' if \sfive\, is in the > 84\% percentile of overdensities, and as `underdense galaxies' if in the < 16\% percentile, given by:
\begin{equation}     
    \begin{cases}
    \Sigma_{5,i} \geq \Sigma_{5, \ 84\%}(z_\text{bin}) & \text{overdense galaxies}  \\ 
    \Sigma_{5,i} \leq \Sigma_{5, \ 16\%}(z_\text{bin}) & \text{underdense galaxies}
    \end{cases}.
    \label{eq: over dense galaxies}
\end{equation}
where $\Sigma_{5, 84\%}(z_\text{bin})$ and $\Sigma_{5, 16\%}(z_\text{bin})$ are the 84\% and 16\% confidence interval on \sfive\, at a given redshift bin respectively.

Here we consider several key factors influencing this measurement, and we apply corresponding masks to remove certain biases, which are as follows: 
\begin{itemize}
    \item $\mathcal{M}^{z}(z)$: the mask of redshift selection bias, described in \S\ref{sec: number counts}.
    \item $\mathcal{M}^{v}(z)$: the mask of the velocity offset (redshift offset), which restricts galaxies that are too distant along the line of sight to avoid contamination from foreground/background objects, described in \S\ref{sec: Galaxy density estimates}.
    \item $\mathcal{M}^\text{edge}(z)$: the mask for the galaxies close to image boundaries or masked regions, see \S\ref{sec: Completeness Due to Image Borders}.
    \item $\mathcal{M}^\text{CP}(z, M_\star)$: the mask of the selection completeness, described in \S\ref{sec: Completeness Cut from JAGUAR simulation}.
\end{itemize}

\noindent Finally, we calculate the local galaxy density $\Sigma_5(z,M_\star)$ for each galaxy at a given redshift and stellar mass. 
\begin{equation}
    \Sigma_{5,i}(z, M_\star) = \Sigma_{5,i} \times \mathcal{M}^{z}(z) \times \mathcal{M}^{v}(z) \times \mathcal{M}^\text{edge}(z) \times \mathcal{M}^\text{CP}(z, M_\star).
    \label{eq: Sigma_5}
\end{equation}

Following the classification criteria in Equation~\ref{eq: over dense galaxies}, we select 1006 (34\%) `overdense galaxies' in overdense regions, and `underdense galaxies' in underdense regions combined from all redshift bins from $z=4$ to 10, which we define as the parent sample. After applying the masking criteria (Eq.~\ref{eq: Sigma_5}, we end up with 376 overdense galaxies and 199 underdense galaxies in our final catalog across the CEERS, JADES, and NEP fields.

\begin{figure}
\centering
 \includegraphics[width=8.5cm]{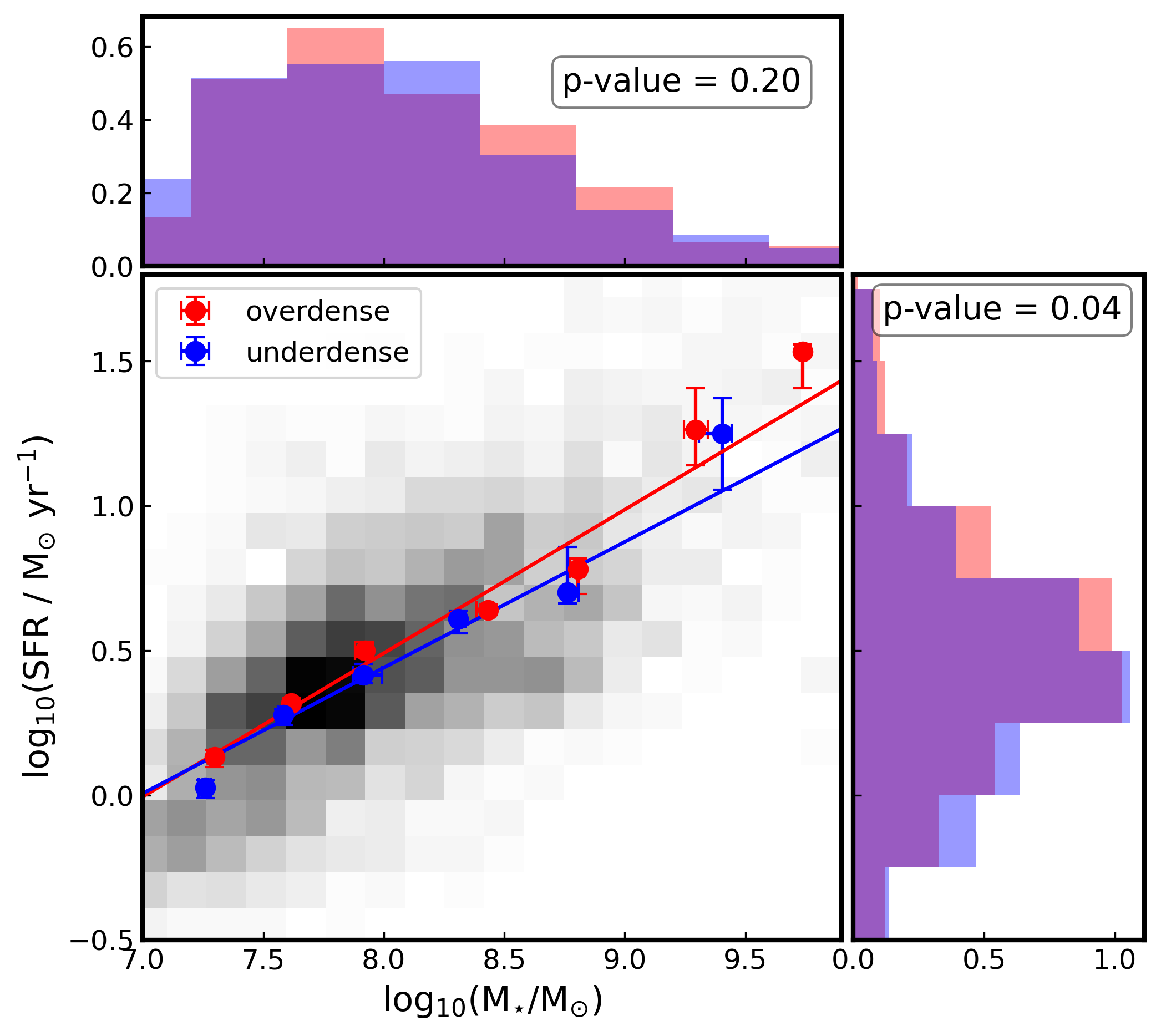} 
    \caption{The relationship between the SFR and stellar mass for our whole sample at $z=4.5-10$. The grayscale background on the (SFR, $M_{\star}$) plane represents the source density on a linear scale. The red and blue represent the overdense and underdense galaxies, respectively. The circles are the median values and the errors in the stellar mass bins are derived through a bootstrapping method. The solid colored lines are the best linear least-squares regression fits, with the parameters shown in the Table~\ref{tab:sfr}. The top and right panels show the probability density of each stellar mass and SFR bin, respectively. The $p$ value of K-S test for SFR is 0.04, showing different distributions; while $p$ value for stellar mass is 0.20, showing that these are probably not significantly different. 
    }
    \label{fig:m_sfr}
\end{figure}

{}


\begin{figure*}
\centering
 \includegraphics[width=0.9\textwidth]{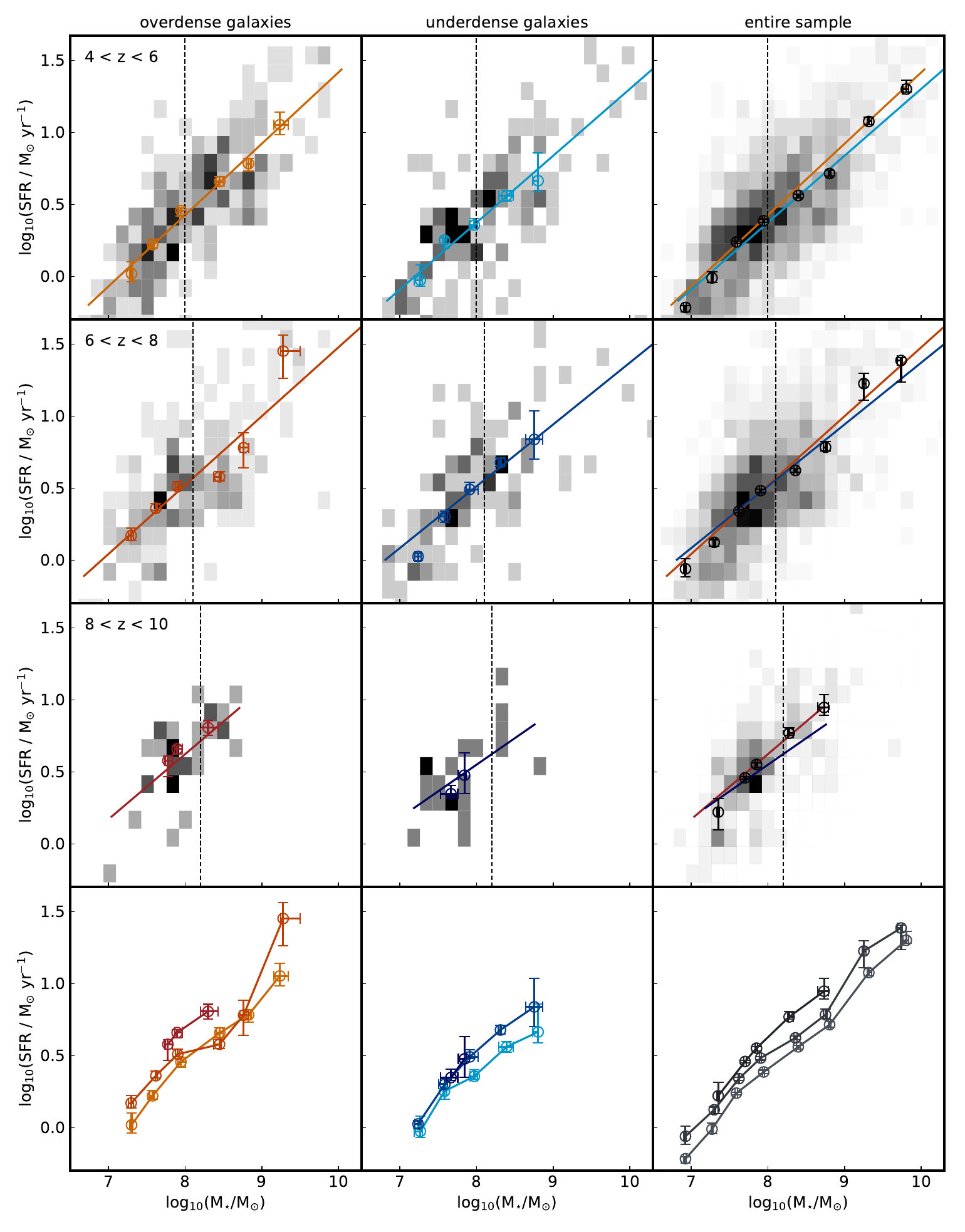} 
    \caption{The relationship between log$_{10}$SFR and log$_{10}$M$_{\star}$ in different redshift bins. The grayscale background on the (SFR, $M_{\star}$) plane represents the source density on a linear scale.  The open boxes are the median values in the stellar mass bins, with error bars determined through a bootstrapping method. The solid colored lines are the best linear fits. The dashed vertical lines are the observational mass completeness limit for the sample at the central redshift in each bin. The bottom panel: the median values and main sequence relations are displayed for all redshift bins, with color-coding corresponding to each bin.
    }
    \label{fig:m_sfr_z}
\end{figure*}

\subsection{SFR and stellar mass}
We use \bagpipes\, to perform Bayesian SED fitting for each individual galaxy and derive its stellar mass, SFR and sSFR, see \S \ref{sec:photo_measure} and \cite{Harvey2024}.  Additionally, \bagpipes\ provides the probability distribution for each parameter of interest for every galaxy in the analysis.

However, \bagpipes\ gives rise to issues with a maximum sSFR upper limit. This is a consequence of the timescale on which \bagpipes\  measures star formation rates, which by default is approximately 100 Myr. We attempted to address this issue by changing the timescale on which \bagpipes\, calculates the SFR, with ages = [10, 50, 100] million years to provide more flexibility. However, setting the timescale too small is impractical, as all galaxies would need to rapidly form within this timescale, resembling a `burst,' which is unreasonable.

We also derived SFRs from observed UV rest-frame photometry by employing a power-law approximation to fit the photometric points within the 1250–3000 Å spectral range. We applied the \citet{Meurer1999} relation to estimate the extinction necessary for correcting the 1600 Å luminosity. Subsequently, the dust-corrected luminosity is transformed into an SFR estimate using the factor in \citealt{kennicutt2012star}. This approach mitigates the degeneracy of SED parameters.

Figure~\ref{fig:m_sfr} shows the relationship between SFR and stellar mass for both overdense and underdense galaxies. The histograms of stellar mass and SFR are displayed in the top and right panels, respectively. We use the bootstrapping method to determine the median values and errors represented by the data points in the plot. We apply a linear fit to the relation from the full sample and in three separate redshift bins, and the resulting parameters are presented in Table~\ref{tab:sfr}. While the slope of overdense galaxies appears slightly steeper, the difference is not significant within the error bars. We conducted a Kolmogorov-Smirnov (K-S) test, yielding a $p$-value of 0.04 for SFR, indicating distinct distributions. However, the $p$-value for stellar mass is 0.20, suggesting no statistically significant difference in the stellar mass distribution between overdense and underdense galaxies. These findings suggest that overdense galaxies exhibit more active star-forming activity and tend to be more massive, although the latter trend is less pronounced.

Figure~\ref{fig:m_sfr_z} shows the relation between SFR and stellar mass across various redshift bins in our sample, ranging from $z=4$ to $z=10$. The dataset undergoes a 2$\sigma$ clipping process to exclude outliers, and within each stellar mass bin, we compute the median and its corresponding error using the bootstrap method. We fit the data with the linear relation:
\begin{equation}\label{eq: sfr-m}
    \text{log (SFR)} = \alpha \text{log} (\text{$M_{\star}$}) + \beta
\end{equation}
The best fits are shown in the colored lines in Figure~\ref{fig:m_sfr_z} and Table~\ref{tab:sfr}.

\begin{table}
\centering
\begin{tabular}{c|cc}
\hline
\hline
\textbf{redshift bins} & $\alpha$ & $\beta$ \\
\hline
  & overdense galaxies &\\
\hline
 4-6   &0.50$\pm$0.02 & -3.54$\pm$0.17 \\
 6-8   &0.48 $\pm$0.03 & -3.30$\pm$0.22 \\
 8-10  &0.45 $\pm$0.10 & -3.00$\pm$0.79  \\
 all   &0.49 $\pm$0.01 & -3.39$\pm$0.11 \\
\hline
  & underdense galaxies &\\
\hline
 4-6   &0.46 $\pm$0.03 & -3.34$\pm$0.24 \\
 6-8   &0.43 $\pm$0.04 & -2.92$\pm$0.32 \\
 8-10  &0.37$\pm$0.12 & -2.40$\pm$0.96  \\
 all   &0.43$\pm$0.02 & -3.00$\pm$0.17 \\
\hline
\end{tabular}
\caption{The best fits of the slope ($\alpha$) and intercept ($\beta$) from the linear least-squares regression (see Eq.~\ref{eq: sfr-m}) for the relationship between between SFR and stellar mass across various redshift bins, ranging from $z=4$ to $z=10$.}
\label{tab:sfr}
\end{table}

The lower panel of Figure~\ref{fig:m_sfr_z} displays the median SFRs within various stellar mass bins at different redshifts. Our findings reveal that galaxies at higher redshifts demonstrate elevated levels of star formation activity when compared to their lower-redshift counterparts, as has long been known. We observe a subtle rise with increasing redshift and a gentle flattening of the slope. It is important to note that the highest redshift bin has limitations and is more susceptible to poor number statistics. Overall, we observe this trend extended up to $z=10$, which is rarely probed in previous studies (e.g. \citealt{Tasca2015,Tomczak2016,Santini2017,Magnelli2014,Schreiber2015,Whitaker2014,Popesso2023}).

The left and the middle panels show these relationships for overdense and underdense galaxies. Notably, overdense galaxies consistently exhibit higher SFRs across the entire redshift range studied. Furthermore, the overdense galaxies display around $>1\sigma$ slightly steeper slope (Table~\ref{tab:sfr}), suggesting that the environment may indeed influence the stellar formation activity of galaxies, although this impact does not appear to be very pronounced, also see Figure~\ref{fig:m_sfr}.


\subsection{Star formation activities and quenching}

\begin{figure}
    \includegraphics[width=8.5cm]{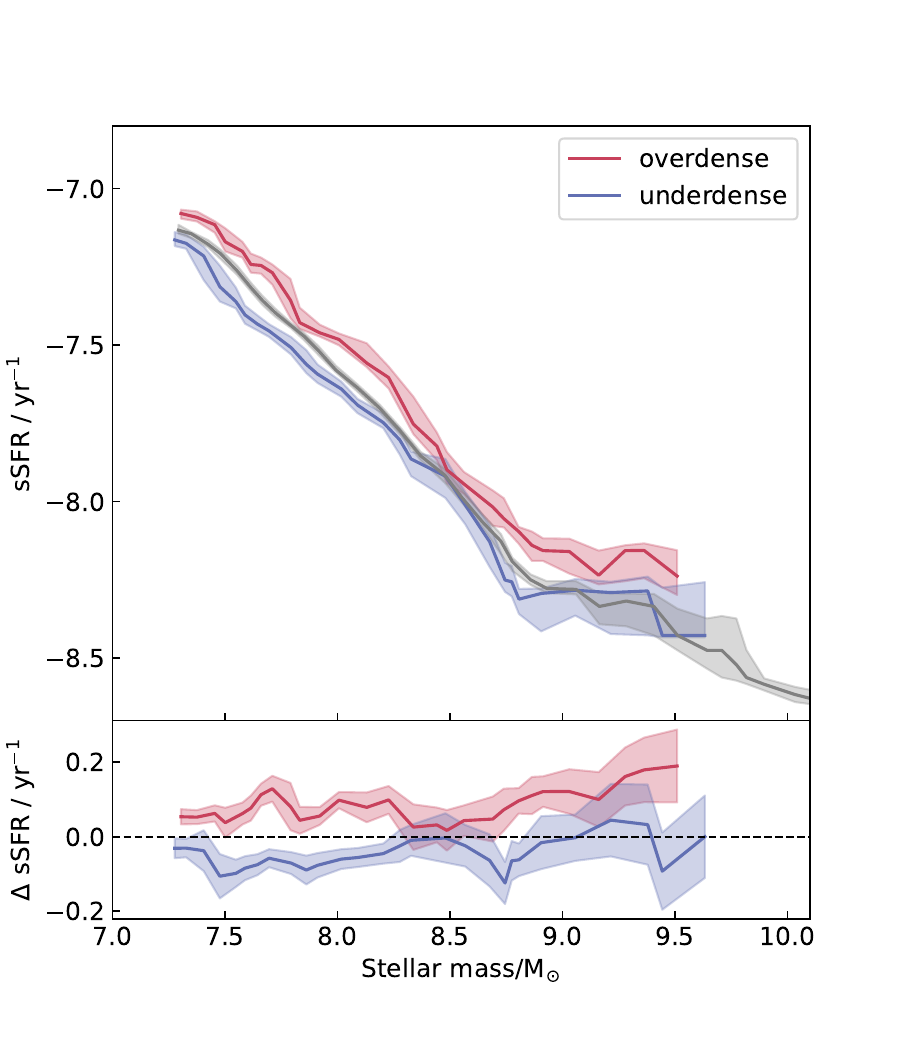}
    \caption{The sSFR of the overdense and underdense galaxies as a function of stellar mass at $z=4-10$. The red and blue represent the overdense and underdense galaxies, respectively. The grey line is for the whole sample. The value of the differences in sSFR ($\Delta$sSFR) is shown in the bottom panel, which is calculated as $\Delta$sSFR = (sSFR$_{\rm over | under}-$<sSFR>). Error bars are calculated with the bootstrap method.}
    \label{fig:M_ssfr_over_under}
\end{figure}

\begin{figure}
    \includegraphics[width=8.5cm]{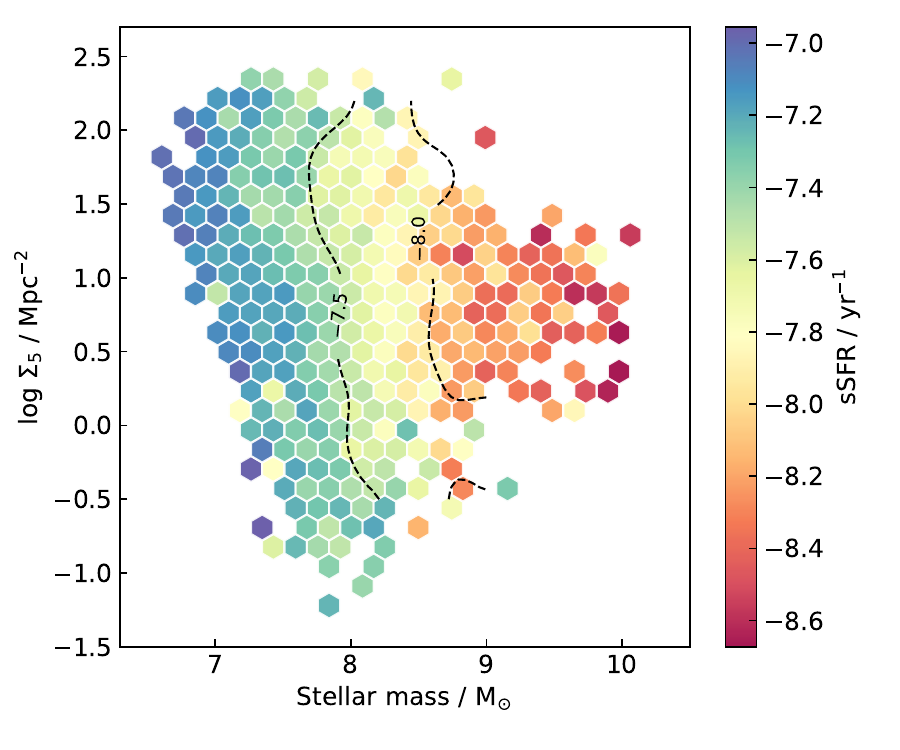}
    \caption{The sSFR of all $z=4-10$ galaxies on the \sfive\, - stellar mass plane. The colour of the various bins show the value of the sSFR in that region of this parameter space. The black dashed lines show the locations of the sSFR = [-8.0,-7.5] contour lines. From this there is a clear indication that stellar mass is the driving mechanism for producing galaxies with low sSFR values.}
    \label{fig:M_s5}
\end{figure}

Cluster galaxies are often associated with quiescence at low redshift, but at high-$z$ a significant portion of systems in clusters/groups continue to display active star-forming behaviour. In our sample, the specific star formation rates (sSFRs) of all selected galaxies fall within the range [log(sSFR/yr$^{-1}) = -9.5, -7$], while their stellar masses span the range log($M_\star/M_{\odot}$) = [6.5, 10.6]. Clearly, our sample only comprises star-forming galaxies. Within the redshift range of $z=$[4.5, 10] in the CEERS field, we do not detect any massive galaxies ($M_\star > 10^{10} M_{\odot}$) exhibiting quiescent behavior (sSFR $< 10^{-11.8}$ yr$^{-1}$, \citealt{Salim2014}). This finding aligns with predictions from models of cosmic evolution \citep{Somerville2015}. During this epoch, which is near the time of the epoch of reionization, when the universe was only less than 1-2 Gyr, galaxies both within galaxy overdensities and in the field were highly active in terms of star formation. 

Figure~\ref{fig:M_ssfr_over_under} displays the sSFR of the overdense and underdense galaxies as a function of stellar mass.
Remarkably, from Figure~\ref{fig:M_ssfr_over_under} bottom panel, $\Delta$sSFR = sSFR$_{\rm over | under}-$<sSFR>, we find that overdense galaxies exhibited significantly higher sSFRs, exceeding those of field galaxies by $\sim 0.13$ dex at all stellar masses. The median sSFR of overdense galaxies is 3$\sigma$ higher than that of underdense galaxies using MCMC method. Statistical tests, such as the Kolmogorov-Smirnov (KS) test, confirm the differences between these populations, with a p-value of 0.01, showing a highly statistically significant. 


During cosmic youth when the universe was less populated by massive galaxies, likely nearly all galaxies are star forming galaxies and the sSFRs are well above the quenching value. Figure~\ref{fig:M_s5} shows the sSFR of all high-$z$ galaxies in CEERS on the \sfive\, - stellar mass plane. We find that the mass factor is the one dominating whether a system is star forming or not. The quenching mechanisms related to environmental factors do not seem as pronounced at these redshifts. In contrast, cluster galaxies exhibited higher sSFRs, due to the perhaps higher mass accretion rate of gas which is likely present at these epochs \citep{Overzier2016}. The gravitational mass assembly of baryonic matter within some massive dark matter halos seems to enhance star formation activity among overdense galaxies. For example, \citet{Chiang2017} explored cluster evolution using N-body simulations and semi-analytic models. They point out that from $z \sim 5$ to $z \sim 1.5$, there exists a prolonged star formation phase responsible for about 65\% of the final stellar mass in present-day clusters. Protoclusters might harbor massive cores by the end of this epoch, marking the onset of widespread galaxy quenching and the build up of the intracluster medium (ICM). Additionally, several studies at high redshifts also indicate that star formation in dense environments is enhanced compared to field galaxies, attributed to the increased rates of gas accretion and mergers \citep{Alberts2014,Monson2021,Lemaux2022,Perez-Martinez2024}. However, we can show that this increase continues to increase at the highest redshifts where these proto-clusters can be identified.

\begin{figure}
    \includegraphics[width=8.5cm]{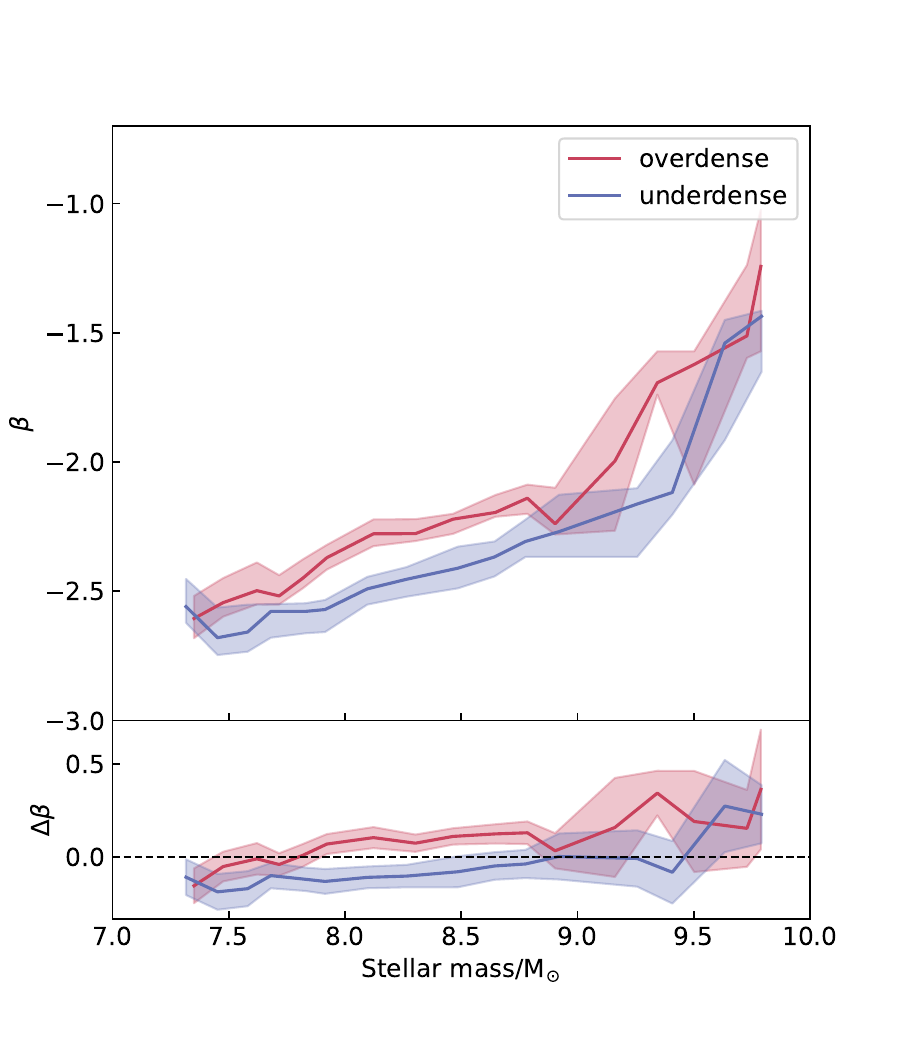}
    \caption{The UV slope $\beta$ of under and overdense galaxies in different stellar mass bins. The lines and the shallow regions is their median and the median errors, derived from the bootstrap method. The bottom panel is $\Delta {\beta} = \beta_{\rm over | under}-<\beta>$.}
    \label{fig:beta_M}
\end{figure}

\subsection{Effect of the Environment on UV slope}\label{sec: beta}
The rest-frame UV continuum slope, $\beta$, is commonly used as a diagnostic tool to probe the characteristics of the stellar continuum \citep{Calzetti1994}. It serves as an indicator of massive O/B-type main sequence stars and the associated HII nebular regions, and it is also a main indicator of dust in high redshift galaxies as well as for any unusal stellar populations, such as Pop III star dominated galaxies.  We use a Bayesian Markov Chain Monte Carlo (MCMC) methods to fit a power-law form of the UV continuum, as $f_{\lambda}\propto\lambda^{\beta}$. This fitting is conducted using the rest-frame UV photometric data within the wavelength range $1250<\lambda_{\text{rest}}~/~\text{\AA}<3000$ \citep[e.g.][]{Rogers2014, Cullen2023}. We use the \citet{Meurer1999} relation to infer the UV dust extinction as $A_{\mathrm{UV}}=4.43+1.99\beta$, which assumes an intrinsic $\beta_{\mathrm{int}}=-2.23$. Following a visual inspection of sources with extreme UV slope values, we remove sources with $\beta > -1$ or $\beta < -3.5$, primarily attributed to noisy photometry. For additional details on this fitting, refer to EPOCHS-III \citep{Austin2024}.

Figure~\ref{fig:beta_M} shows the variation of $\beta$ in different stellar mass bins for galaxies in diverse environments. The value of $\beta$, denoted as $\beta(z, M_{\star})$, correlates with both redshift ($z$) and stellar mass ($M_{\star}$), with a predominant strong dependence on stellar mass. Our analysis distinctly reveals that galaxies located in overdense regions consistently exhibit higher and thus redder $\beta$ values across all stellar mass bins. The average deviation is substantial, with $(\beta_{\rm over} - \beta_{\rm under}) \sim 0.15$. Using MCMC method, the median slope $\beta$ of overdense galaxies are 2$\sigma$ higher than the underdense galaxies. Employing the K-S test on the two datasets yields a p-value of 0.03, indicating a high likelihood that they originate from distinct distributions.

$\beta$ values depend on the stellar age and stellar population as well as on the dust extinction.  Since young stars are mostly located in proximity of dusty molecular clouds, star formation activity is usually associated with high extinction values. Galaxies in overdensities are characterised by enhanced star formation activity, it is thus plausible that their redder UV slope is due to higher extinction rather than to an older stellar population, indicating that the cause may be dust. This is consistent with the presence of neighboring galaxies that generate dust, which might be expelled from the host galaxy through outflows but persists within the halo, eventually falling back onto other galaxies in proximity. Another plausible explanation is the influence of Ly$\alpha$ emission. Large Ly$\alpha$ EWs result in a decrease in the UV slope $\beta$ due to the biases present in wide-band photometric studies \citep{Rogers2013,Austin2024}, suggesting that galaxies in underdense environments may exhibit stronger Ly$\alpha$ emission. There is also an alternative explanation involving Damped Lyman-$\alpha$ systems (DLAs). \citet{Austin2024} suggested that galaxies with high neutral Hydrogen column densities ($N_{\mathrm{HI}}>2\times10^{20}~\mathrm{cm}^{-2}$) along the line-of-sight could result in higher $\beta$ values in wide-band photometric studies \citep{Heintz2023}, indicating that galaxies in denser regions may exhibit stronger DLA absorption (e.g. \citealt{Heintz2023,Cen1999}). It appears that overdense regions tend to host a greater abundance of metal-rich objects such as damped Ly$\alpha$ systems or exhibit stronger absorption \citep{Heintz2024,Chen2024}. Further spectroscopic observations will be essential to investigate this idea in the future.

\begin{figure}
    \centering
    \begin{subfigure}{\linewidth}
        \centering
        \includegraphics[width=\linewidth]{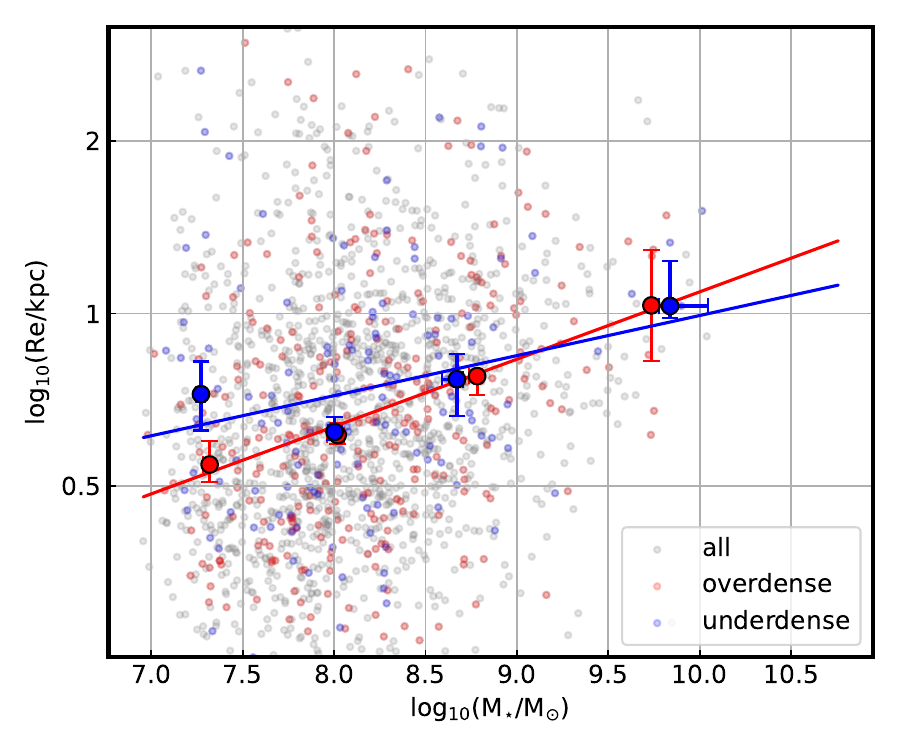} 
        \caption{Median stellar mass–size relation for overdense and underdense galaxies at different stellar masses. Error bars are computed through the bootstrapping method.}
        \label{fig: figure_s5_morpy2}
    \end{subfigure}
    
    \begin{subfigure}{\linewidth}
        \centering
        \includegraphics[width=\linewidth]{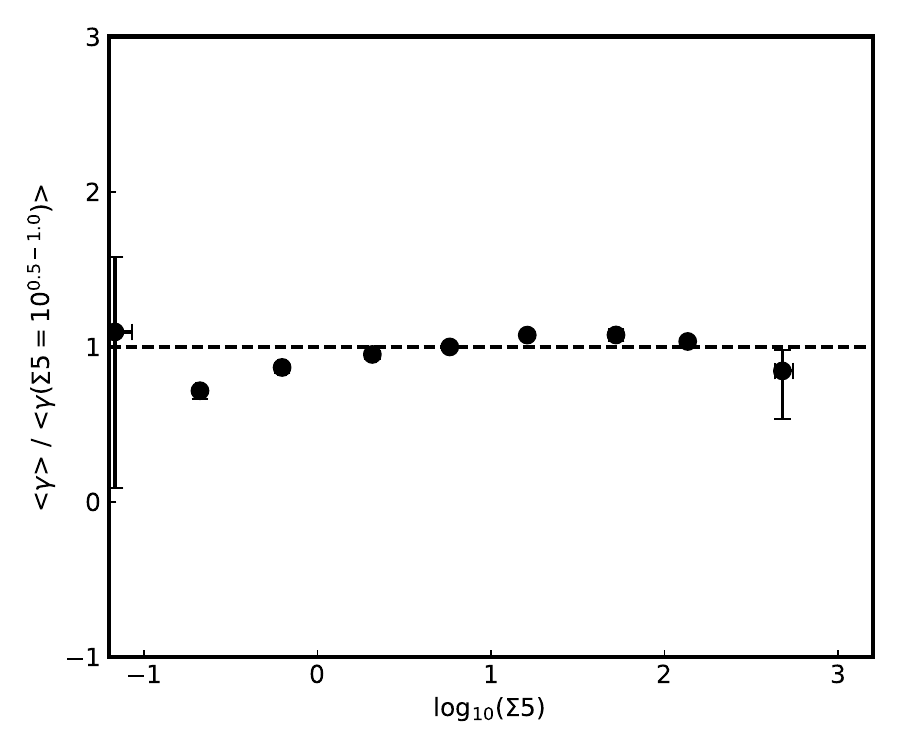}
        \caption{The normalized size ($\gamma$) of galaxies as a function of $\Sigma_5$. <$\gamma$> is defined as Equation~\ref{eq: gamma}. Values have been normalized to the median size in the $\Sigma_5$ bin $\Sigma_5 = 10^{0.5-1.0}$. Errors on the median values are computed through bootstrapping.}
        \label{fig: figure_s5_morpy5}
    \end{subfigure}

    \caption{Mass–size relations in different environments.}
    \label{fig:morph_M}
\end{figure}

\subsection{Effect of the Environment on the mass-size relation}

\begin{table}
\centering
\begin{tabular}{c|cc}
\hline
\hline
\textbf{redshift bins (total)} & $\gamma$ & $\beta$ \\
\hline
 $z<6$   &0.102$\pm$0.055 & 0.087$\pm$0.006 \\
 $z>6$   &0.024$\pm$0.055 & 0.086$\pm$0.006 \\
\hline
\textbf{Environment subsets} & $\gamma$ & $\beta$ \\
\hline
 overdense   &0.155$\pm$0.079 & 0.117$\pm$0.009 \\
 underdense   &0.074$\pm$0.333 & 0.073$\pm$0.039 \\
 all   &0.075$\pm$0.033 & 0.089$\pm$0.003 \\
\hline
\end{tabular}
\caption{The best linear fits of the Mass-Size relation in Eq.~\ref{eq: gamma}.}
\label{tab: mass-size}
\end{table}

To investigate the mass-size relation in different environments, we firstly use \galfit\, version 3.0.5 \citep{Peng2002Galfit1,Peng2010Galfit2} to fit a single Sersic light profile to each galaxy. \galfit\, in a least-squares-fitting algorithm that finds the best-fitting model by minimizing the reduced chi-squared statistic ($\chi^2$). We choose filters that closely match the rest-frame optical wavelength of the sources to minimize the morphological $k$-correction effect \citep{TaylorMager2007}. Extreme cases are filtered out based on the criteria that the half-light radius should be within $0.01 < R_e {\rm (pixels)} < 100$ and the fitted Sersic index lies within the range $0.05 < n < 10$. For more details, please refer to \citet{ormerod2023} and Westcott et al. 2024 (in prep).

Figure~\ref{fig: figure_s5_morpy2} shows the median of the stellar mass-size relation for the underdense and overdense galaxies. We find the stellar mass-size relation is universal within the mass ranges we probe (log$M_{\star}/M_{\odot}=7-10$). Here we perform a linear fitting on their respective medians in each stellar mass bins using a procedure similar to that described by \citet{Newman2012}:
\begin{equation}
    R_e = \gamma \left ( \frac{M_{\star}}{ M_{11}} \right )^{\beta} 
\end{equation}
where $M_{11}$ is $10^{11} M_{\odot}$. This expression can be written as:
\begin{equation}
    \text{log ($\gamma$)} = \text{log$_{10}(R_e)$}+\text{$\beta(11$}-\text{log$_{10} M_{\star}$)}
\label{eq: gamma}
\end{equation}
The linear fitting results in a slope of $\beta=0.12\pm0.01$ and an intercept of $C=-1.14\pm0.08$ for the overdense galaxies, and a slope of $\beta=0.07\pm0.04$ and an intercept of $C=-0.74\pm0.33$ for the underdense galaxies. These best fitting parameters are list in Table~\ref{tab: mass-size}. We find that the mass–size relation does not show a significant trend with environment; there are similar mass–size relations regardless of the host's mass. Our results are consistent with the findings of the study in the local universe, indicating no evidence of a dependence of the mass-size relation of galaxies on environment. \citep{Huertas-Company2013,Weinmann2009}.

Despite the observation from Figure~\ref{fig: figure_s5_morpy2} that galaxies of similar mass exhibit similar size distributions regardless of their environment, there remains the possibility of some intrinsic environmental dependence \citep{Bernardi2010}. To investigate this further, we examine the $\Sigma_5$–$R_e$ relation, which characterizes the median size at a fixed stellar mass as a function of environment. Given the correlation between environment and stellar mass, it is expected that more massive galaxies, and therefore larger galaxies, are more likely to be found in overdense regions. To account for this effect, we normalize the sizes ($\gamma$) using a procedure similar to that described by \citet{Newman2012}. Here, $\beta$ is the slope of the stellar mass–size relation within the specified mass range, and $R_e$ denotes the effective radius. We adopt a standard value of $\beta = 0.83$, which corresponds to the slope measured across the entire galaxy sample.

In Figure~\ref{fig: figure_s5_morpy5}, we present the relative differences across different environments by normalizing all sizes to the median size in the $\Sigma_5$ bin $\Sigma_5 = 10^{0.5}$–$10^{1.0}$. The uncertainties associated with the median values are estimated using bootstrapping. The most notable finding is that the $\gamma$–$\Sigma_5$ relation appears essentially flat, indicating that galaxy sizes remain consistent across all environments within the margins of error.

\section{Galaxy protocluster candidates }\label{sec:Group galaxy membership probabilities}

To further describe the global environment of our galaxy sample, we search for protocluster candidates in the JWST photometric data using the DETECTIFz algorithm \footnote{https://github.com/fsarron/detectifz}  \citep{Sarron2021} to detect extended overdensities. In particular we asses the probability of individual galaxies to belong to protocluster candidates and study the most probable members properties. This is related to but different from the efforts earlier in the paper where we have determined the properties of galaxies as a function of local density. Here we investigate whether any of these overdense regions would qualify as being clusters or proto-clusters. 

The DETECTIFz algorithm utilizes the Delaunay Tessellation Field Estimator (DTFE), previously used to detect cosmic web filaments \citep{Sousbie2011}, to find extended galaxy overdensities in redshift slices. The approach is empirical and model-free, relying solely on galaxy sky coordinates and samples from joint photometric redshift and stellar mass probability distribution. We review the main steps of the algorithm relevant to this work in the next sections, focusing on the adjustment made for JWST data. We refer to \citet{Sarron2021} for a more detailed description of the algorithm and the establishment of the DETECTIFz method.

\subsection{Overdensity detection}

\subsubsection{Overview of the detection process}

To search for galaxy protocluster candidates, we proceed with the estimation of the 2D galaxy density in redshift slices. The  slice width is based directly on the uncertainty associated with photometric redshifts such that the galaxy population sampled in the slice is representative of the underlying galaxy population at the slice central redshift.





In each slice centered at $z_i$, we estimate the 2D galaxy density using 100 Monte Carlo realisation of each galaxy's redshift probability distribution. For each realization ${\rm mc}_j$, we build the 2D density $\Sigma_{\rm gal} (\text{mc}_j, z_i)$ distribution using the DTFE method. The final surface density map is taken as the logarithmic average over all  Monte Carlo realisations:
\begin{equation}
\log \Sigma_{\rm gal}^{\text{mc}}(z_i) = \frac{1}{N_{\text{mc}}} \sum_j \log \Sigma_{\rm gal} (\text{mc}_j, z_i).
\end{equation}

\noindent The 2D overdensity maps in each slice can then be written in \citet{Cucciati2018} as:
\begin{equation}
\log(1 + \delta_{\rm gal})(z_i) =  \frac{\Sigma_{\rm gal}^{\text{mc}}(z_i)}{\left\langle\Sigma_{\rm gal}^{\text{mc}}(z_i)\right\rangle} ,
\end{equation}
where ${\left\langle\Sigma_{\rm gal}^{\text{mc}}(z_i)\right\rangle}$ is the mean 3$\sigma$ clipped mean surface density map. After obtaining a overdensity map for each slice $z_i$, we smooth the 3D data cube (ra, dec, $z$) along the redshift dimension. This is achieved through convolution with a one-dimensional normal distribution with a standard deviation of $\sigma_z$ = 0.02. 

We then search for peaks in each slice with a minimum signal-to-noise ratio $(S/N)_{\rm min}>1.5$ in the 2D overdensity maps using the {\tt photutils} Python package \citep{Bradley2019}, as:
\begin{equation}
S/N = \frac{\log(1 + \delta^{\rm mc}_{\rm gal}) - \mu_{\delta}}{\sigma_{\delta}},
\end{equation}
where $\mu_{\delta}$ and $\sigma_{\delta}$ are respectively the mean and standard deviation of the 3$\sigma$ clipped distribution of $\log(1 + \delta^{\rm mc}_{\rm gal})$. For each detection, we record the bounding-box $({\rm RA_{min}, RA_{max}, Dec_{min}, Dec_{max}})$ encompassing the region with pixel values above the $(S/N)_{\rm min}$ threshold.

To remove multiple detection of the same protocluster in different adjacent slices and/or with multiple substructures, we merge detected peaks that are contiguous in the line-of-sight direction (i.e., redshift direction) and whose peaks are separated by less than $2$ comoving Mpc or have overlapping bounding boxes.  These selected regions are our final galaxy protocluster candidates. The sky coordinates of the peak with the highest $S/N$ ratio, with its corresponding redshift, are recorded as the centre of the galaxy protocluster candidate are listed in Table~\ref{tab:cat_cluster}.

\subsubsection{Adjustment made for protocluster detection}

Due to the scarcity of galaxies at high redshifts ($z > 7$), we are interested in detecting protocluster candidates at $5 < z < 7$, we had to make a few adjustments to the DETECTIFz algorithm that was originally devised to detect galaxy groups at $0 < z < 3$.

First of all, given the lower spatial density of the JWST data compared to lower redshift deep photometric surveys, we elect to detect the overdensities using the galaxy density rather than the stellar-mass density. Indeed, our test with stellar mass densities turned out to be biased towards detecting overdentities centered on very massive galaxies, even with relatively few neighbours. The galaxy density is more robust in this regard. Thus, in this set-up, we did not use samples from the 2D probability density PDF$(M_\star, z)$ to compute the stellar mass surface density, but instead worked with the galaxy surface density, and thus only used samples from the 1D $p(z)$. 

We also modified the way we merge individual peaks detected in adjacent slices to get rid of multiple detections of the same overdensity. Here, we are interested in detecting protoclusters i.e., very extended unvirialized structures, rather than more compact galaxy groups at lower redshift. Protoclusters also typically present several substructures. We focused on larger scales by merging peaks further away and considered the extent of each individual peak in the merging process. Thus, while the DETECTIFz method merges overdensities closer than $500$ physical kpc, here we merge peaks closer than $2$ comoving Mpc, or having overlapping bounding-boxes. The bounding-box of each candidate protocluster is defined as the union of the bounding-boxes of all merged individual detections.

Finally, in the original DETECTIFz, we made sure to have a 90\% stellar-mass complete sample of galaxies in each slice in order to have a very robust estimate of the expected density fluctuations and in turn of the $S/N$. For this study, we fixed the lower limiting stellar mass of the sample to be close to the 50\% stellar-mass completeness estimated from the JAGUAR simulations. This lower limiting was chosen to be redshift independent and fixed respectively to $M_{\rm lim} = 10^{8.5} M_\odot$ for CEERS and NEP and $M_{\rm lim} = 10^{7.5} M_\odot$ for JADES. This choice was a necessary trade-off in order to have a sufficient galaxy density, but this could impact the purity of the output protocluster catalogues.

We note that we were not able to test these different changes on simulations to assess how they impact in detail the selection function of our JWST protocluster catalogue. Given the novelty and exploratory nature of our results, we however defer this effort to future work.

\subsection{Group membership probability and cluster mass}\label{sec: Group membership probability and cluster mass}
For individual galaxies within the bounding-box of each protocluster candidate, we compute their probability to be protocluster members. This probability is calculated in a Bayesian method following \citet{Sarron2021}, similar to the approaches outlined in \citet{George2011} and \citet{Castignani2016}. It involves the convolution of the redshift PDFs of galaxies and that of the protocluster with a prior based on excess number counts in the protocluster region. We refer to  \citet{Sarron2021} for a more thorough description of the method and outline here the main modifications that we adopted for JWST data. 

We write the probability membership $P_{\text{mem}} \equiv P(\text{gal} \in {\rm PC} | \text{PDF}_{\text{gal}}(z), \text{PDF}_{\text{PC}}(z), M_{\star}^{{\rm gal}})$ in a given protocluster (PC) and use Bayes rule

\begin{equation}
\begin{split}
P_{\text{mem}} \propto \int P(&\text{PDF}_{\text{gal}}(z) | \text{gal} \in G, \, \text{PDF}_{\text{PC}}(z), \, M_{\star}^{\text{gal}})(z) \\
&\hfill \times P(\text{gal} \in G \,|\, \text{PDF}_{\text{PC}}(z), \,M_{\star}^{\text{gal}})(z) \, dz,
\end{split}
\end{equation}
where $P(\text{PDF}_{\text{gal}}(z) | \text{gal} \in G, \, \text{PDF}_{\text{PC}}(z), \, M_{\star}^{\text{gal}})(z)$ is the likelihood of observing the data (PDF$_{\text{gal}}(z)$) evaluated at redshift $z$ given the galaxy belong to the protocluster, while $P(\text{gal} \in G \,|\, \text{PDF}_{\text{PC}}(z), \, M_{\star}^{\text{gal}})(z)$ is the prior probability that a galaxy with stellar mass $M_{\star}^{\text{gal}}$ belongs to the protocluster evaluated at a redshift $z$.

The main difference with \citet{Sarron2021} resides in our use of the stellar mass information in the computation of the prior. Due to number counts in the JWST data being quite low, the method proposed in \citet{Sarron2021}, where the prior was based on number counts excess in bins of redshift and stellar-mass was simplified by using only bins of redshifts, to prevent number counts excess estimate to be shot noise dominated. Thus we were able to use directly the DETECTIFz overdensity maps that allows us to compute number count excess at any (ra, dec, $z$) and deduce our prior as:
\begin{equation}
    P_{\rm Prior}(z) = 1 - \frac{\langle n^{\rm DETECTIFz}(z) \rangle_{\rm not PC}}{n^{\rm DETECTIFz}(z)(\rm RA_{gal}, Dec_{gal})},
\end{equation}
\noindent where $n^{\rm DETECTIFz}$ are surface densities given by DETECTIFz 3D overdensity maps $n \propto 10^{(1 + \delta_{\rm gal}^{\rm DETECTIFz})}$. In the numerator, the surface density estimate are averaged over pixels outside any protocluster bounding-box at redshift $z$ ("notPC"). In the numerator, we take the surface densities at the pixel in which the galaxy falls.

Further details of the definition of likelihood and the rescaling of probabilities can be found in \citet{Sarron2021}. We note that while probabilities calibration was thoroughly checked using simulations in \citet{Sarron2021}, there is no guarantee here that the absolute $P_{\rm mem}$ values are good estimators of the probability of a galaxy to belong to the protocluster. In particular, since protoclusters are unvirialized clusters, we expect $P_{\rm mem}$ to behave differently than what was presented in \citet{Sarron2021} for the case of groups and clusters at lower redshifts. However, our probability membership definition is well motivated by probability theory, so we expect our probability memberships to be informative in a relative sense. This is reflected in the following, where we use $P_{\rm mem}$ to select a quantile of the most probable protocluster galaxies for further analysis of their properties.

We use the $P_{\rm mem}$ values to compute the cluster's total stellar mass, which can be written as:
\begin{equation}
M_{\rm \star, PC}= \sum_{
\begin{array}{c}
{\rm gal} \in {\rm PC}\\
P_{\rm mem} \in Q_4
\end{array}}
M_{\star}^{\rm gal}
\end{equation}
where $Q_4$ is the upper quartile of probability memberships, $M_{\star}^{\rm gal}$ is the galaxy's stellar mass. The final protocluster candidate catalogue is listed in Table~\ref{tab:cat_cluster}.

We estimated the cluster size $R_{200}$ in our catalogue following the methodology of \citet{Hansen2005} and \citet{Sarron2021}. This size is the radius of a sphere in which an average mass density is 200 times the critical density of the Universe.

\subsection{Protocluster detected in DETECTIFz}
In total, we find 26 protocluster candidates  using DETECTIFz at $5<z<7$, of which 16 are at $z>6$. There are 11, 10 and 6 in CEERS, JADES and NEP, respectively. The total stellar mass of these protoclusters ranges log($M_{\star}/M_{\odot}$) = 9.2 - 10.6. 
Figure~\ref{fig:cluster_JADES_C0} shows examples of the largest photometric protocluster candidate in JADES field and its probable member galaxies ( with $P_{\rm mem} \in Q_4$). The member probabilities $P_{\rm mem}$ are coloured in the right panel. The point size shows the galaxy's stellar mass at the group redshift. The overdensity log $\delta_{\rm gal}=0.59$, is shown in the blue shaded regions.

We tested whether the results obtained from different SED fitting codes would impact our search for protocluster candidates. Thus, we also run the \texttt{EAZY} Larson PDF($z$) + \texttt{Bagpipes} best stellar-mass fitting at the \texttt{EAZY}  photo-$z$ set-up on CEERS, as the `alternative' run. We identified a total of 6 overdensities, consistent with the initial EAZY run (`reference' run). Remarkably, 5 out of 6 overdensities are in common. Two overdensities observed in the reference run have amalgamated into a single entity, likely owing to the implementation of a smoothed version of the slice width, a refinement that enhances the accuracy of the detection process. 

Encouragingly, we have detected a novel overdensity located at the periphery, absent in the previous run. However, in the alternative run, its significance is at $S/N=1.8$, indicating that it may have narrowly missed the $S/N=1.5$ threshold in the reference run. On average, the $S/N$ for detections are higher in the recent run. This uptrend could be attributed to the smoothing of the slice width and/or the consistent stellar-mass input across all 100 Monte Carlo realizations of the redshift. Nonetheless, this difference does not cause significant concern and does not affect our main detections and conclusions here.

\begin{figure*}
    \centering
    \includegraphics[width=0.7\textwidth]{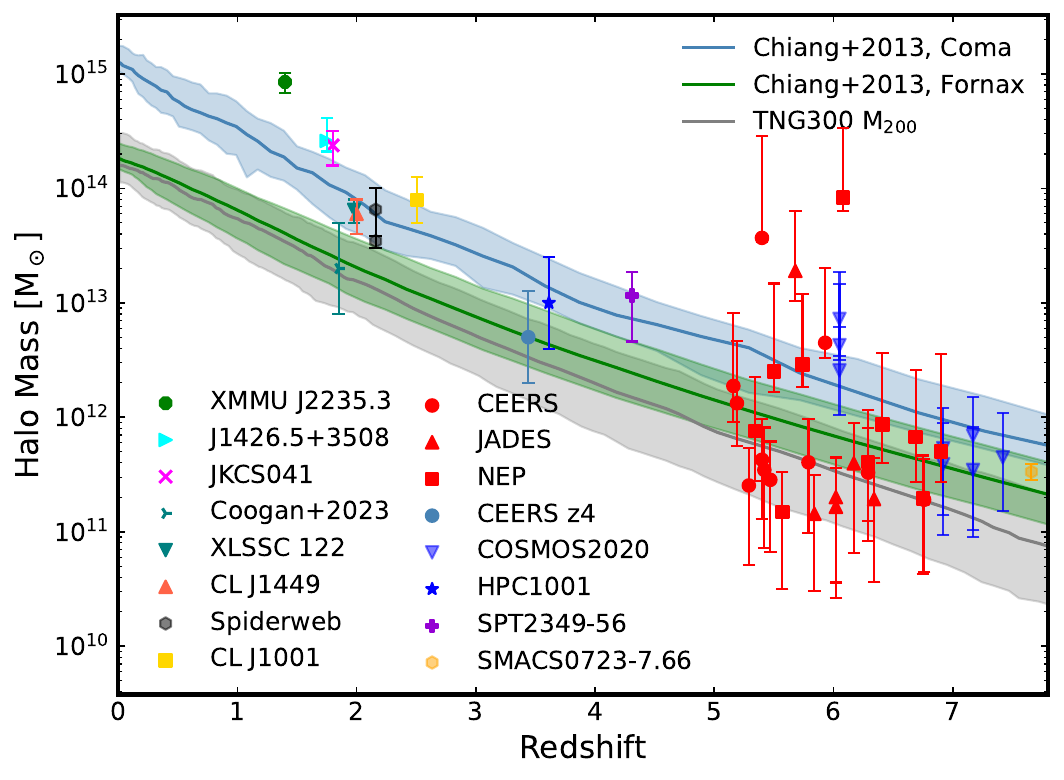}
    \caption{The dark matter halo mass distribution for protocluster candidates at $z=1.3-8.0$, compared with simulations. The red points are the photometric confirmed protocluster candidates from this work. The blue and green lines are the simulated halo mass evolution of `Coma' ($>10^{15} M_{\odot}$) and `Fornax' (($1.37-3)\times10^{14} M_{\odot}$) type clusters in \citet{Chiang2013}, respectively. The gray shaded region is from the TNG300 simulation of $M_{200} \le 10^{14} M_{\odot}$ systems at $z = 0$ in \citet{Montenegro-Taborda2023}. }
    \label{fig:halomass_z}
\end{figure*}

\subsection{Dark Matter Halo Mass}\label{sec: halo mass}
We estimated dark matter halo masses for these protoclusters and compared these with simulations to infer their properties and evolution. This estimate assumes a scenario where each galaxy originates within the same dark matter halo. Consequently, we derived the dark matter halo mass from the total stellar mass of the clusters, leveraging the $M_{\rm halo}-M_{\star}$ scaling relation outlined by \citet{Behroozi2013}. By summing the stellar masses of individual galaxies constituting the overdensity and converting them to a halo mass, our analysis yields a total halo mass of log10 ($M_{\rm halo}/M_{\odot}$) = $(11.2-13.9) \pm 0.6$ for most systems.

Figure~\ref{fig:halomass_z} illustrates the distribution of dark matter halo masses across various representative groups and protoclusters at $z < 8$, with simulation data from \citet{Chiang2013} and TNG300 simulations from \citet{Montenegro-Taborda2023}. \citet{Chiang2013} employed a semi-analytical galaxy formation model \citep{Guo2011} applied to the dark matter-only N-body simulation Millennium \citep{Springel2005}, tracking the evolution of both dark matter and galaxies across approximately 3000 clusters from $z = 7$ to $z = 0$. The blue and green shaded regions denote the simulated halo mass evolution of `Coma' ($>10^{15} M_{\odot}$) and `Fornax' (($1.37-3)\times10^{14} M_{\odot}$) type clusters in \citet{Chiang2013}, respectively. In contrast, \citet{Montenegro-Taborda2023} identified 280 systems with $M_{200} \le 10^{14} M_{\odot}$ at $z = 0$ in the TNG300 simulation \citep{Pillepich2018} and traced their progenitors at high redshift, represented by the gray shaded areas.

Our findings indicate that the confirmed halo mass aligns with  the theoretical prediction in \citet{Chiang2013} and the progenitor of the TNG300 simulations, even when considering the uncertainty in halo mass. This suggests that most of the confirmed clusters would evolve into clusters with $M_{\rm halo} > 10^{14}$ at $z = 0$, with some (8/27) potentially evolving into Coma-like clusters with log10($M_{\rm halo}/M_{\odot}$) > 15 by $z = 0$ (e.g. \citealt{Ruel2014,Buddendiek2015}). We also plot the halo and stellar masses of other massive proto-clusters at $1.3 < z < 8.0$ that also exhibit consistency with simulations (e.g., \citealt{Jin2024,Rosati2009,Stanford2012,Gobat2013,Andreon2014,Mantz2018,WangTao2016,Miller2018,Sillassen2022,Coogan2023,Shimakawa2018,Perez-Martinez2024,DiMascolo2023,Laporte2022}), further supporting the forming cluster scenario of the theoretical and simulation prediction.

It is worth noting that our sample may lack additional overdensity members characterized by low levels of star formation or heavy obscuration (e.g., DSFGs and/or obscured AGNs). Consequently, our total halo mass range is likely an underestimation of the true halo mass. However, considering the uncertainties in our halo mass calculations, these factors do not significantly impact the main conclusions of our study.

\subsection{Galaxy environment of protocluster candidates}\label{sec: pc and sigma_5}
To validate whether the identified protocluster candidate members align with the findings on galaxy environments discussed in Section~\ref{sec:over_under}, we compare the members of protocluster candidates with a control sample. The control sample are selected from the lower quartile of probability membership within each protocluster candidate. Some galaxies, although having a low probability within a specific protocluster, may be associated with another protocluster; these are also excluded from the sample. Ultimately, 401 galaxies are selected, providing a representation of galaxies in underdense regions.

We examine the $\Sigma_{5}$ values, where the median $\Sigma_{5}$ of the protocluster members was found to be 2.6 times higher than that of the control sample, with errors estimated using the bootstrap method. Employing the MCMC approach, the median $\Sigma_{5}$ for cluster members is determined to be significantly higher than that of the control sample, supporting the consistency between different methods in identifying overdense and underdense galaxies. We conduct the further analysis on their physical characteristics such as mass, SFR, and UV slope. We find that the median UV slope ($\beta$) values for cluster members and the control sample were $-2.28 \pm 0.04$ and $-2.38 \pm 0.05$, respectively, indicating that cluster members were significantly redder, consistent with our previous findings. Similarly, the SFR for cluster members was also marginally higher by $\sim$2 sigma compared to the control sample. The median stellar masses for cluster members and the control sample are $8.22 \pm 0.08 M_{\odot}$ and $7.95 \pm 0.03 M_{\odot}$, respectively, showing a significant difference of about 3 sigma; the difference in sSFR is not as pronounced. 

We note that 32\% of the overdense galaxies identified by $\Sigma_{5}$ values overlap with cluster members identified by the DETECTIFz algorithm. The selection by $\Sigma_{5}$ is based on individual redshift bins, hence it still identifies relatively overdense galaxies at high redshifts. However, due to the limitations in the number of high-redshift galaxies available, it is very challenging for the DETECTIFz algorithm at $z>7$ to produce meaningful density maps. Furthermore, while the proxy may work in a relative sense (i.e., a higher number indicates larger mass structures), it may be a poor absolute measure of the number of protocluster galaxies. These factors can lead to biases in the selection of overdense galaxies.

\section{Discussion}\label{sec: discussion}

In this study we have conducted an analysis of high-$z$ galaxies within the CEERS, JADES and NEP field at $z>4$, utilizing high-resolution broadband imaging data from the HST and using JWST NIRCam data. Our investigation quantitatively examines the members of protoclusters within the range $z=4.5-10$. By leveraging the three wide-field datasets, we have illuminated the crucial role of environment in galaxy evolution at $z>4$. In this section, we compare our findings with relevant studies on the clustering properties of galaxies at these redshifts. We explore how environmental factors influence the formation and evolution of galaxy clusters at these early cosmic epochs. By integrating simulations and previous observational data at lower redshifts, we can analyze the process of cluster assembly. 

\subsection{Association Between Photometric and Spectroscopic Overdensity}

\begin{figure*}
\centering
 \includegraphics[width=0.95\textwidth]{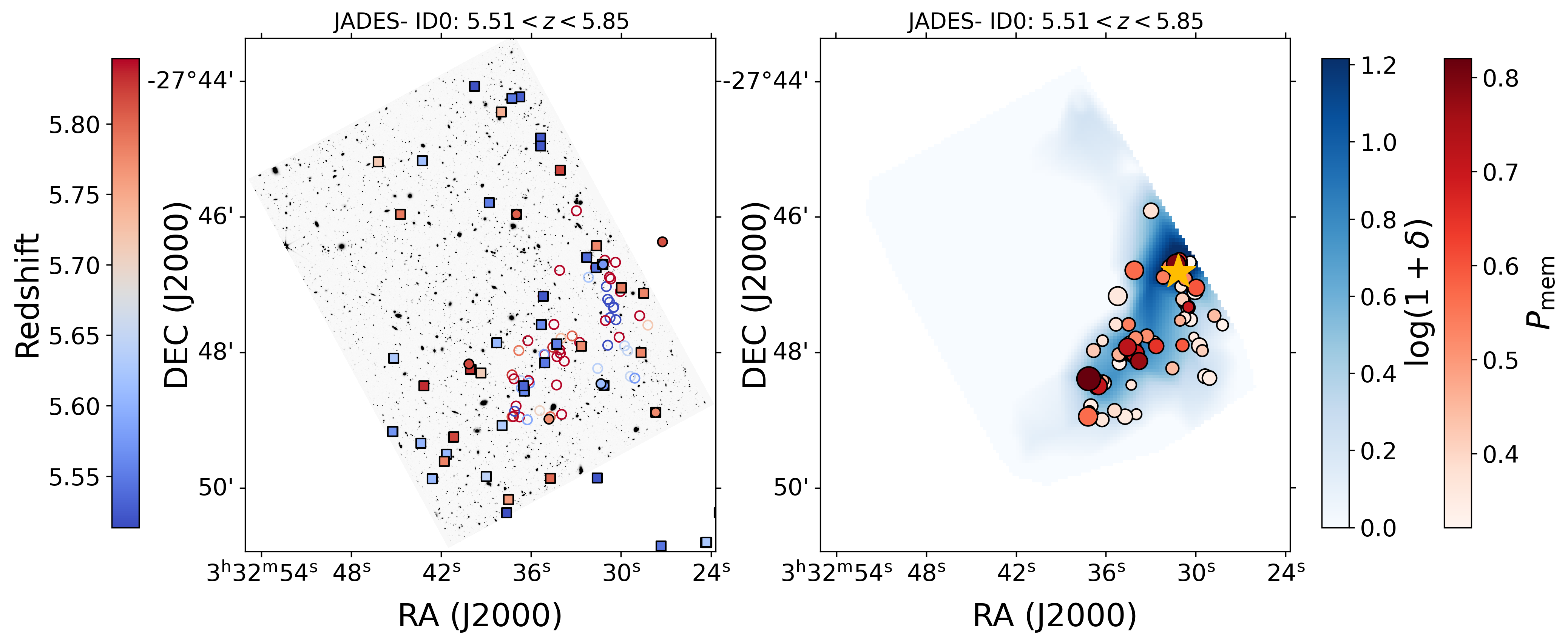} 
    \caption{Left: The photometric and spectroscopic galaxies identified in a protocluster at $z=5.51-5.85$ in the JADES field as shown by open circles and filled squares, respectively. The colors for each system show their redshifts. Right: the member galaxies in the protocluster candidate with $P_{\rm mem}\in Q_4$ colour-coded by their probability of being at the group redshift. The point size is proportional to the galaxy stellar mass at the group redshift. The blue shaded regions present the overdensity level as indicated by the color bars.
    }
    \label{fig:cluster_JADES_C0}
\end{figure*}

\begin{figure*}
\centering
 \includegraphics[width=0.95\textwidth]{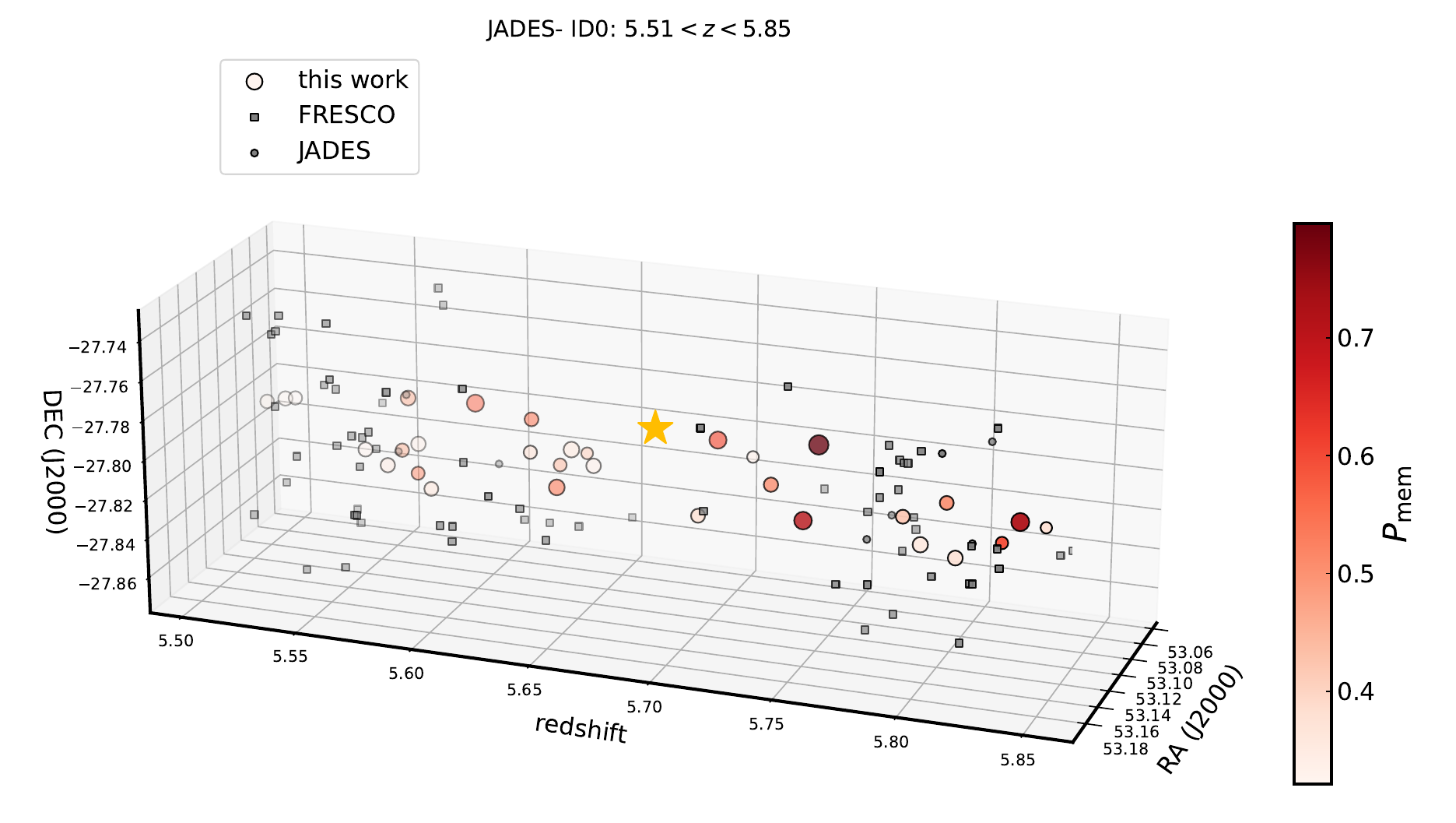} 
    \caption{A 3D plot of the largest galaxy overdense environment at $5.51 < z_{\rm photo} < 5.85$ in the JADES dataset. Each photometric selected galaxy is color-coded based on its probability of cluster membership, with the point size indicating the galaxy's stellar mass, the same as in Figure~\ref{fig:cluster_JADES_C0}. Spectroscopy-confirmed cluster members are denoted by the gray squares and circles in the plot. The yellow star is the center of the protocluster candidate. This figure shows that these galaxy overdensities are situated within a complex environment characterized by possible interconnected filamentary structures. 
    }
    \label{fig:cluster_JADES_C0_3D}
\end{figure*}

In theory, there should be a correlation between photometric and spectroscopic overdensities because they trace the same dark matter halo or structure. However, previous studies on overdensities at $z=2-5$ have shown that different tracer galaxies (e.g., LAE, LBG, or SMG) exhibit different distributions when describing the same dark matter halo. In this section, we compare the overdensities identified through our photometric analysis with those confirmed by spectroscopic observations to assess their consistency and provide a more comprehensive study of large-scale structure.

Figure~\ref{fig:cluster_JADES_C0} shows one of our protocluster candidates in the GOODS-S field, JADES-ID1-5.68, with a total of 66 galaxies ($P_{\rm mem}>0.32$). This overdensity holds the distinction of being the largest in terms of the total number of galaxies identified in JADES DR1, and it also ranks as the most massive inferred halo mass within these fields. The average overdensity value within a central circle of 250 kpc radius is $\delta$gal = 3.9. These photometric sources exhibit potential associations with the overdensity studies of JADES-GS-OD-5.386 reported in \cite{Helton2023}. The latter identified 42 spectroscopic sources at redshifts $z = 5.2 - 5.5$, although they were labeled as `field galaxies' potentially due to limitations in spectroscopic data coverage. Spectroscopic follow-up observations are imperative to confirm the true association of the photometric sources with the overdensity. Notably, JADES-ID1-5.68 is in close proximity to JADES-GS-OD-5.386.  The latter exhibits an average overdensity value of $\delta$gal = 9.134 \citep{Helton2023}. This makes JADES-GS-OD-5.386 the most significant overdensity characterized by H$\alpha$ emitters within the redshift range of $4.9 < z_{\rm spec} < 6.6$. It is plausible that our photometric sources and these overdensities are situated within a larger, intricate filamentary structure.

While some photometrical confirmed overdensities exhibit concordance with spectroscopically confirmed counterparts, others do not. This discrepancy can be attributed to various factors. Spectroscopic surveys frequently prioritize the observation of brighter sources to ensure robust detection rates, often focusing on galaxies at $m<20$ AB mag, as exemplified by the CEERS NIRSpec survey and the JADES NIRSpec survey. However, this strategy may inadvertently neglect fainter sources. Such dimmer sources may nevertheless be detected through JWST broad-band surveys, particularly at higher redshifts, where galaxies with magnitudes around $m\sim24$ become accessible.

Even overdensities confirmed using narrow-band selected emission line galaxies such as H$\alpha$ and \oiii\,, or even Ly$\alpha$, may also contain bias. This is because some high-$z$ galaxies may exhibit weak emission lines, which can introduce uncertainties in the overdensity identification process (e.g. \citealt{Lemaux2018}). Additionally, insights from UNCOVER spectra shed light on why Ly$\alpha$ emission is not detected in many galaxies within the overdensity identified by \citet{Morishita2023}. Many observations have indicated evidence of damped Ly$\alpha$ absorption, which can significantly attenuate any Ly$\alpha$ emission that may be present. Consequently, the density of Ly$\alpha$ emitters may be underestimated, potentially leading to the oversight of genuinely dense regions within the overdensity.

It is worth noting that the uncertainties in redshift measurements for galaxies confirmed through photometry, typically ranging from 0.1 to 0.3. At a redshift of approximately $z\sim6$, this corresponds to uncertainties of about 30 to 90 h$^{-1}$ Mpc. This can result a lack of sensitivity in identifying structures with scales smaller than 30 h$^{-1}$  Mpc in the overdensity regions. Another significant effect of redshift measurement uncertainty is the contamination introduced by $\Delta z$ when measuring excess surface density. As $\Delta z$ increases, it smoothies the density field, reduces scatter within each bin, and largely underestimates the proto-cluster overdensity. This effect was also pointed by \citet{Chiang2013} in Figure~13, that find at $z=3$ a drop of approximately 40\% in the surface density of protoclusters ($1 + \delta_{\rm gal}$), when using a narrow-band filter with a typical FWHM of $\sim60$\AA\  (corresponding to a $\Delta z\sim 0.05$ and the comoving length $d_c \sim 50$ h$^{-1}$ Mpc) compared to a 15 Mpc cubic window. Therefore, protoclusters identified through our photometric methods may underestimate the surface density and provide the lower limit densities. Follow-up spectroscopic data from JWST will be instrumental in confirming their densities.


\subsection{When were the first mature clusters virialized?}

The evidence for galaxy clusters typically relies on the observation that these entities are virialized, distinguishing them from protoclusters. This virialization requirement necessitates detection through emission from the intracluster medium (ICM), either in X-rays (e.g., \citealt{Rosati2002,Rosati2004,Rosati2009,Gobat2011,WangTao2016}) or at submillimeter wavelengths via the S-Z effect (e.g., \citealt{Reichardt2013,Staniszewski2009,Bleem2015,Williamson2011,Hilton2021}). But current sensitivity in X-ray and sub-millimeter observations remains insufficient to detect galaxy clusters at early cosmic epochs, beyond $z = 2-3$ (e.g., \citealt{DiMascolo2023,Tozzi2022}). Additionally, early galaxy clusters may not have undergone virialization, as their halos are undergoing rapid mass accretion. We aim to facilitate subsequent multi-wavelength studies to better understand when these structures begin to virialize, a crucial aspect of cosmic evolution.

In Figure~\ref{fig:halomass_z}, we find that some extremely large-scale structures with halo masses exceeding $10^{13} M_\odot$ are predicted to have already formed as early as $z\sim6$. Similar structures have also been identified at high redshift in previous studies (e.g., \citealt{Shimasaku2003,Brinch2024,Matsuda2005}). Although the central regions of these structures may evolve into massive virialized clusters by $z = 0$, the outer structures are not necessarily destined to virialize, it is improbable that the collapse of the entire structure will be completed by that time. This can explain why their mass is larger compared to clusters or simulations evolved to $z=0$.

Most galaxies within $z =  4.5 - 10$ exhibit significant star formation activity regardless of their environment. Notably, as shown in Figure~\ref{fig:M_ssfr_over_under}, the galaxies in high-density environments exhibit even stronger star formation activity compared to those in low-density environments. In contrast, observational studies have demonstrated that during the epoch of $1 < z < 4$ certain galaxies in high-density environments generally display suppressed star formation activity relative to their counterparts in underdense regions (e.g., \citealt{Tomczak2019,vonderLinden2010,Gomez2003,Lemaux2019,Nantais2017,Sarron2021}). 

In protoclusters at $z\sim2$ there is evidence of most members being on the main sequence, so they are not more star forming than typical galaxies \citep{Polletta2021,Perez-Martinez2024}. The observation of an overabundance of massive quiescent galaxies in overdense environments during these redshift epochs (e.g., \citealt{Davidzon2016,Tomczak2017}) suggests that rapid stellar mass (SM) growth occurred during the early stages of cluster assembly, likely through a combination of in situ star formation processes and ex situ galaxy-galaxy merging activity. We therefore expect there to be a heightened levels of star formation activity in the high redshift to align with the characteristics of low- and intermediate-redshift cluster populations.


\section{Conclusions}\label{sec:conclusions}

In this paper we investigate the environment of galaxies at $z=4.5-10$ using JWST photometric data. Utilizing datasets from JWST and HST wide-field broadband surveys (CEERS, JADES, and NEP), totalling $\sim185$ arcmin$^2$ in area, we carefully selected a sample of 2946 high-redshift galaxies falling within the redshift range of $4.5 < z < 10$. The photometric data obtained from these surveys exhibit well constraints and exhibit close agreement with spectroscopic redshift measurements.

Our primary scientific goal is to analyze whether galaxies inhabiting distinct environments exhibit varying physical characteristics and whether these attributes undergo evolutionary changes with redshifts. Furthermore, we conduct a comprehensive survey for the first time utilizing JWST photometry data to find galaxy protocluster candidates that emerged during the cosmic dawn. Our findings include:

I. We employed the KDTree data structure to facilitate the nearest neighbor measurement, to effectively estimate the local environment of galaxies and to select potential galaxy cluster members. Simultaneously, we curated a control sample consisting of underdense galaxies for comparative analysis. Our selection process meticulously accounts for various factors that could influence sample selection, such as redshift selection bias, velocity offsets along the line of sight, image boundaries, and selection completeness.

II. We find that overdense galaxies consistently have higher star formation activity across all studied redshifts at $z > 4.5$. These galaxies also display a slightly steeper slope in the relation of SFR and $M_{\star}$, suggesting a possible influence of the environment on their stellar formation activity, though this is not very pronounced. Galaxies at higher redshifts in overdensities exhibit increased star formation activity compared to their lower-redshift counterparts. It has a subtle rise with increasing redshift, followed by a gentle flattening of the $M_{\star}$-SFR slope.

III. The sSFR of both overdense and underdense galaxies indicates that overdense galaxies have higher sSFRs, surpassing those of field galaxies by 0-0.5 dex. We find, during the early universe when massive galaxies were less common, the effect of mass dominates the sSFR, while the quenching mechanisms related to environmental factors appeared less pronounced. In contrast, overdense galaxies exhibit higher sSFRs due to their higher mass accretion rates.

IV. Our study also reveals that galaxies situated in dense regions of consistently redder in terms of UV slope $\beta$ values across all stellar mass bins. These observations imply that galaxies in dense environments may exhibit weaker or even absent Ly$\alpha$ emission, or alternatively, that galaxies in dense regions may possess stronger damped Ly$\alpha$ (DLA) absorption or a higher prevalence of DLAs.

V. We use the DETECTIFz algorithm, a Bayesian approach tailored for quantitatively identifying cosmic web filaments and galaxy clusters/groups using photometric data, to evaluate the probability of a galaxy's membership in a galaxy group. Our analysis yields a total of 26 candidate overdensities, with several being effectively associated through spectroscopic observations. Additionally, we conduct tests to ensure that variations in the results obtained from different SED fitting codes did not significantly impact our cluster detection efforts.

VI. we estimate the dark matter halo mass of our proto-clusters and compared these with other protoclusters identified at $z=1.3-7.0$, as well as with N-body simulations and TNG300 simulation. The close correspondence observed lends support that all confirmed clusters are likely to evolve into clusters with $M_{\rm halo} > 10^{14}$ at $z = 0$, thereby reinforcing the theoretical and simulation predictions of forming clusters. 

Our study represents one of the first searches for protocluster candidates in JWST wide field regions based on photometric data, offering valuable insights into the early phases of cosmic structure formation. Looking ahead, future work with advanced telescopes such as new JWST, ALMA, or similar instruments offers the potential to advance our understanding of protoclusters in the early Universe by providing in-depth, multiple-band and/or spectroscopy observations and analyses. Utilising the spectroscopic capabilities of JWST, for instance, would allow for detailed investigations of the chemical composition and physical properties of galaxies within these protocluster candidates. Additionally, ALMA's sensitivity to submillimeter wavelengths could unveil crucial information about the cold gas and dust content, shedding light on the early stages of galaxy formation within these cosmic structures. This work serves as a crucial reference for future observations of galaxy clusters. As we move forward, combining the power of such advanced tools with spectrum observations promises to enhance our comprehension of the intricate processes shaping the universe and structure formation at the earliest stages.

\section*{Acknowledgements}
We acknowledge support from the ERC Advanced Investigator Grant EPOCHS (788113) and support from STFC studentships. 
This work is based on observations made with the NASA/ESA \textit{Hubble Space Telescope} (HST) and NASA/ESA/CSA \textit{James Webb Space Telescope} (JWST) obtained from the \texttt{Mikulski Archive for Space Telescopes} (\texttt{MAST}) at the \textit{Space Telescope Science Institute} (STScI), which is operated by the Association of Universities for Research in Astronomy, Inc., under NASA contract NAS 5-03127 for JWST, and NAS 5–26555 for HST. The observations used in this work are associated with JWST program 1345, 1180 1176, and 2738. The authors thank all involved in the construction and operations of the telescope as well as those who designed and executed these observations. RAW acknowledges support from NASA JWST Interdisciplinary Scientist grants
NAG5-12460, NNX14AN10G and 80NSSC18K0200 from GSFC. MP acknowledges financial support from INAF mini-grant 2023 (1.05.23.04.01).

The authors thank Anthony Holloway and Sotirios Sanidas for their providing their expertise in high performance computing and other IT support throughout this work. The authors also thank Shuowen Jin and Yi-Kuan Chiang for their support with the data and simulations. This work makes use of {\tt astropy} \citep{Astropy2013,Astropy2018,Astropy2022}, {\tt matplotlib} \citep{Hunter2007}, {\tt reproject}, {\tt DrizzlePac} \citep{Hoffmann2021}, {\tt SciPy} \citep{2020SciPy-NMeth} and {\tt photutils} \citep{Bradley2022}.

\section*{Data Availability}

The JWST data used in this work are available in the Cosmic Evolution Early Release Science Survey (ID: 1345, PI: S. Finkelstein, \citealt{Bagley2023}), the JWST Advanced Deep Extragalactic Survey ( ID: 1180 and 1210, PI: Eisenstein, N. Lützgendorf, \citealt{Rieke2023} ) and the North Ecliptic Pole Time-Domain Fields (NEP-TDF) \citep{Windhorst2023} in JWST PEARLS observational program (JWST GTO program, PID 1176, 2738, PI: Rogier Windhorst, \citealt{Diego2023,Frye2023,Windhorst2023}), through the Mikulski Archive for Space Telescopes (https://mast.stsci.edu/). HST data is from the the Cosmic Assembly Near-Infrared Deep Extragalactic Legacy Survey (CANDELS; \citealt{Grogin2011, Koekemoer2011}), which be shared on reasonable request to the CANDELS team. Additional data products will be shared on reasonable request to the first author.



\bibliographystyle{mnras}
\bibliography{example} 

\begin{thebibliography}{}
\makeatletter
\relax
\def\mn@urlcharsother{\let\do\@makeother \do\$\do\&\do\#\do\^\do\_\do\%\do\~}
\def\mn@doi{\begingroup\mn@urlcharsother \@ifnextchar [ {\mn@doi@}
  {\mn@doi@[]}}
\def\mn@doi@[#1]#2{\def\@tempa{#1}\ifx\@tempa\@empty \href
  {http://dx.doi.org/#2} {doi:#2}\else \href {http://dx.doi.org/#2} {#1}\fi
  \endgroup}
\def\mn@eprint#1#2{\mn@eprint@#1:#2::\@nil}
\def\mn@eprint@arXiv#1{\href {http://arxiv.org/abs/#1} {{\tt arXiv:#1}}}
\def\mn@eprint@dblp#1{\href {http://dblp.uni-trier.de/rec/bibtex/#1.xml}
  {dblp:#1}}
\def\mn@eprint@#1:#2:#3:#4\@nil{\def\@tempa {#1}\def\@tempb {#2}\def\@tempc
  {#3}\ifx \@tempc \@empty \let \@tempc \@tempb \let \@tempb \@tempa \fi \ifx
  \@tempb \@empty \def\@tempb {arXiv}\fi \@ifundefined
  {mn@eprint@\@tempb}{\@tempb:\@tempc}{\expandafter \expandafter \csname
  mn@eprint@\@tempb\endcsname \expandafter{\@tempc}}}

\bibitem[\protect\citeauthoryear{{Adams} et~al.,}{{Adams}
  et~al.}{2023}]{Adams2023}
{Adams} N.~J.,  et~al., 2023, \mn@doi [\mnras] {10.1093/mnras/stac3347}, \href
  {https://ui.adsabs.harvard.edu/abs/2023MNRAS.518.4755A} {518, 4755}

\bibitem[\protect\citeauthoryear{{Adams} et~al.,}{{Adams}
  et~al.}{2024}]{Adams2024}
{Adams} N.~J.,  et~al., 2024, \mn@doi [\apj] {10.3847/1538-4357/ad2a7b}, \href
  {https://ui.adsabs.harvard.edu/abs/2024ApJ...965..169A} {965, 169}

\bibitem[\protect\citeauthoryear{{Alberts} \& {Noble}}{{Alberts} \&
  {Noble}}{2022}]{Alberts2022}
{Alberts} S.,  {Noble} A.,  2022, \mn@doi [Universe] {10.3390/universe8110554},
  \href {https://ui.adsabs.harvard.edu/abs/2022Univ....8..554A} {8, 554}

\bibitem[\protect\citeauthoryear{{Alberts} et~al.,}{{Alberts}
  et~al.}{2014}]{Alberts2014}
{Alberts} S.,  et~al., 2014, \mn@doi [\mnras] {10.1093/mnras/stt1897}, \href
  {https://ui.adsabs.harvard.edu/abs/2014MNRAS.437..437A} {437, 437}

\bibitem[\protect\citeauthoryear{{Andreon}, {Newman}, {Trinchieri}, {Raichoor},
  {Ellis}  \& {Treu}}{{Andreon} et~al.}{2014}]{Andreon2014}
{Andreon} S.,  {Newman} A.~B.,  {Trinchieri} G.,  {Raichoor} A.,  {Ellis}
  R.~S.,   {Treu} T.,  2014, \mn@doi [\aap] {10.1051/0004-6361/201323077},
  \href {https://ui.adsabs.harvard.edu/abs/2014A&A...565A.120A} {565, A120}

\bibitem[\protect\citeauthoryear{{Astropy Collaboration} et~al.,}{{Astropy
  Collaboration} et~al.}{2013}]{Astropy2013}
{Astropy Collaboration} et~al., 2013, \mn@doi [\aap]
  {10.1051/0004-6361/201322068}, \href
  {https://ui.adsabs.harvard.edu/abs/2013A&A...558A..33A} {558, A33}

\bibitem[\protect\citeauthoryear{{Astropy Collaboration} et~al.,}{{Astropy
  Collaboration} et~al.}{2018}]{Astropy2018}
{Astropy Collaboration} et~al., 2018, \mn@doi [\aj] {10.3847/1538-3881/aabc4f},
  \href {https://ui.adsabs.harvard.edu/abs/2018AJ....156..123A} {156, 123}

\bibitem[\protect\citeauthoryear{{Astropy Collaboration} et~al.,}{{Astropy
  Collaboration} et~al.}{2022}]{Astropy2022}
{Astropy Collaboration} et~al., 2022, \mn@doi [\apj]
  {10.3847/1538-4357/ac7c74}, \href
  {https://ui.adsabs.harvard.edu/abs/2022ApJ...935..167A} {935, 167}

\bibitem[\protect\citeauthoryear{{Austin} et~al.,}{{Austin}
  et~al.}{2024}]{Austin2024}
{Austin} D.,  et~al., 2024, \mn@doi [arXiv e-prints]
  {10.48550/arXiv.2404.10751}, \href
  {https://ui.adsabs.harvard.edu/abs/2024arXiv240410751A} {p. arXiv:2404.10751}

\bibitem[\protect\citeauthoryear{{Bagley} et~al.,}{{Bagley}
  et~al.}{2023}]{Bagley2023}
{Bagley} M.~B.,  et~al., 2023, \mn@doi [\apjl] {10.3847/2041-8213/acbb08},
  \href {https://ui.adsabs.harvard.edu/abs/2023ApJ...946L..12B} {946, L12}

\bibitem[\protect\citeauthoryear{{Barro} et~al.,}{{Barro}
  et~al.}{2023}]{2023yCat..22430022B}
{Barro} G.,  et~al., 2023, \mn@doi [VizieR Online Data Catalog]
  {10.26093/cds/vizier.22430022}, \href
  {https://ui.adsabs.harvard.edu/abs/2023yCat..22430022B} {p. J/ApJS/243/22}

\bibitem[\protect\citeauthoryear{{Behroozi}, {Wechsler}  \&
  {Conroy}}{{Behroozi} et~al.}{2013}]{Behroozi2013}
{Behroozi} P.~S.,  {Wechsler} R.~H.,   {Conroy} C.,  2013, \mn@doi [\apj]
  {10.1088/0004-637X/770/1/57}, \href
  {https://ui.adsabs.harvard.edu/abs/2013ApJ...770...57B} {770, 57}

\bibitem[\protect\citeauthoryear{{Bernardi}, {Shankar}, {Hyde}, {Mei},
  {Marulli}  \& {Sheth}}{{Bernardi} et~al.}{2010}]{Bernardi2010}
{Bernardi} M.,  {Shankar} F.,  {Hyde} J.~B.,  {Mei} S.,  {Marulli} F.,
  {Sheth} R.~K.,  2010, \mn@doi [\mnras] {10.1111/j.1365-2966.2010.16425.x},
  \href {https://ui.adsabs.harvard.edu/abs/2010MNRAS.404.2087B} {404, 2087}

\bibitem[\protect\citeauthoryear{{Bertin} \& {Arnouts}}{{Bertin} \&
  {Arnouts}}{1996}]{Bertin1996}
{Bertin} E.,  {Arnouts} S.,  1996, \mn@doi [\aaps] {10.1051/aas:1996164}, \href
  {https://ui.adsabs.harvard.edu/abs/1996A&AS..117..393B} {117, 393}

\bibitem[\protect\citeauthoryear{{Bleem} et~al.,}{{Bleem}
  et~al.}{2015}]{Bleem2015}
{Bleem} L.~E.,  et~al., 2015, \mn@doi [\apjs] {10.1088/0067-0049/216/2/27},
  \href {https://ui.adsabs.harvard.edu/abs/2015ApJS..216...27B} {216, 27}

\bibitem[\protect\citeauthoryear{Bouwens et~al.,}{Bouwens
  et~al.}{2015}]{Bouwens2015}
Bouwens R.~J.,  et~al., 2015, \mn@doi [\apj] {10.1088/0004-637X/803/1/34}, 803,
  1

\bibitem[\protect\citeauthoryear{{Bradley} et~al.,}{{Bradley}
  et~al.}{2019}]{Bradley2019}
{Bradley} L.,  et~al., 2019, {astropy/photutils: v0.7.2},
  \mn@doi{10.5281/zenodo.3568287}

\bibitem[\protect\citeauthoryear{{Bradley} et~al.,}{{Bradley}
  et~al.}{2022}]{Bradley2022}
{Bradley} L.,  et~al., 2022, {astropy/photutils: 1.5.0}, Zenodo,
  \mn@doi{10.5281/zenodo.6825092}

\bibitem[\protect\citeauthoryear{{Brammer}, {van Dokkum}  \& {Coppi}}{{Brammer}
  et~al.}{2008}]{brammer2008eazy}
{Brammer} G.~B.,  {van Dokkum} P.~G.,   {Coppi} P.,  2008, \mn@doi [\apj]
  {10.1086/591786}, \href
  {https://ui.adsabs.harvard.edu/abs/2008ApJ...686.1503B} {686, 1503}

\bibitem[\protect\citeauthoryear{{Brinch} et~al.,}{{Brinch}
  et~al.}{2024}]{Brinch2024}
{Brinch} M.,  et~al., 2024, \mn@doi [\mnras] {10.1093/mnras/stad3409}, \href
  {https://ui.adsabs.harvard.edu/abs/2024MNRAS.527.6591B} {527, 6591}

\bibitem[\protect\citeauthoryear{{Bruzual} \& {Charlot}}{{Bruzual} \&
  {Charlot}}{2003}]{Bruzual2003}
{Bruzual} G.,  {Charlot} S.,  2003, \mn@doi [\mnras]
  {10.1046/j.1365-8711.2003.06897.x}, \href
  {https://ui.adsabs.harvard.edu/abs/2003MNRAS.344.1000B} {344, 1000}

\bibitem[\protect\citeauthoryear{{Buddendiek} et~al.,}{{Buddendiek}
  et~al.}{2015}]{Buddendiek2015}
{Buddendiek} A.,  et~al., 2015, \mn@doi [\mnras] {10.1093/mnras/stv783}, \href
  {https://ui.adsabs.harvard.edu/abs/2015MNRAS.450.4248B} {450, 4248}

\bibitem[\protect\citeauthoryear{{Bunker} et~al.,}{{Bunker}
  et~al.}{2023}]{bunker2023jades}
{Bunker} A.~J.,  et~al., 2023, \mn@doi [arXiv e-prints]
  {10.48550/arXiv.2306.02467}, \href
  {https://ui.adsabs.harvard.edu/abs/2023arXiv230602467B} {p. arXiv:2306.02467}

\bibitem[\protect\citeauthoryear{{Bushouse} et~al.,}{{Bushouse}
  et~al.}{2022}]{Bushouse2022}
{Bushouse} H.,  et~al., 2022, {JWST Calibration Pipeline},
  \mn@doi{10.5281/zenodo.7325378}

\bibitem[\protect\citeauthoryear{Calzetti, Kinney  \&
  Storchi-Bergmann}{Calzetti et~al.}{1994}]{Calzetti1994}
Calzetti D.,  Kinney A.~L.,   Storchi-Bergmann T.,  1994, \mn@doi [The
  Astrophysical Journal] {10.1086/174346}, 429, 582

\bibitem[\protect\citeauthoryear{{Calzetti}, {Armus}, {Bohlin}, {Kinney},
  {Koornneef}  \& {Storchi-Bergmann}}{{Calzetti} et~al.}{2000}]{calzetti2000}
{Calzetti} D.,  {Armus} L.,  {Bohlin} R.~C.,  {Kinney} A.~L.,  {Koornneef} J.,
   {Storchi-Bergmann} T.,  2000, \mn@doi [\apj] {10.1086/308692}, \href
  {https://ui.adsabs.harvard.edu/abs/2000ApJ...533..682C} {533, 682}

\bibitem[\protect\citeauthoryear{{Carnall}, {McLure}, {Dunlop}  \&
  {Dav{\'e}}}{{Carnall} et~al.}{2018}]{carnall2018inferring}
{Carnall} A.~C.,  {McLure} R.~J.,  {Dunlop} J.~S.,   {Dav{\'e}} R.,  2018,
  \mn@doi [\mnras] {10.1093/mnras/sty2169}, \href
  {https://ui.adsabs.harvard.edu/abs/2018MNRAS.480.4379C} {480, 4379}

\bibitem[\protect\citeauthoryear{{Carnall} et~al.,}{{Carnall}
  et~al.}{2023a}]{carnall2023surprising}
{Carnall} A.~C.,  et~al., 2023a, \mn@doi [\mnras] {10.1093/mnras/stad369},
  \href {https://ui.adsabs.harvard.edu/abs/2023MNRAS.520.3974C} {520, 3974}

\bibitem[\protect\citeauthoryear{{Carnall} et~al.,}{{Carnall}
  et~al.}{2023b}]{2023Natur.619..716C}
{Carnall} A.~C.,  et~al., 2023b, \mn@doi [\nat] {10.1038/s41586-023-06158-6},
  \href {https://ui.adsabs.harvard.edu/abs/2023Natur.619..716C} {619, 716}

\bibitem[\protect\citeauthoryear{{Castignani} \& {Benoist}}{{Castignani} \&
  {Benoist}}{2016}]{Castignani2016}
{Castignani} G.,  {Benoist} C.,  2016, \mn@doi [\aap]
  {10.1051/0004-6361/201528009}, \href
  {https://ui.adsabs.harvard.edu/abs/2016A&A...595A.111C} {595, A111}

\bibitem[\protect\citeauthoryear{{Cen} \& {Ostriker}}{{Cen} \&
  {Ostriker}}{1999}]{Cen1999}
{Cen} R.,  {Ostriker} J.~P.,  1999, \mn@doi [\apjl] {10.1086/312123}, \href
  {https://ui.adsabs.harvard.edu/abs/1999ApJ...519L.109C} {519, L109}

\bibitem[\protect\citeauthoryear{{Chabrier}}{{Chabrier}}{2003}]{Chabrier2003}
{Chabrier} G.,  2003, \mn@doi [\pasp] {10.1086/376392}, \href
  {https://ui.adsabs.harvard.edu/abs/2003PASP..115..763C} {115, 763}

\bibitem[\protect\citeauthoryear{{Chapman}, {Blain}, {Ibata}, {Ivison}, {Smail}
   \& {Morrison}}{{Chapman} et~al.}{2009}]{Chapman2009}
{Chapman} S.~C.,  {Blain} A.,  {Ibata} R.,  {Ivison} R.~J.,  {Smail} I.,
  {Morrison} G.,  2009, \mn@doi [\apj] {10.1088/0004-637X/691/1/560}, \href
  {https://ui.adsabs.harvard.edu/abs/2009ApJ...691..560C} {691, 560}

\bibitem[\protect\citeauthoryear{{Chen}, {Stark}, {Mason}, {Topping},
  {Whitler}, {Tang}, {Endsley}  \& {Charlot}}{{Chen} et~al.}{2024}]{Chen2024}
{Chen} Z.,  {Stark} D.~P.,  {Mason} C.,  {Topping} M.~W.,  {Whitler} L.,
  {Tang} M.,  {Endsley} R.,   {Charlot} S.,  2024, \mn@doi [\mnras]
  {10.1093/mnras/stae455}, \href
  {https://ui.adsabs.harvard.edu/abs/2024MNRAS.tmp..523C} {}

\bibitem[\protect\citeauthoryear{{Chiang}, {Overzier}  \& {Gebhardt}}{{Chiang}
  et~al.}{2013}]{Chiang2013}
{Chiang} Y.-K.,  {Overzier} R.,   {Gebhardt} K.,  2013, \mn@doi [\apj]
  {10.1088/0004-637X/779/2/127}, \href
  {https://ui.adsabs.harvard.edu/abs/2013ApJ...779..127C} {779, 127}

\bibitem[\protect\citeauthoryear{{Chiang} et~al.,}{{Chiang}
  et~al.}{2015}]{Chiang2015}
{Chiang} Y.-K.,  et~al., 2015, \mn@doi [\apj] {10.1088/0004-637X/808/1/37},
  \href {https://ui.adsabs.harvard.edu/abs/2015ApJ...808...37C} {808, 37}

\bibitem[\protect\citeauthoryear{{Chiang}, {Overzier}, {Gebhardt}  \&
  {Henriques}}{{Chiang} et~al.}{2017}]{Chiang2017}
{Chiang} Y.-K.,  {Overzier} R.~A.,  {Gebhardt} K.,   {Henriques} B.,  2017,
  \mn@doi [\apjl] {10.3847/2041-8213/aa7e7b}, \href
  {https://ui.adsabs.harvard.edu/abs/2017ApJ...844L..23C} {844, L23}

\bibitem[\protect\citeauthoryear{{Coogan} et~al.,}{{Coogan}
  et~al.}{2023}]{Coogan2023}
{Coogan} R.~T.,  et~al., 2023, \mn@doi [\aap] {10.1051/0004-6361/202346172},
  \href {https://ui.adsabs.harvard.edu/abs/2023A&A...677A...3C} {677, A3}

\bibitem[\protect\citeauthoryear{{Cucciati} et~al.,}{{Cucciati}
  et~al.}{2014}]{Cucciati2014}
{Cucciati} O.,  et~al., 2014, \mn@doi [\aap] {10.1051/0004-6361/201423811},
  \href {https://ui.adsabs.harvard.edu/abs/2014A&A...570A..16C} {570, A16}

\bibitem[\protect\citeauthoryear{{Cucciati} et~al.,}{{Cucciati}
  et~al.}{2018}]{Cucciati2018}
{Cucciati} O.,  et~al., 2018, \mn@doi [\aap] {10.1051/0004-6361/201833655},
  \href {https://ui.adsabs.harvard.edu/abs/2018A&A...619A..49C} {619, A49}

\bibitem[\protect\citeauthoryear{{Cullen} et~al.,}{{Cullen}
  et~al.}{2023}]{Cullen2023}
{Cullen} F.,  et~al., 2023, \mn@doi [\mnras] {10.1093/mnras/stad073}, \href
  {https://ui.adsabs.harvard.edu/abs/2023MNRAS.520...14C} {520, 14}

\bibitem[\protect\citeauthoryear{{Davidzon} et~al.,}{{Davidzon}
  et~al.}{2016}]{Davidzon2016}
{Davidzon} I.,  et~al., 2016, \mn@doi [\aap] {10.1051/0004-6361/201527129},
  \href {https://ui.adsabs.harvard.edu/abs/2016A&A...586A..23D} {586, A23}

\bibitem[\protect\citeauthoryear{{Dey}, {Lee}, {Reddy}, {Cooper}, {Inami},
  {Hong}, {Gonzalez}  \& {Jannuzi}}{{Dey} et~al.}{2016}]{Dey2016}
{Dey} A.,  {Lee} K.-S.,  {Reddy} N.,  {Cooper} M.,  {Inami} H.,  {Hong} S.,
  {Gonzalez} A.~H.,   {Jannuzi} B.~T.,  2016, \mn@doi [\apj]
  {10.3847/0004-637X/823/1/11}, \href
  {https://ui.adsabs.harvard.edu/abs/2016ApJ...823...11D} {823, 11}

\bibitem[\protect\citeauthoryear{{Di Mascolo} et~al.,}{{Di Mascolo}
  et~al.}{2023}]{DiMascolo2023}
{Di Mascolo} L.,  et~al., 2023, \mn@doi [\nat] {10.1038/s41586-023-05761-x},
  \href {https://ui.adsabs.harvard.edu/abs/2023Natur.615..809D} {615, 809}

\bibitem[\protect\citeauthoryear{{Diego} et~al.,}{{Diego}
  et~al.}{2023}]{Diego2023}
{Diego} J.~M.,  et~al., 2023, \mn@doi [\aap] {10.1051/0004-6361/202245238},
  \href {https://ui.adsabs.harvard.edu/abs/2023A&A...672A...3D} {672, A3}

\bibitem[\protect\citeauthoryear{{Diener} et~al.,}{{Diener}
  et~al.}{2015}]{Diener2015}
{Diener} C.,  et~al., 2015, \mn@doi [\apj] {10.1088/0004-637X/802/1/31}, \href
  {https://ui.adsabs.harvard.edu/abs/2015ApJ...802...31D} {802, 31}

\bibitem[\protect\citeauthoryear{{Ferland} et~al.,}{{Ferland}
  et~al.}{2013}]{2013RMxAA..49..137F}
{Ferland} G.~J.,  et~al., 2013, \rmxaa, \href
  {https://ui.adsabs.harvard.edu/abs/2013RMxAA..49..137F} {49, 137}

\bibitem[\protect\citeauthoryear{{Ferruit} et~al.,}{{Ferruit}
  et~al.}{2022}]{ferruit2022near}
{Ferruit} P.,  et~al., 2022, \mn@doi [\aap] {10.1051/0004-6361/202142673},
  \href {https://ui.adsabs.harvard.edu/abs/2022A&A...661A..81F} {661, A81}

\bibitem[\protect\citeauthoryear{{Frye} et~al.,}{{Frye}
  et~al.}{2023}]{Frye2023}
{Frye} B.~L.,  et~al., 2023, \mn@doi [\apj] {10.3847/1538-4357/acd929}, \href
  {https://ui.adsabs.harvard.edu/abs/2023ApJ...952...81F} {952, 81}

\bibitem[\protect\citeauthoryear{{Gardner} et~al.,}{{Gardner}
  et~al.}{2006}]{Gardner2006}
{Gardner} J.~P.,  et~al., 2006, \mn@doi [\ssr] {10.1007/s11214-006-8315-7},
  \href {https://ui.adsabs.harvard.edu/abs/2006SSRv..123..485G} {123, 485}

\bibitem[\protect\citeauthoryear{{George} et~al.,}{{George}
  et~al.}{2011}]{George2011}
{George} M.~R.,  et~al., 2011, \mn@doi [\apj] {10.1088/0004-637X/742/2/125},
  \href {https://ui.adsabs.harvard.edu/abs/2011ApJ...742..125G} {742, 125}

\bibitem[\protect\citeauthoryear{{Gobat} et~al.,}{{Gobat}
  et~al.}{2011}]{Gobat2011}
{Gobat} R.,  et~al., 2011, \mn@doi [\aap] {10.1051/0004-6361/201016084}, \href
  {https://ui.adsabs.harvard.edu/abs/2011A&A...526A.133G} {526, A133}

\bibitem[\protect\citeauthoryear{{Gobat} et~al.,}{{Gobat}
  et~al.}{2013}]{Gobat2013}
{Gobat} R.,  et~al., 2013, \mn@doi [\apj] {10.1088/0004-637X/776/1/9}, \href
  {https://ui.adsabs.harvard.edu/abs/2013ApJ...776....9G} {776, 9}

\bibitem[\protect\citeauthoryear{{G{\'o}mez} et~al.,}{{G{\'o}mez}
  et~al.}{2003}]{Gomez2003}
{G{\'o}mez} P.~L.,  et~al., 2003, \mn@doi [\apj] {10.1086/345593}, \href
  {https://ui.adsabs.harvard.edu/abs/2003ApJ...584..210G} {584, 210}

\bibitem[\protect\citeauthoryear{{Grogin} et~al.,}{{Grogin}
  et~al.}{2011}]{Grogin2011}
{Grogin} N.~A.,  et~al., 2011, \mn@doi [\apjs] {10.1088/0067-0049/197/2/35},
  \href {https://ui.adsabs.harvard.edu/abs/2011ApJS..197...35G} {197, 35}

\bibitem[\protect\citeauthoryear{{Guo} et~al.,}{{Guo} et~al.}{2011}]{Guo2011}
{Guo} Q.,  et~al., 2011, \mn@doi [\mnras] {10.1111/j.1365-2966.2010.18114.x},
  \href {https://ui.adsabs.harvard.edu/abs/2011MNRAS.413..101G} {413, 101}

\bibitem[\protect\citeauthoryear{{Guo} et~al.,}{{Guo}
  et~al.}{2013}]{2013ApJS..207...24G}
{Guo} Y.,  et~al., 2013, \mn@doi [\apjs] {10.1088/0067-0049/207/2/24}, \href
  {https://ui.adsabs.harvard.edu/abs/2013ApJS..207...24G} {207, 24}

\bibitem[\protect\citeauthoryear{{Hansen}, {McKay}, {Wechsler}, {Annis},
  {Sheldon}  \& {Kimball}}{{Hansen} et~al.}{2005}]{Hansen2005}
{Hansen} S.~M.,  {McKay} T.~A.,  {Wechsler} R.~H.,  {Annis} J.,  {Sheldon}
  E.~S.,   {Kimball} A.,  2005, \mn@doi [\apj] {10.1086/444554}, \href
  {https://ui.adsabs.harvard.edu/abs/2005ApJ...633..122H} {633, 122}

\bibitem[\protect\citeauthoryear{{Harikane} et~al.,}{{Harikane}
  et~al.}{2019}]{Harikane2019}
{Harikane} Y.,  et~al., 2019, \mn@doi [\apj] {10.3847/1538-4357/ab2cd5}, \href
  {https://ui.adsabs.harvard.edu/abs/2019ApJ...883..142H} {883, 142}

\bibitem[\protect\citeauthoryear{{Harikane} et~al.,}{{Harikane}
  et~al.}{2023}]{Harikane2023}
{Harikane} Y.,  et~al., 2023, \mn@doi [\apjs] {10.3847/1538-4365/acaaa9}, \href
  {https://ui.adsabs.harvard.edu/abs/2023ApJS..265....5H} {265, 5}

\bibitem[\protect\citeauthoryear{{Harvey} et~al.,}{{Harvey}
  et~al.}{2024}]{Harvey2024}
{Harvey} T.,  et~al., 2024, \mn@doi [arXiv e-prints]
  {10.48550/arXiv.2403.03908}, \href
  {https://ui.adsabs.harvard.edu/abs/2024arXiv240303908H} {p. arXiv:2403.03908}

\bibitem[\protect\citeauthoryear{{Heintz} et~al.,}{{Heintz}
  et~al.}{2023}]{Heintz2023}
{Heintz} K.~E.,  et~al., 2023, \mn@doi [arXiv e-prints]
  {10.48550/arXiv.2306.00647}, \href
  {https://ui.adsabs.harvard.edu/abs/2023arXiv230600647H} {p. arXiv:2306.00647}

\bibitem[\protect\citeauthoryear{{Heintz} et~al.,}{{Heintz}
  et~al.}{2024}]{Heintz2024}
{Heintz} K.~E.,  et~al., 2024, \mn@doi [arXiv e-prints]
  {10.48550/arXiv.2404.02211}, \href
  {https://ui.adsabs.harvard.edu/abs/2024arXiv240402211H} {p. arXiv:2404.02211}

\bibitem[\protect\citeauthoryear{{Helton} et~al.,}{{Helton}
  et~al.}{2023}]{Helton2023}
{Helton} J.~M.,  et~al., 2023, \mn@doi [arXiv e-prints]
  {10.48550/arXiv.2311.04270}, \href
  {https://ui.adsabs.harvard.edu/abs/2023arXiv231104270H} {p. arXiv:2311.04270}

\bibitem[\protect\citeauthoryear{{Herard-Demanche} et~al.,}{{Herard-Demanche}
  et~al.}{2023}]{Herard-Demanche2023}
{Herard-Demanche} T.,  et~al., 2023, \mn@doi [arXiv e-prints]
  {10.48550/arXiv.2309.04525}, \href
  {https://ui.adsabs.harvard.edu/abs/2023arXiv230904525H} {p. arXiv:2309.04525}

\bibitem[\protect\citeauthoryear{{Hilton} et~al.,}{{Hilton}
  et~al.}{2021}]{Hilton2021}
{Hilton} M.,  et~al., 2021, \mn@doi [\apjs] {10.3847/1538-4365/abd023}, \href
  {https://ui.adsabs.harvard.edu/abs/2021ApJS..253....3H} {253, 3}

\bibitem[\protect\citeauthoryear{{Hoffmann}, {Mack}, {Avila}, {Martlin},
  {Cohen}  \& {Bajaj}}{{Hoffmann} et~al.}{2021}]{Hoffmann2021}
{Hoffmann} S.~L.,  {Mack} J.,  {Avila} R.,  {Martlin} C.,  {Cohen} Y.,
  {Bajaj} V.,  2021, in American Astronomical Society Meeting Abstracts. p.
  216.02

\bibitem[\protect\citeauthoryear{{Huertas-Company}, {Shankar}, {Mei},
  {Bernardi}, {Aguerri}, {Meert}  \& {Vikram}}{{Huertas-Company}
  et~al.}{2013}]{Huertas-Company2013}
{Huertas-Company} M.,  {Shankar} F.,  {Mei} S.,  {Bernardi} M.,  {Aguerri}
  J.~A.~L.,  {Meert} A.,   {Vikram} V.,  2013, \mn@doi [\apj]
  {10.1088/0004-637X/779/1/29}, \href
  {https://ui.adsabs.harvard.edu/abs/2013ApJ...779...29H} {779, 29}

\bibitem[\protect\citeauthoryear{Hunter}{Hunter}{2007}]{Hunter2007}
Hunter J.~D.,  2007, \mn@doi [Computing in Science \& Engineering]
  {10.1109/MCSE.2007.55}, 9, 90

\bibitem[\protect\citeauthoryear{{Jin} et~al.,}{{Jin} et~al.}{2024}]{Jin2024}
{Jin} S.,  et~al., 2024, \mn@doi [\aap] {10.1051/0004-6361/202348540}, \href
  {https://ui.adsabs.harvard.edu/abs/2024A&A...683L...4J} {683, L4}

\bibitem[\protect\citeauthoryear{{Kashino}, {Lilly}, {Matthee}, {Eilers},
  {Mackenzie}, {Bordoloi}  \& {Simcoe}}{{Kashino} et~al.}{2023}]{Kashino2023}
{Kashino} D.,  {Lilly} S.~J.,  {Matthee} J.,  {Eilers} A.-C.,  {Mackenzie} R.,
  {Bordoloi} R.,   {Simcoe} R.~A.,  2023, \mn@doi [\apj]
  {10.3847/1538-4357/acc588}, \href
  {https://ui.adsabs.harvard.edu/abs/2023ApJ...950...66K} {950, 66}

\bibitem[\protect\citeauthoryear{{Kennicutt} \& {Evans}}{{Kennicutt} \&
  {Evans}}{2012}]{kennicutt2012star}
{Kennicutt} R.~C.,  {Evans} N.~J.,  2012, \mn@doi [\araa]
  {10.1146/annurev-astro-081811-125610}, \href
  {https://ui.adsabs.harvard.edu/abs/2012ARA&A..50..531K} {50, 531}

\bibitem[\protect\citeauthoryear{{Koekemoer} et~al.,}{{Koekemoer}
  et~al.}{2011}]{Koekemoer2011}
{Koekemoer} A.~M.,  et~al., 2011, \mn@doi [\apjs] {10.1088/0067-0049/197/2/36},
  \href {https://ui.adsabs.harvard.edu/abs/2011ApJS..197...36K} {197, 36}

\bibitem[\protect\citeauthoryear{{Kriek} et~al.,}{{Kriek}
  et~al.}{2015}]{Kriek2015}
{Kriek} M.,  et~al., 2015, \mn@doi [\apjs] {10.1088/0067-0049/218/2/15}, \href
  {https://ui.adsabs.harvard.edu/abs/2015ApJS..218...15K} {218, 15}

\bibitem[\protect\citeauthoryear{{Kroupa}}{{Kroupa}}{2001}]{kroupa2001MNRAS.322..231K}
{Kroupa} P.,  2001, \mn@doi [\mnras] {10.1046/j.1365-8711.2001.04022.x}, \href
  {https://ui.adsabs.harvard.edu/abs/2001MNRAS.322..231K} {322, 231}

\bibitem[\protect\citeauthoryear{{Laporte}, {Zitrin}, {Dole},
  {Roberts-Borsani}, {Furtak}  \& {Witten}}{{Laporte}
  et~al.}{2022}]{Laporte2022}
{Laporte} N.,  {Zitrin} A.,  {Dole} H.,  {Roberts-Borsani} G.,  {Furtak} L.~J.,
    {Witten} C.,  2022, \mn@doi [\aap] {10.1051/0004-6361/202244719}, \href
  {https://ui.adsabs.harvard.edu/abs/2022A&A...667L...3L} {667, L3}

\bibitem[\protect\citeauthoryear{{Larson} et~al.,}{{Larson}
  et~al.}{2023}]{larson2022spectral}
{Larson} R.~L.,  et~al., 2023, \mn@doi [\apj] {10.3847/1538-4357/acfed4}, \href
  {https://ui.adsabs.harvard.edu/abs/2023ApJ...958..141L} {958, 141}

\bibitem[\protect\citeauthoryear{Lee et~al.,}{Lee et~al.}{2014}]{Lee2014}
Lee J.,  et~al., 2014, \mn@doi [\mnras] {10.1093/mnras/stu2039}, 445, 4197

\bibitem[\protect\citeauthoryear{{Lemaux} et~al.,}{{Lemaux}
  et~al.}{2014}]{Lemaux2014}
{Lemaux} B.~C.,  et~al., 2014, \mn@doi [\aap] {10.1051/0004-6361/201423828},
  \href {https://ui.adsabs.harvard.edu/abs/2014A&A...572A..41L} {572, A41}

\bibitem[\protect\citeauthoryear{{Lemaux} et~al.,}{{Lemaux}
  et~al.}{2018}]{Lemaux2018}
{Lemaux} B.~C.,  et~al., 2018, \mn@doi [\aap] {10.1051/0004-6361/201730870},
  \href {https://ui.adsabs.harvard.edu/abs/2018A&A...615A..77L} {615, A77}

\bibitem[\protect\citeauthoryear{{Lemaux} et~al.,}{{Lemaux}
  et~al.}{2019}]{Lemaux2019}
{Lemaux} B.~C.,  et~al., 2019, \mn@doi [\mnras] {10.1093/mnras/stz2661}, \href
  {https://ui.adsabs.harvard.edu/abs/2019MNRAS.490.1231L} {490, 1231}

\bibitem[\protect\citeauthoryear{{Lemaux} et~al.,}{{Lemaux}
  et~al.}{2022}]{Lemaux2022}
{Lemaux} B.~C.,  et~al., 2022, \mn@doi [\aap] {10.1051/0004-6361/202039346},
  \href {https://ui.adsabs.harvard.edu/abs/2022A&A...662A..33L} {662, A33}

\bibitem[\protect\citeauthoryear{{Li} et~al.,}{{Li} et~al.}{2021}]{LiQiong2021}
{Li} Q.,  et~al., 2021, \mn@doi [\apj] {10.3847/1538-4357/ac29c6}, \href
  {https://ui.adsabs.harvard.edu/abs/2021ApJ...922..236L} {922, 236}

\bibitem[\protect\citeauthoryear{{Li} et~al.,}{{Li} et~al.}{2023}]{QiongLi2023}
{Li} Q.,  et~al., 2023, \mn@doi [\apj] {10.3847/1538-4357/acd7f3}, \href
  {https://ui.adsabs.harvard.edu/abs/2023ApJ...954..174L} {954, 174}

\bibitem[\protect\citeauthoryear{{Lopes}, {Rembold}, {Ribeiro}, {Nascimento}
  \& {Vajgel}}{{Lopes} et~al.}{2016}]{Lopes2016}
{Lopes} P.~A.~A.,  {Rembold} S.~B.,  {Ribeiro} A.~L.~B.,  {Nascimento} R.~S.,
  {Vajgel} B.,  2016, \mn@doi [\mnras] {10.1093/mnras/stw1497}, \href
  {https://ui.adsabs.harvard.edu/abs/2016MNRAS.461.2559L} {461, 2559}

\bibitem[\protect\citeauthoryear{{Madau}}{{Madau}}{1995}]{1995ApJ...441...18M}
{Madau} P.,  1995, \mn@doi [\apj] {10.1086/175332}, \href
  {https://ui.adsabs.harvard.edu/abs/1995ApJ...441...18M} {441, 18}

\bibitem[\protect\citeauthoryear{{Magnelli} et~al.,}{{Magnelli}
  et~al.}{2014}]{Magnelli2014}
{Magnelli} B.,  et~al., 2014, \mn@doi [\aap] {10.1051/0004-6361/201322217},
  \href {https://ui.adsabs.harvard.edu/abs/2014A&A...561A..86M} {561, A86}

\bibitem[\protect\citeauthoryear{{Mantz} et~al.,}{{Mantz}
  et~al.}{2018}]{Mantz2018}
{Mantz} A.~B.,  et~al., 2018, \mn@doi [\aap] {10.1051/0004-6361/201630096},
  \href {https://ui.adsabs.harvard.edu/abs/2018A&A...620A...2M} {620, A2}

\bibitem[\protect\citeauthoryear{{Matsuda} et~al.,}{{Matsuda}
  et~al.}{2005}]{Matsuda2005}
{Matsuda} Y.,  et~al., 2005, \mn@doi [\apjl] {10.1086/499071}, \href
  {https://ui.adsabs.harvard.edu/abs/2005ApJ...634L.125M} {634, L125}

\bibitem[\protect\citeauthoryear{{Meurer}, {Heckman}  \& {Calzetti}}{{Meurer}
  et~al.}{1999}]{Meurer1999}
{Meurer} G.~R.,  {Heckman} T.~M.,   {Calzetti} D.,  1999, \mn@doi [\apj]
  {10.1086/307523}, \href
  {https://ui.adsabs.harvard.edu/abs/1999ApJ...521...64M} {521, 64}

\bibitem[\protect\citeauthoryear{{Miley} \& {De Breuck}}{{Miley} \& {De
  Breuck}}{2008}]{Miley2008}
{Miley} G.,  {De Breuck} C.,  2008, \mn@doi [\aapr]
  {10.1007/s00159-007-0008-z}, \href
  {https://ui.adsabs.harvard.edu/abs/2008A&ARv..15...67M} {15, 67}

\bibitem[\protect\citeauthoryear{{Miller} et~al.,}{{Miller}
  et~al.}{2018}]{Miller2018}
{Miller} T.~B.,  et~al., 2018, \mn@doi [\nat] {10.1038/s41586-018-0025-2},
  \href {https://ui.adsabs.harvard.edu/abs/2018Natur.556..469M} {556, 469}

\bibitem[\protect\citeauthoryear{{Momcheva} et~al.,}{{Momcheva}
  et~al.}{2016}]{Momcheva2016}
{Momcheva} I.~G.,  et~al., 2016, \mn@doi [\apjs] {10.3847/0067-0049/225/2/27},
  \href {https://ui.adsabs.harvard.edu/abs/2016ApJS..225...27M} {225, 27}

\bibitem[\protect\citeauthoryear{{Monson} et~al.,}{{Monson}
  et~al.}{2021}]{Monson2021}
{Monson} E.~B.,  et~al., 2021, \mn@doi [\apj] {10.3847/1538-4357/ac0f84}, \href
  {https://ui.adsabs.harvard.edu/abs/2021ApJ...919...51M} {919, 51}

\bibitem[\protect\citeauthoryear{{Montenegro-Taborda}, {Rodriguez-Gomez},
  {Pillepich}, {Avila-Reese}, {Sales}, {Rodr{\'\i}guez-Puebla}  \&
  {Hernquist}}{{Montenegro-Taborda} et~al.}{2023}]{Montenegro-Taborda2023}
{Montenegro-Taborda} D.,  {Rodriguez-Gomez} V.,  {Pillepich} A.,  {Avila-Reese}
  V.,  {Sales} L.~V.,  {Rodr{\'\i}guez-Puebla} A.,   {Hernquist} L.,  2023,
  \mn@doi [\mnras] {10.1093/mnras/stad586}, \href
  {https://ui.adsabs.harvard.edu/abs/2023MNRAS.521..800M} {521, 800}

\bibitem[\protect\citeauthoryear{{Morishita} et~al.,}{{Morishita}
  et~al.}{2023}]{Morishita2023}
{Morishita} T.,  et~al., 2023, \mn@doi [\apjl] {10.3847/2041-8213/acb99e},
  \href {https://ui.adsabs.harvard.edu/abs/2023ApJ...947L..24M} {947, L24}

\bibitem[\protect\citeauthoryear{{Morselli} et~al.,}{{Morselli}
  et~al.}{2014}]{Morselli2014}
{Morselli} L.,  et~al., 2014, \mn@doi [\aap] {10.1051/0004-6361/201423853},
  \href {https://ui.adsabs.harvard.edu/abs/2014A&A...568A...1M} {568, A1}

\bibitem[\protect\citeauthoryear{{Muldrew}, {Hatch}  \& {Cooke}}{{Muldrew}
  et~al.}{2015}]{Muldrew2015}
{Muldrew} S.~I.,  {Hatch} N.~A.,   {Cooke} E.~A.,  2015, \mn@doi [\mnras]
  {10.1093/mnras/stv1449}, \href
  {https://ui.adsabs.harvard.edu/abs/2015MNRAS.452.2528M} {452, 2528}

\bibitem[\protect\citeauthoryear{{Nantais} et~al.,}{{Nantais}
  et~al.}{2017}]{Nantais2017}
{Nantais} J.~B.,  et~al., 2017, \mn@doi [\mnras] {10.1093/mnrasl/slw224}, \href
  {https://ui.adsabs.harvard.edu/abs/2017MNRAS.465L.104N} {465, L104}

\bibitem[\protect\citeauthoryear{{Nayyeri} et~al.,}{{Nayyeri}
  et~al.}{2017}]{2017ApJS..228....7N}
{Nayyeri} H.,  et~al., 2017, \mn@doi [\apjs] {10.3847/1538-4365/228/1/7}, \href
  {https://ui.adsabs.harvard.edu/abs/2017ApJS..228....7N} {228, 7}

\bibitem[\protect\citeauthoryear{{Newman}, {Ellis}, {Bundy}  \&
  {Treu}}{{Newman} et~al.}{2012}]{Newman2012}
{Newman} A.~B.,  {Ellis} R.~S.,  {Bundy} K.,   {Treu} T.,  2012, \mn@doi [\apj]
  {10.1088/0004-637X/746/2/162}, \href
  {https://ui.adsabs.harvard.edu/abs/2012ApJ...746..162N} {746, 162}

\bibitem[\protect\citeauthoryear{{Newman} et~al.,}{{Newman}
  et~al.}{2013}]{Newman2013}
{Newman} J.~A.,  et~al., 2013, \mn@doi [\apjs] {10.1088/0067-0049/208/1/5},
  \href {https://ui.adsabs.harvard.edu/abs/2013ApJS..208....5N} {208, 5}

\bibitem[\protect\citeauthoryear{{Noirot} et~al.,}{{Noirot}
  et~al.}{2023}]{Noirot2023}
{Noirot} G.,  et~al., 2023, \mn@doi [\mnras] {10.1093/mnras/stad1019}, \href
  {https://ui.adsabs.harvard.edu/abs/2023MNRAS.525.1867N} {525, 1867}

\bibitem[\protect\citeauthoryear{{Oke} \& {Gunn}}{{Oke} \&
  {Gunn}}{1983}]{Oke1983}
{Oke} J.~B.,  {Gunn} J.~E.,  1983, \mn@doi [\apj] {10.1086/160817}, \href
  {https://ui.adsabs.harvard.edu/abs/1983ApJ...266..713O} {266, 713}

\bibitem[\protect\citeauthoryear{{Ormerod} et~al.,}{{Ormerod}
  et~al.}{2023}]{ormerod2023}
{Ormerod} K.,  et~al., 2023, \mn@doi [arXiv e-prints]
  {10.48550/arXiv.2309.04377}, \href
  {https://ui.adsabs.harvard.edu/abs/2023arXiv230904377O} {p. arXiv:2309.04377}

\bibitem[\protect\citeauthoryear{{Oteo} et~al.,}{{Oteo}
  et~al.}{2018}]{Oteo2018}
{Oteo} I.,  et~al., 2018, \mn@doi [\apj] {10.3847/1538-4357/aaa1f1}, \href
  {https://ui.adsabs.harvard.edu/abs/2018ApJ...856...72O} {856, 72}

\bibitem[\protect\citeauthoryear{{Ouchi} et~al.,}{{Ouchi}
  et~al.}{2005}]{Ouchi2005}
{Ouchi} M.,  et~al., 2005, \mn@doi [\apjl] {10.1086/499519}, \href
  {https://ui.adsabs.harvard.edu/abs/2005ApJ...635L.117O} {635, L117}

\bibitem[\protect\citeauthoryear{{Overzier}}{{Overzier}}{2016}]{Overzier2016}
{Overzier} R.~A.,  2016, \mn@doi [\aapr] {10.1007/s00159-016-0100-3}, \href
  {https://ui.adsabs.harvard.edu/abs/2016A&ARv..24...14O} {24, 14}

\bibitem[\protect\citeauthoryear{{Peng}, {Ho}, {Impey}  \& {Rix}}{{Peng}
  et~al.}{2002}]{Peng2002Galfit1}
{Peng} C.~Y.,  {Ho} L.~C.,  {Impey} C.~D.,   {Rix} H.-W.,  2002, \mn@doi [\aj]
  {10.1086/340952}, \href
  {https://ui.adsabs.harvard.edu/abs/2002AJ....124..266P} {124, 266}

\bibitem[\protect\citeauthoryear{{Peng}, {Ho}, {Impey}  \& {Rix}}{{Peng}
  et~al.}{2010}]{Peng2010Galfit2}
{Peng} C.~Y.,  {Ho} L.~C.,  {Impey} C.~D.,   {Rix} H.-W.,  2010, \mn@doi [\aj]
  {10.1088/0004-6256/139/6/2097}, \href
  {https://ui.adsabs.harvard.edu/abs/2010AJ....139.2097P} {139, 2097}

\bibitem[\protect\citeauthoryear{{P{\'e}rez-Mart{\'\i}nez}
  et~al.,}{{P{\'e}rez-Mart{\'\i}nez} et~al.}{2024}]{Perez-Martinez2024}
{P{\'e}rez-Mart{\'\i}nez} J.~M.,  et~al., 2024, \mn@doi [\mnras]
  {10.1093/mnras/stad3805}, \href
  {https://ui.adsabs.harvard.edu/abs/2024MNRAS.52710221P} {527, 10221}

\bibitem[\protect\citeauthoryear{{Perrin}, {Soummer}, {Elliott}, {Lallo}  \&
  {Sivaramakrishnan}}{{Perrin} et~al.}{2012}]{Perrin2012}
{Perrin} M.~D.,  {Soummer} R.,  {Elliott} E.~M.,  {Lallo} M.~D.,
  {Sivaramakrishnan} A.,  2012, in {Clampin} M.~C.,  {Fazio} G.~G.,  {MacEwen}
  H.~A.,   {Oschmann} Jacobus~M. J.,  eds,  Society of Photo-Optical
  Instrumentation Engineers (SPIE) Conference Series Vol. 8442, Space
  Telescopes and Instrumentation 2012: Optical, Infrared, and Millimeter Wave.
  p. 84423D, \mn@doi{10.1117/12.925230}

\bibitem[\protect\citeauthoryear{{Perrin}, {Sivaramakrishnan}, {Lajoie},
  {Elliott}, {Pueyo}, {Ravindranath}  \& {Albert}}{{Perrin}
  et~al.}{2014}]{Perrin2014}
{Perrin} M.~D.,  {Sivaramakrishnan} A.,  {Lajoie} C.-P.,  {Elliott} E.,
  {Pueyo} L.,  {Ravindranath} S.,   {Albert} L.,  2014, in {Oschmann}
  Jacobus~M. J.,  {Clampin} M.,  {Fazio} G.~G.,   {MacEwen} H.~A.,  eds,
  Society of Photo-Optical Instrumentation Engineers (SPIE) Conference Series
  Vol. 9143, Space Telescopes and Instrumentation 2014: Optical, Infrared, and
  Millimeter Wave. p. 91433X, \mn@doi{10.1117/12.2056689}

\bibitem[\protect\citeauthoryear{{Pillepich} et~al.,}{{Pillepich}
  et~al.}{2018}]{Pillepich2018}
{Pillepich} A.,  et~al., 2018, \mn@doi [\mnras] {10.1093/mnras/stx3112}, \href
  {https://ui.adsabs.harvard.edu/abs/2018MNRAS.475..648P} {475, 648}

\bibitem[\protect\citeauthoryear{{Polletta} et~al.,}{{Polletta}
  et~al.}{2021}]{Polletta2021}
{Polletta} M.,  et~al., 2021, \mn@doi [\aap] {10.1051/0004-6361/202140612},
  \href {https://ui.adsabs.harvard.edu/abs/2021A&A...654A.121P} {654, A121}

\bibitem[\protect\citeauthoryear{{Polletta} et~al.,}{{Polletta}
  et~al.}{2024}]{Polletta2024}
{Polletta} M.,  et~al., 2024, \mn@doi [arXiv e-prints]
  {10.48550/arXiv.2405.07986}, \href
  {https://ui.adsabs.harvard.edu/abs/2024arXiv240507986P} {p. arXiv:2405.07986}

\bibitem[\protect\citeauthoryear{{Popesso} et~al.,}{{Popesso}
  et~al.}{2023}]{Popesso2023}
{Popesso} P.,  et~al., 2023, \mn@doi [\mnras] {10.1093/mnras/stac3214}, \href
  {https://ui.adsabs.harvard.edu/abs/2023MNRAS.519.1526P} {519, 1526}

\bibitem[\protect\citeauthoryear{{Reichardt} et~al.,}{{Reichardt}
  et~al.}{2013}]{Reichardt2013}
{Reichardt} C.~L.,  et~al., 2013, \mn@doi [\apj] {10.1088/0004-637X/763/2/127},
  \href {https://ui.adsabs.harvard.edu/abs/2013ApJ...763..127R} {763, 127}

\bibitem[\protect\citeauthoryear{{Rieke} et~al.,}{{Rieke}
  et~al.}{2023}]{Rieke2023}
{Rieke} M.~J.,  et~al., 2023, \mn@doi [\apjs] {10.3847/1538-4365/acf44d}, \href
  {https://ui.adsabs.harvard.edu/abs/2023ApJS..269...16R} {269, 16}

\bibitem[\protect\citeauthoryear{{Rogers}, {McLure}  \& {Dunlop}}{{Rogers}
  et~al.}{2013}]{Rogers2013}
{Rogers} A.~B.,  {McLure} R.~J.,   {Dunlop} J.~S.,  2013, \mn@doi [\mnras]
  {10.1093/mnras/sts515}, \href
  {https://ui.adsabs.harvard.edu/abs/2013MNRAS.429.2456R} {429, 2456}

\bibitem[\protect\citeauthoryear{{Rogers} et~al.,}{{Rogers}
  et~al.}{2014}]{Rogers2014}
{Rogers} A.~B.,  et~al., 2014, \mn@doi [\mnras] {10.1093/mnras/stu558}, \href
  {https://ui.adsabs.harvard.edu/abs/2014MNRAS.440.3714R} {440, 3714}

\bibitem[\protect\citeauthoryear{{Rosati}, {Borgani}  \& {Norman}}{{Rosati}
  et~al.}{2002}]{Rosati2002}
{Rosati} P.,  {Borgani} S.,   {Norman} C.,  2002, \mn@doi [\araa]
  {10.1146/annurev.astro.40.120401.150547}, \href
  {https://ui.adsabs.harvard.edu/abs/2002ARA&A..40..539R} {40, 539}

\bibitem[\protect\citeauthoryear{{Rosati} et~al.,}{{Rosati}
  et~al.}{2004}]{Rosati2004}
{Rosati} P.,  et~al., 2004, \mn@doi [\aj] {10.1086/379857}, \href
  {https://ui.adsabs.harvard.edu/abs/2004AJ....127..230R} {127, 230}

\bibitem[\protect\citeauthoryear{{Rosati} et~al.,}{{Rosati}
  et~al.}{2009}]{Rosati2009}
{Rosati} P.,  et~al., 2009, \mn@doi [\aap] {10.1051/0004-6361/200913099}, \href
  {https://ui.adsabs.harvard.edu/abs/2009A&A...508..583R} {508, 583}

\bibitem[\protect\citeauthoryear{{Ruel} et~al.,}{{Ruel}
  et~al.}{2014}]{Ruel2014}
{Ruel} J.,  et~al., 2014, \mn@doi [\apj] {10.1088/0004-637X/792/1/45}, \href
  {https://ui.adsabs.harvard.edu/abs/2014ApJ...792...45R} {792, 45}

\bibitem[\protect\citeauthoryear{{Salim}}{{Salim}}{2014}]{Salim2014}
{Salim} S.,  2014, \mn@doi [Serbian Astronomical Journal]
  {10.2298/SAJ1489001S}, \href
  {https://ui.adsabs.harvard.edu/abs/2014SerAJ.189....1S} {189, 1}

\bibitem[\protect\citeauthoryear{{Santini} et~al.,}{{Santini}
  et~al.}{2017}]{Santini2017}
{Santini} P.,  et~al., 2017, \mn@doi [\apj] {10.3847/1538-4357/aa8874}, \href
  {https://ui.adsabs.harvard.edu/abs/2017ApJ...847...76S} {847, 76}

\bibitem[\protect\citeauthoryear{{Santos} et~al.,}{{Santos}
  et~al.}{2014}]{Santos2014}
{Santos} J.~S.,  et~al., 2014, \mn@doi [\mnras] {10.1093/mnras/stt2376}, \href
  {https://ui.adsabs.harvard.edu/abs/2014MNRAS.438.2565S} {438, 2565}

\bibitem[\protect\citeauthoryear{{Sarron} \& {Conselice}}{{Sarron} \&
  {Conselice}}{2021}]{Sarron2021}
{Sarron} F.,  {Conselice} C.~J.,  2021, \mn@doi [\mnras]
  {10.1093/mnras/stab1844}, \href
  {https://ui.adsabs.harvard.edu/abs/2021MNRAS.506.2136S} {506, 2136}

\bibitem[\protect\citeauthoryear{{Scholtz} et~al.,}{{Scholtz}
  et~al.}{2023}]{Scholtz2023}
{Scholtz} J.,  et~al., 2023, \mn@doi [arXiv e-prints]
  {10.48550/arXiv.2306.09142}, \href
  {https://ui.adsabs.harvard.edu/abs/2023arXiv230609142S} {p. arXiv:2306.09142}

\bibitem[\protect\citeauthoryear{{Schreiber} et~al.,}{{Schreiber}
  et~al.}{2015}]{Schreiber2015}
{Schreiber} C.,  et~al., 2015, \mn@doi [\aap] {10.1051/0004-6361/201425017},
  \href {https://ui.adsabs.harvard.edu/abs/2015A&A...575A..74S} {575, A74}

\bibitem[\protect\citeauthoryear{{Shimakawa} et~al.,}{{Shimakawa}
  et~al.}{2018}]{Shimakawa2018}
{Shimakawa} R.,  et~al., 2018, \mn@doi [\mnras] {10.1093/mnras/stx2494}, \href
  {https://ui.adsabs.harvard.edu/abs/2018MNRAS.473.1977S} {473, 1977}

\bibitem[\protect\citeauthoryear{{Shimasaku} et~al.,}{{Shimasaku}
  et~al.}{2003}]{Shimasaku2003}
{Shimasaku} K.,  et~al., 2003, \mn@doi [\apjl] {10.1086/374880}, \href
  {https://ui.adsabs.harvard.edu/abs/2003ApJ...586L.111S} {586, L111}

\bibitem[\protect\citeauthoryear{{Sillassen} et~al.,}{{Sillassen}
  et~al.}{2022}]{Sillassen2022}
{Sillassen} N.~B.,  et~al., 2022, \mn@doi [\aap] {10.1051/0004-6361/202244661},
  \href {https://ui.adsabs.harvard.edu/abs/2022A&A...665L...7S} {665, L7}

\bibitem[\protect\citeauthoryear{{Smail} et~al.,}{{Smail}
  et~al.}{2023}]{Smail2023}
{Smail} I.,  et~al., 2023, \mn@doi [\apj] {10.3847/1538-4357/acf931}, \href
  {https://ui.adsabs.harvard.edu/abs/2023ApJ...958...36S} {958, 36}

\bibitem[\protect\citeauthoryear{{Somerville} \& {Dav{\'e}}}{{Somerville} \&
  {Dav{\'e}}}{2015}]{Somerville2015}
{Somerville} R.~S.,  {Dav{\'e}} R.,  2015, \mn@doi [\araa]
  {10.1146/annurev-astro-082812-140951}, \href
  {https://ui.adsabs.harvard.edu/abs/2015ARA&A..53...51S} {53, 51}

\bibitem[\protect\citeauthoryear{{Sousbie}}{{Sousbie}}{2011}]{Sousbie2011}
{Sousbie} T.,  2011, \mn@doi [\mnras] {10.1111/j.1365-2966.2011.18394.x}, \href
  {https://ui.adsabs.harvard.edu/abs/2011MNRAS.414..350S} {414, 350}

\bibitem[\protect\citeauthoryear{{Springel} et~al.,}{{Springel}
  et~al.}{2005}]{Springel2005}
{Springel} V.,  et~al., 2005, \mn@doi [\nat] {10.1038/nature03597}, \href
  {https://ui.adsabs.harvard.edu/abs/2005Natur.435..629S} {435, 629}

\bibitem[\protect\citeauthoryear{{Stanford} et~al.,}{{Stanford}
  et~al.}{2012}]{Stanford2012}
{Stanford} S.~A.,  et~al., 2012, \mn@doi [\apj] {10.1088/0004-637X/753/2/164},
  \href {https://ui.adsabs.harvard.edu/abs/2012ApJ...753..164S} {753, 164}

\bibitem[\protect\citeauthoryear{{Staniszewski} et~al.,}{{Staniszewski}
  et~al.}{2009}]{Staniszewski2009}
{Staniszewski} Z.,  et~al., 2009, \mn@doi [\apj] {10.1088/0004-637X/701/1/32},
  \href {https://ui.adsabs.harvard.edu/abs/2009ApJ...701...32S} {701, 32}

\bibitem[\protect\citeauthoryear{{Stefanon} et~al.,}{{Stefanon}
  et~al.}{2017a}]{2017ApJS..229...32S}
{Stefanon} M.,  et~al., 2017a, \mn@doi [\apjs] {10.3847/1538-4365/aa66cb},
  \href {https://ui.adsabs.harvard.edu/abs/2017ApJS..229...32S} {229, 32}

\bibitem[\protect\citeauthoryear{{Stefanon}, {Bouwens}, {Labb{\'e}}, {Muzzin},
  {Marchesini}, {Oesch}  \& {Gonzalez}}{{Stefanon}
  et~al.}{2017b}]{stefanon2017b}
{Stefanon} M.,  {Bouwens} R.~J.,  {Labb{\'e}} I.,  {Muzzin} A.,  {Marchesini}
  D.,  {Oesch} P.,   {Gonzalez} V.,  2017b, \mn@doi [\apj]
  {10.3847/1538-4357/aa72d8}, \href
  {https://ui.adsabs.harvard.edu/abs/2017ApJ...843...36S} {843, 36}

\bibitem[\protect\citeauthoryear{{Steidel}, {Adelberger}, {Shapley}, {Erb},
  {Reddy}  \& {Pettini}}{{Steidel} et~al.}{2005}]{Steidel2005}
{Steidel} C.~C.,  {Adelberger} K.~L.,  {Shapley} A.~E.,  {Erb} D.~K.,  {Reddy}
  N.~A.,   {Pettini} M.,  2005, \mn@doi [\apj] {10.1086/429989}, \href
  {https://ui.adsabs.harvard.edu/abs/2005ApJ...626...44S} {626, 44}

\bibitem[\protect\citeauthoryear{{Sun}, {Helton}, {Egami}, {Hainline}, {Rieke},
  {Rieke}, {Willmer}  \& {Jades Collaboration}}{{Sun} et~al.}{2023}]{Sun2023}
{Sun} F.,  {Helton} J.,  {Egami} E.,  {Hainline} K.,  {Rieke} G.,  {Rieke} M.,
  {Willmer} C.,   {Jades Collaboration} 2023, in American Astronomical Society
  Meeting Abstracts. p. 206.02

\bibitem[\protect\citeauthoryear{{Tacchella} et~al.,}{{Tacchella}
  et~al.}{2023}]{Tacchella2023}
{Tacchella} S.,  et~al., 2023, \mn@doi [\apj] {10.3847/1538-4357/acdbc6}, \href
  {https://ui.adsabs.harvard.edu/abs/2023ApJ...952...74T} {952, 74}

\bibitem[\protect\citeauthoryear{{Tasca} et~al.,}{{Tasca}
  et~al.}{2015}]{Tasca2015}
{Tasca} L.~A.~M.,  et~al., 2015, \mn@doi [\aap] {10.1051/0004-6361/201425379},
  \href {https://ui.adsabs.harvard.edu/abs/2015A&A...581A..54T} {581, A54}

\bibitem[\protect\citeauthoryear{{Taylor-Mager}, {Conselice}, {Windhorst}  \&
  {Jansen}}{{Taylor-Mager} et~al.}{2007}]{TaylorMager2007}
{Taylor-Mager} V.~A.,  {Conselice} C.~J.,  {Windhorst} R.~A.,   {Jansen} R.~A.,
   2007, \mn@doi [\apj] {10.1086/511806}, \href
  {https://ui.adsabs.harvard.edu/abs/2007ApJ...659..162T} {659, 162}

\bibitem[\protect\citeauthoryear{{Tomczak} et~al.,}{{Tomczak}
  et~al.}{2014}]{2014ApJ...783...85T}
{Tomczak} A.~R.,  et~al., 2014, \mn@doi [\apj] {10.1088/0004-637X/783/2/85},
  \href {https://ui.adsabs.harvard.edu/abs/2014ApJ...783...85T} {783, 85}

\bibitem[\protect\citeauthoryear{{Tomczak} et~al.,}{{Tomczak}
  et~al.}{2016}]{Tomczak2016}
{Tomczak} A.~R.,  et~al., 2016, \mn@doi [\apj] {10.3847/0004-637X/817/2/118},
  \href {https://ui.adsabs.harvard.edu/abs/2016ApJ...817..118T} {817, 118}

\bibitem[\protect\citeauthoryear{{Tomczak} et~al.,}{{Tomczak}
  et~al.}{2017}]{Tomczak2017}
{Tomczak} A.~R.,  et~al., 2017, \mn@doi [\mnras] {10.1093/mnras/stx2245}, \href
  {https://ui.adsabs.harvard.edu/abs/2017MNRAS.472.3512T} {472, 3512}

\bibitem[\protect\citeauthoryear{{Tomczak} et~al.,}{{Tomczak}
  et~al.}{2019}]{Tomczak2019}
{Tomczak} A.~R.,  et~al., 2019, \mn@doi [\mnras] {10.1093/mnras/stz342}, \href
  {https://ui.adsabs.harvard.edu/abs/2019MNRAS.484.4695T} {484, 4695}

\bibitem[\protect\citeauthoryear{{Toshikawa} et~al.,}{{Toshikawa}
  et~al.}{2012}]{Toshikawa2012}
{Toshikawa} J.,  et~al., 2012, \mn@doi [\apj] {10.1088/0004-637X/750/2/137},
  \href {https://ui.adsabs.harvard.edu/abs/2012ApJ...750..137T} {750, 137}

\bibitem[\protect\citeauthoryear{{Toshikawa} et~al.,}{{Toshikawa}
  et~al.}{2016}]{Toshikawa2016}
{Toshikawa} J.,  et~al., 2016, \mn@doi [\apj] {10.3847/0004-637X/826/2/114},
  \href {https://ui.adsabs.harvard.edu/abs/2016ApJ...826..114T} {826, 114}

\bibitem[\protect\citeauthoryear{{Toshikawa} et~al.,}{{Toshikawa}
  et~al.}{2018}]{Toshikawa2018}
{Toshikawa} J.,  et~al., 2018, \mn@doi [\pasj] {10.1093/pasj/psx102}, \href
  {https://ui.adsabs.harvard.edu/abs/2018PASJ...70S..12T} {70, S12}

\bibitem[\protect\citeauthoryear{{Tozzi} et~al.,}{{Tozzi}
  et~al.}{2022}]{Tozzi2022}
{Tozzi} P.,  et~al., 2022, \mn@doi [\aap] {10.1051/0004-6361/202142333}, \href
  {https://ui.adsabs.harvard.edu/abs/2022A&A...662A..54T} {662, A54}

\bibitem[\protect\citeauthoryear{{Umehata} et~al.,}{{Umehata}
  et~al.}{2015}]{Umehata2015}
{Umehata} H.,  et~al., 2015, \mn@doi [\apjl] {10.1088/2041-8205/815/1/L8},
  \href {https://ui.adsabs.harvard.edu/abs/2015ApJ...815L...8U} {815, L8}

\bibitem[\protect\citeauthoryear{Virtanen et~al.,}{Virtanen
  et~al.}{2020}]{2020SciPy-NMeth}
Virtanen P.,  et~al., 2020, \mn@doi [Nature Methods]
  {10.1038/s41592-019-0686-2}, \href {https://rdcu.be/b08Wh} {17, 261}

\bibitem[\protect\citeauthoryear{{Wang} et~al.,}{{Wang}
  et~al.}{2016}]{WangTao2016}
{Wang} T.,  et~al., 2016, \mn@doi [\apj] {10.3847/0004-637X/828/1/56}, \href
  {https://ui.adsabs.harvard.edu/abs/2016ApJ...828...56W} {828, 56}

\bibitem[\protect\citeauthoryear{{Wang} et~al.,}{{Wang}
  et~al.}{2023}]{wangfg2023}
{Wang} F.,  et~al., 2023, \mn@doi [\apjl] {10.3847/2041-8213/accd6f}, \href
  {https://ui.adsabs.harvard.edu/abs/2023ApJ...951L...4W} {951, L4}

\bibitem[\protect\citeauthoryear{{Weinmann}, {Kauffmann}, {van den Bosch},
  {Pasquali}, {McIntosh}, {Mo}, {Yang}  \& {Guo}}{{Weinmann}
  et~al.}{2009}]{Weinmann2009}
{Weinmann} S.~M.,  {Kauffmann} G.,  {van den Bosch} F.~C.,  {Pasquali} A.,
  {McIntosh} D.~H.,  {Mo} H.,  {Yang} X.,   {Guo} Y.,  2009, \mn@doi [\mnras]
  {10.1111/j.1365-2966.2009.14412.x}, \href
  {https://ui.adsabs.harvard.edu/abs/2009MNRAS.394.1213W} {394, 1213}

\bibitem[\protect\citeauthoryear{{Whitaker} et~al.,}{{Whitaker}
  et~al.}{2014}]{Whitaker2014}
{Whitaker} K.~E.,  et~al., 2014, \mn@doi [\apj] {10.1088/0004-637X/795/2/104},
  \href {https://ui.adsabs.harvard.edu/abs/2014ApJ...795..104W} {795, 104}

\bibitem[\protect\citeauthoryear{{Whitler}, {Stark}, {Endsley}, {Leja},
  {Charlot}  \& {Chevallard}}{{Whitler} et~al.}{2023}]{2023MNRAS.519.5859W}
{Whitler} L.,  {Stark} D.~P.,  {Endsley} R.,  {Leja} J.,  {Charlot} S.,
  {Chevallard} J.,  2023, \mn@doi [\mnras] {10.1093/mnras/stad004}, \href
  {https://ui.adsabs.harvard.edu/abs/2023MNRAS.519.5859W} {519, 5859}

\bibitem[\protect\citeauthoryear{{Williams} et~al.,}{{Williams}
  et~al.}{2018}]{2018ApJS..236...33W}
{Williams} C.~C.,  et~al., 2018, \mn@doi [\apjs] {10.3847/1538-4365/aabcbb},
  \href {https://ui.adsabs.harvard.edu/abs/2018ApJS..236...33W} {236, 33}

\bibitem[\protect\citeauthoryear{{Williamson} et~al.,}{{Williamson}
  et~al.}{2011}]{Williamson2011}
{Williamson} R.,  et~al., 2011, \mn@doi [\apj] {10.1088/0004-637X/738/2/139},
  \href {https://ui.adsabs.harvard.edu/abs/2011ApJ...738..139W} {738, 139}

\bibitem[\protect\citeauthoryear{{Windhorst} et~al.,}{{Windhorst}
  et~al.}{2023}]{Windhorst2023}
{Windhorst} R.~A.,  et~al., 2023, \mn@doi [\aj] {10.3847/1538-3881/aca163},
  \href {https://ui.adsabs.harvard.edu/abs/2023AJ....165...13W} {165, 13}

\bibitem[\protect\citeauthoryear{{Wylezalek} et~al.,}{{Wylezalek}
  et~al.}{2013}]{Wylezalek2013}
{Wylezalek} D.,  et~al., 2013, \mn@doi [\apj] {10.1088/0004-637X/769/1/79},
  \href {https://ui.adsabs.harvard.edu/abs/2013ApJ...769...79W} {769, 79}

\bibitem[\protect\citeauthoryear{{von der Linden}, {Wild}, {Kauffmann}, {White}
   \& {Weinmann}}{{von der Linden} et~al.}{2010}]{vonderLinden2010}
{von der Linden} A.,  {Wild} V.,  {Kauffmann} G.,  {White} S. D.~M.,
  {Weinmann} S.,  2010, \mn@doi [\mnras] {10.1111/j.1365-2966.2010.16375.x},
  \href {https://ui.adsabs.harvard.edu/abs/2010MNRAS.404.1231V} {404, 1231}

\makeatother
\end{thebibliography}




\appendix

\section{Catalogue of the high-z protocluster in \texttt{DETECTIFz} }
In \S\ref{sec:Group galaxy membership probabilities}, we describe our systematic approach utilizing the \texttt{DETECTIFz} to quantitatively assess the propobality of a galaxy being associated with a particular galaxy cluster. Our investigation spans the contiguous regions of CEERS, JADES, and NEP fields, encompassing a cumulative area of $\sim185$ arcmin$^2$. Within these expanses, we have identified 26 galaxy clusters at $5<z<7$. Below, we present the spatial locations and the properties of the protocluster candidates. Table~\ref{tab:cat_cluster} is the catalogue of the high-z protocluster candidates found in this work. Table~\ref{tab:cat_galaxies} is the catalog illustrating group membership.


\begin{table*}
\caption{The catalog of the clusters candidates found in CEERS, JADES, and NEP using \texttt{DETECTIFz}. The sky coordinates and group redshift are measured at the peak of the highest $S/N$ detection in the group. $z_{\rm group}$ is the redshift of the center of the protocluster. $z_{\rm inf}$ and $z_{\rm sup}$ are the minimum and maximum redshift of detection at $S/N>1.5$. log $\delta_{gal}$ is the number density contrast in a circle of 250 kpc radius. The definition of $R_{200}$ is in \S\ref{sec: Group membership probability and cluster mass}, measured in the highest SNR single slice in Mpc unit. $n_{\rm dets}$ is the number of members with $P_{\rm mem}\in Q_4$ in the galaxy cluster. $M_{\rm \star, PC}$ is the total cluster stellar mass by adding galaxy members. $M_{\rm halo}$ is the total halo mass of clusters using $M_{\rm halo}-M_{\star}$ scaling relation in \citet{Behroozi2013}, described in \S~\ref{sec: halo mass}. }
\label{tab:cat_cluster}
\begin{tabular}{ccccccccccl} 
\hline
\hline
Group ID & RA & DEC & $z_{\rm group}$ & $z_{\rm inf}$ & $z_{\rm sup}$ & log $\delta_{\rm gal}$& $R_{200}$ & $M_{\rm \star, PC}$& $M_{\rm halo}$& $n_{\rm dets}$  \\
&J2000&J2000&&&&&Mpc&&&\\
\hline
CEERS - ID 1 & 214.98271 & 52.87480 & 5.16 & 5.01 & 5.91 & 0.37 & 0.78 & 10.48 $^{+ 0.15 }_{ -0.16 }$ & 12.27 $^{+ 0.52 }_{ -0.29 }$ & 32 \\
CEERS - ID 2 & 214.95095 & 52.91322 & 5.19 & 5.04 & 5.53 & 0.26 & 0.31 & 10.40 $^{+ 0.15 }_{ -0.17 }$ & 12.12 $^{+ 0.40 }_{ -0.24 }$ & 9 \\
CEERS - ID 3 & 214.80541 & 52.73319 & 5.40 & 5.12 & 5.68 & 1.01 & 0.41 & 9.95 $^{+ 0.13 }_{ -0.25 }$ & 11.63 $^{+ 0.10 }_{ -0.16 }$ & 6 \\
CEERS - ID 4 & 214.92576 & 52.94283 & 5.40 & 5.10 & 5.84 & 0.37 & 0.30 & 10.73 $^{+ 0.14 }_{ -0.22 }$ & 13.57 $^{+ 0.83 }_{ -1.07 }$ & 20 \\
CEERS - ID 5 & 214.94564 & 52.91242 & 5.42 & 5.25 & 5.61 & 0.34 & 0.33 & 9.82 $^{+ 0.19 }_{ -0.17 }$ & 11.54 $^{+ 0.13 }_{ -0.10 }$ & 5 \\
CEERS - ID 6 & 214.80659 & 52.79719 & 5.47 & 5.19 & 5.76 & 0.35 & 0.19 & 9.68 $^{+ 0.11 }_{ -0.21 }$ & 11.45 $^{+ 0.06 }_{ -0.12 }$ & 5 \\
CEERS - ID 7 & 214.95629 & 52.93801 & 5.79 & 5.53 & 6.07 & 0.39 & 0.32 & 9.94 $^{+ 0.16 }_{ -0.16 }$ & 11.61 $^{+ 0.15 }_{ -0.12 }$ & 10 \\
CEERS - ID 8 & 214.98669 & 52.87240 & 5.93 & 5.68 & 6.15 & 0.35 & 0.25 & 10.45 $^{+ 0.11 }_{ -0.16 }$ & 12.65 $^{+ 0.55 }_{ -0.58 }$ & 6 \\
CEERS - ID 9 & 214.83161 & 52.87481 & 6.29 & 5.78 & 6.62 & 0.63 & 0.49 & 9.85 $^{+ 0.17 }_{ -0.17 }$ & 11.52 $^{+ 0.17 }_{ -0.13 }$ & 25 \\
CEERS - ID 10 & 214.92044 & 52.91803 & 6.76 & 6.21 & 7.01 & 0.36 & 0.47 & 9.58 $^{+ 0.14 }_{ -0.18 }$ & 11.28 $^{+ 0.11 }_{ -0.12 }$ & 12 \\
\hline
NEP - ID 1 & 260.69501 & 65.77955 & 5.34 & 5.00 & 5.60 & 0.46 & 0.26 & 10.23 $^{+ 0.18 }_{ -0.21 }$ & 11.88 $^{+ 0.29 }_{ -0.20 }$ & 14 \\
NEP - ID 2 & 260.73205 & 65.76115 & 5.50 & 5.00 & 5.88 & 0.46 & 0.41 & 10.46 $^{+ 0.16 }_{ -0.22 }$ & 12.40 $^{+ 0.69 }_{ -0.48 }$ & 37 \\
NEP - ID 3 & 260.69490 & 65.90675 & 5.57 & 5.16 & 5.87 & 0.45 & 0.16 & 9.19 $^{+ 0.18 }_{ -0.20 }$ & 11.17 $^{+ 0.10 }_{ -0.10 }$ & 5 \\
NEP - ID 4 & 260.73008 & 65.72515 & 5.74 & 5.33 & 6.03 & 0.45 & 0.46 & 10.44 $^{+ 0.11 }_{ -0.16 }$ & 12.46 $^{+ 0.50 }_{ -0.45 }$ & 29 \\
NEP - ID 5 & 260.81414 & 65.82672 & 6.08 & 5.61 & 6.34 & 0.37 & 0.40 & 10.68 $^{+ 0.10 }_{ -0.13 }$ & 13.92 $^{+ 0.48 }_{ -0.63 }$ & 41 \\
NEP - ID 6 & 260.43907 & 65.82410 & 6.29 & 6.06 & 6.48 & 0.44 & 0.34 & 9.96 $^{+ 0.19 }_{ -0.18 }$ & 11.61 $^{+ 0.27 }_{ -0.16 }$ & 12 \\
NEP - ID 7 & 260.95675 & 65.82736 & 6.41 & 6.04 & 6.70 & 0.46 & 0.58 & 10.16 $^{+ 0.15 }_{ -0.16 }$ & 11.94 $^{+ 0.51 }_{ -0.27 }$ & 37 \\
NEP - ID 8 & 260.68717 & 65.81155 & 6.69 & 6.47 & 6.85 & 0.19 & 0.33 & 10.07 $^{+ 0.15 }_{ -0.14 }$ & 11.83 $^{+ 0.46 }_{ -0.23 }$ & 13 \\
NEP - ID 9 & 260.65784 & 65.83154 & 6.75 & 6.65 & 6.87 & 0.20 & 0.26 & 9.59 $^{+ 0.18 }_{ -0.17 }$ & 11.29 $^{+ 0.14 }_{ -0.11 }$ & 5 \\
NEP - ID 10 & 260.77117 & 65.83474 & 6.90 & 6.75 & 7.02 & 0.32 & 0.31 & 9.98 $^{+ 0.24 }_{ -0.28 }$ & 11.70 $^{+ 0.78 }_{ -0.34 }$ & 20 \\
\hline
JADES - ID 1 & 53.12985 & -27.78021 & 5.68 & 5.25 & 6.23 & 0.59 & 0.41 & 10.62 $^{+ 0.07 }_{ -0.07 }$ & 13.28 $^{+ 0.37 }_{ -0.34 }$ & 66 \\
JADES - ID 2 & 53.17325 & -27.75462 & 5.84 & 5.69 & 5.99 & 0.21 & 0.11 & 9.21 $^{+ 0.13 }_{ -0.20 }$ & 11.16 $^{+ 0.07 }_{ -0.10 }$ & 4 \\
JADES - ID 3 & 53.11175 & -27.80421 & 6.02 & 5.89 & 6.16 & 0.39 & 0.15 & 9.36 $^{+ 0.13 }_{ -0.14 }$ & 11.22 $^{+ 0.07 }_{ -0.08 }$ & 12 \\
JADES - ID 4 & 53.19043 & -27.77381 & 6.02 & 5.86 & 6.20 & 0.26 & 0.13 & 9.51 $^{+ 0.14 }_{ -0.15 }$ & 11.30 $^{+ 0.09 }_{ -0.09 }$ & 10 \\
JADES - ID 5 & 53.15517 & -27.74182 & 6.17 & 5.99 & 6.38 & 0.31 & 0.22 & 9.94 $^{+ 0.09 }_{ -0.09 }$ & 11.60 $^{+ 0.10 }_{ -0.08 }$ & 19 \\
JADES - ID 6 & 53.16421 & -27.77302 & 6.34 & 6.11 & 6.50 & 0.26 & 0.17 & 9.53 $^{+ 0.16 }_{ -0.15 }$ & 11.29 $^{+ 0.10 }_{ -0.09 }$ & 13 \\
\hline
\end{tabular}
\end{table*}

\begin{table*}
\caption{The example catalog of the group membership found in CEERS using \texttt{DETECTIFz}. The redshift, stellar mass and SFR are from \bagpipes. $M_{\rm UV}$ and UV slope $\beta$ are calculated in \S\ref{sec: beta}. P$_{\rm mem}$ is the group membership probability for each individual galaxies in a group. The full catalog can be accessed in the online version. The complete catalog is available in the online version.}
\label{tab:cat_galaxies}
\begin{tabular}{c|ccccccccc} 
\hline
\hline
Group ID & Galaxy ID & RA & DEC & $z_{\rm photo}$ & log M$^{\rm gal}_{\star}$ & SFR$_{\rm UV}$ & $M_{\rm UV}$ & $\beta$ & $P_{\rm mem}$ \\
& &J2000&J2000&&M$_{\odot}$&M$_{\odot}$/yr& mag &&\\
\hline
CEERS - ID 1 & 5699 & 214.95790 & 52.85781 & 5.47 $^{+ 0.08 }_{ -0.26 }$ & 8.33 $^{+ 0.25 }_{ -0.18 }$ & 2.82 $^{+ 0.53 }_{- 0.39 }$ & -19.35 $^{+ 0.20 }_{- 0.18 }$ & -2.72 $^{+ 0.52 }_{- 0.54 }$ & 0.99 \\
CEERS - ID 1 & 5938 & 214.94327 & 52.84919 & 5.48 $^{+ 0.08 }_{ -0.27 }$ & 8.57 $^{+ 0.18 }_{ -0.13 }$ & 7.02 $^{+ 8.81 }_{- 2.47 }$ & -19.71 $^{+ 0.20 }_{- 0.19 }$ & -1.88 $^{+ 0.54 }_{- 0.54 }$ & 0.86 \\
CEERS - ID 1 & 8811 & 214.98093 & 52.88148 & 5.54 $^{+ 0.14 }_{ -0.51 }$ & 8.89 $^{+ 0.13 }_{ -0.18 }$ & 3.87 $^{+ 0.83 }_{- 0.57 }$ & -19.68 $^{+ 0.21 }_{- 0.20 }$ & -2.81 $^{+ 0.66 }_{- 0.70 }$ & 0.82 \\
CEERS - ID 1 & 7240 & 214.93700 & 52.91066 & 5.35 $^{+ 0.09 }_{ -0.28 }$ & 9.34 $^{+ 0.06 }_{ -0.06 }$ & 3.48 $^{+ 1.83 }_{- 0.51 }$ & -19.43 $^{+ 0.21 }_{- 0.20 }$ & -2.58 $^{+ 0.54 }_{- 0.53 }$ & 0.73 \\
CEERS - ID 1 & 6018 & 214.94368 & 52.85008 & 5.52 $^{+ 0.09 }_{ -0.15 }$ & 9.31 $^{+ 0.14 }_{ -0.18 }$ & 24.58 $^{+ 30.74 }_{- 13.98 }$ & -20.26 $^{+ 0.20 }_{- 0.18 }$ & -1.72 $^{+ 0.53 }_{- 0.54 }$ & 0.71 \\
CEERS - ID 1 & 7653 & 214.96547 & 52.87933 & 5.47 $^{+ 0.22 }_{ -0.14 }$ & 9.44 $^{+ 0.07 }_{ -0.14 }$ & 5.18 $^{+ 1.95 }_{- 0.70 }$ & -19.90 $^{+ 0.20 }_{- 0.19 }$ & -2.28 $^{+ 0.54 }_{- 0.53 }$ & 0.71 \\
CEERS - ID 1 & 8812 & 214.98087 & 52.88110 & 5.59 $^{+ 0.13 }_{ -0.08 }$ & 9.46 $^{+ 0.11 }_{ -0.13 }$ & 12.49 $^{+ 2.42 }_{- 1.65 }$ & -20.94 $^{+ 0.19 }_{- 0.18 }$ & -2.77 $^{+ 0.54 }_{- 0.54 }$ & 0.70 \\
CEERS - ID 1 & 8505 & 214.95318 & 52.91579 & 5.40 $^{+ 0.21 }_{ -0.25 }$ & 8.81 $^{+ 0.08 }_{ -0.09 }$ & 4.36 $^{+ 0.91 }_{- 0.77 }$ & -19.85 $^{+ 0.21 }_{- 0.21 }$ & -3.90 $^{+ 0.60 }_{- 0.63 }$ & 0.70 \\
CEERS - ID 1 & 5719 & 214.88850 & 52.90485 & 5.62 $^{+ 0.08 }_{ -0.25 }$ & 7.87 $^{+ 0.13 }_{ -0.20 }$ & 4.36 $^{+ 1.06 }_{- 0.54 }$ & -19.76 $^{+ 0.19 }_{- 0.17 }$ & -2.42 $^{+ 0.52 }_{- 0.53 }$ & 0.69 \\
CEERS - ID 1 & 8933 & 214.94735 & 52.91899 & 5.46 $^{+ 0.22 }_{ -0.11 }$ & 8.41 $^{+ 0.17 }_{ -0.26 }$ & 3.63 $^{+ 0.69 }_{- 0.62 }$ & -19.65 $^{+ 0.20 }_{- 0.19 }$ & -3.49 $^{+ 0.54 }_{- 0.53 }$ & 0.68 \\
CEERS - ID 1 & 6255 & 214.94998 & 52.85578 & 5.48 $^{+ 0.12 }_{ -0.21 }$ & 8.96 $^{+ 0.25 }_{ -0.26 }$ & 4.36 $^{+ 1.50 }_{- 0.60 }$ & -19.72 $^{+ 0.20 }_{- 0.19 }$ & -2.37 $^{+ 0.54 }_{- 0.53 }$ & 0.66 \\
CEERS - ID 1 & 5844 & 214.94942 & 52.90788 & 5.32 $^{+ 0.14 }_{ -0.41 }$ & 9.47 $^{+ 0.18 }_{ -0.16 }$ & 18.18 $^{+ 3.62 }_{- 2.92 }$ & -21.39 $^{+ 0.21 }_{- 0.20 }$ & -3.09 $^{+ 0.55 }_{- 0.53 }$ & 0.65 \\
CEERS - ID 1 & 8984 & 214.98532 & 52.95874 & 4.94 $^{+ 0.47 }_{ -0.03 }$ & 7.64 $^{+ 0.28 }_{ -0.19 }$ & 2.46 $^{+ 0.64 }_{- 0.45 }$ & -19.20 $^{+ 0.26 }_{- 0.25 }$ & -3.02 $^{+ 0.60 }_{- 0.62 }$ & 0.64 \\
 & ... & ... & ... & ... & ... \\
\hline
CEERS - ID 2 & ... & ... & ... & ... & ... \\
... & ... & ... & ... & ... & ... \\
\hline\hline
NEP - ID 1 & 1716 & 260.69784 & 65.78069 & 5.46 $^{+ 0.44 }_{ -0.11 }$ & 9.84 $^{+ 0.18 }_{ -0.22 }$ & 34.71 $^{+ 51.00 }_{- 19.15 }$ & -20.12 $^{+ 0.21 }_{- 0.17 }$ & -1.34 $^{+ 0.58 }_{- 0.51 }$ & 0.52 \\
NEP - ID 1 & 8048 & 260.72580 & 65.77226 & 5.44 $^{+ 0.46 }_{ -0.31 }$ & 8.10 $^{+ 0.17 }_{ -0.19 }$ & 13.24 $^{+ 26.43 }_{- 9.12 }$ & -18.02 $^{+ 0.28 }_{- 0.27 }$ & -0.90 $^{+ 0.73 }_{- 0.76 }$ & 0.50 \\
NEP - ID 1 & 1912 & 260.68502 & 65.76349 & 5.63 $^{+ 0.25 }_{ -0.18 }$ & 9.42 $^{+ 0.10 }_{ -0.25 }$ & 9.41 $^{+ 12.38 }_{- 3.17 }$ & -20.08 $^{+ 0.19 }_{- 0.16 }$ & -1.97 $^{+ 0.54 }_{- 0.52 }$ & 0.46 \\
NEP - ID 1 & 7527 & 260.71876 & 65.76629 & 5.50 $^{+ 0.41 }_{ -0.23 }$ & 8.77 $^{+ 0.18 }_{ -0.31 }$ & 3.32 $^{+ 2.27 }_{- 0.45 }$ & -19.37 $^{+ 0.19 }_{- 0.17 }$ & -2.31 $^{+ 0.53 }_{- 0.54 }$ & 0.45 \\
NEP - ID 1 & 1988 & 260.68919 & 65.76916 & 5.20 $^{+ 0.29 }_{ -0.10 }$ & 7.71 $^{+ 0.12 }_{ -0.15 }$ & 0.96 $^{+ 0.80 }_{- 0.13 }$ & -18.06 $^{+ 0.11 }_{- 0.09 }$ & -2.42 $^{+ 0.42 }_{- 0.42 }$ & 0.45 \\
NEP - ID 1 & 9678 & 260.73132 & 65.76976 & 6.29 $^{+ 0.17 }_{ -0.17 }$ & 7.12 $^{+ 0.13 }_{ -0.08 }$ & 8.38 $^{+ 13.98 }_{- 5.42 }$ & -18.54 $^{+ 0.18 }_{- 0.15 }$ & -1.44 $^{+ 0.61 }_{- 0.63 }$ & 0.40 \\
NEP - ID 1 & 5891 & 260.72108 & 65.77939 & 6.31 $^{+ 0.09 }_{ -0.09 }$ & 8.17 $^{+ 0.07 }_{ -0.08 }$ & 6.80 $^{+ 2.56 }_{- 0.67 }$ & -20.25 $^{+ 0.13 }_{- 0.12 }$ & -2.40 $^{+ 0.56 }_{- 0.51 }$ & 0.38 \\
NEP - ID 1 & 1969 & 260.68427 & 65.76288 & 5.94 $^{+ 0.15 }_{ -0.17 }$ & 7.26 $^{+ 0.18 }_{ -0.14 }$ & 1.58 $^{+ 1.16 }_{- 0.21 }$ & -18.60 $^{+ 0.16 }_{- 0.15 }$ & -2.37 $^{+ 0.59 }_{- 0.62 }$ & 0.35 \\
NEP - ID 1 & 2249 & 260.69600 & 65.77503 & 5.47 $^{+ 0.29 }_{ -0.10 }$ & 8.51 $^{+ 0.26 }_{ -0.23 }$ & 3.66 $^{+ 1.04 }_{- 0.46 }$ & -19.55 $^{+ 0.21 }_{- 0.18 }$ & -2.44 $^{+ 0.53 }_{- 0.55 }$ & 0.29 \\
NEP - ID 1 & 3021 & 260.71007 & 65.78709 & 6.38 $^{+ 0.10 }_{ -0.10 }$ & 7.66 $^{+ 0.10 }_{ -0.06 }$ & 5.66 $^{+ 5.16 }_{- 0.68 }$ & -19.98 $^{+ 0.13 }_{- 0.11 }$ & -2.09 $^{+ 0.50 }_{- 0.52 }$ & 0.25 \\
NEP - ID 1 & 9512 & 260.74154 & 65.78315 & 5.50 $^{+ 0.34 }_{ -0.09 }$ & 8.51 $^{+ 0.22 }_{ -0.20 }$ & 5.88 $^{+ 6.94 }_{- 1.23 }$ & -19.80 $^{+ 0.19 }_{- 0.18 }$ & -2.16 $^{+ 0.51 }_{- 0.53 }$ & 0.24 \\
NEP - ID 1 & 8426 & 260.73253 & 65.77678 & 6.37 $^{+ 0.10 }_{ -0.18 }$ & 9.74 $^{+ 0.22 }_{ -0.19 }$ & 85.83 $^{+ 122.07 }_{- 51.79 }$ & -20.28 $^{+ 0.15 }_{- 0.11 }$ & -1.02 $^{+ 0.53 }_{- 0.55 }$ & 0.24 \\
& ... & ... & ... & ... & ... \\
& ... & ... & ... & ... & ... \\
\hline\hline
JADES - ID 1 & 17226 & 53.15477 & -27.80651 & 6.21 $^{+ 0.08 }_{ -0.30 }$ & 10.54 $^{+ 0.05 }_{ -0.05 }$ & 30.93 $^{+ 37.83 }_{- 17.82 }$ & -18.57 $^{+ 0.14 }_{- 0.12 }$ & -0.38 $^{+ 0.50 }_{- 0.52 }$ & 0.82 \\
JADES - ID 1 & 35742 & 53.13047 & -27.77824 & 5.75 $^{+ 0.17 }_{ -0.16 }$ & 8.92 $^{+ 0.10 }_{ -0.10 }$ & 2.32 $^{+ 1.49 }_{- 0.27 }$ & -19.03 $^{+ 0.18 }_{- 0.14 }$ & -2.27 $^{+ 0.52 }_{- 0.52 }$ & 0.80 \\
JADES - ID 1 & 24313 & 53.14078 & -27.80217 & 5.95 $^{+ 0.05 }_{ -0.05 }$ & 8.23 $^{+ 0.19 }_{ -0.16 }$ & 3.49 $^{+ 0.47 }_{- 0.40 }$ & -19.58 $^{+ 0.16 }_{- 0.14 }$ & -2.63 $^{+ 0.54 }_{- 0.54 }$ & 0.77 \\
JADES - ID 1 & 23748 & 53.14398 & -27.79874 & 5.83 $^{+ 0.11 }_{ -0.13 }$ & 8.47 $^{+ 0.13 }_{ -0.14 }$ & 3.17 $^{+ 3.24 }_{- 0.47 }$ & -19.29 $^{+ 0.17 }_{- 0.15 }$ & -2.22 $^{+ 0.56 }_{- 0.54 }$ & 0.72 \\
JADES - ID 1 & 18093 & 53.15205 & -27.80828 & 5.75 $^{+ 0.14 }_{ -0.15 }$ & 8.37 $^{+ 0.12 }_{ -0.20 }$ & 1.37 $^{+ 0.32 }_{- 0.16 }$ & -18.51 $^{+ 0.18 }_{- 0.16 }$ & -2.56 $^{+ 0.56 }_{- 0.53 }$ & 0.71 \\
JADES - ID 1 & 37202 & 53.12705 & -27.78887 & 5.48 $^{+ 0.02 }_{ -0.03 }$ & 6.93 $^{+ 0.14 }_{ -0.11 }$ & 0.77 $^{+ 0.64 }_{- 0.12 }$ & -17.75 $^{+ 0.19 }_{- 0.18 }$ & -2.48 $^{+ 0.56 }_{- 0.54 }$ & 0.70 \\
JADES - ID 1 & 24376 & 53.14189 & -27.80025 & 5.94 $^{+ 0.13 }_{ -0.24 }$ & 8.63 $^{+ 0.11 }_{ -0.12 }$ & 4.93 $^{+ 6.52 }_{- 2.91 }$ & -17.36 $^{+ 0.17 }_{- 0.15 }$ & -1.08 $^{+ 0.55 }_{- 0.55 }$ & 0.70 \\
JADES - ID 1 & 35460 & 53.12954 & -27.77736 & 5.99 $^{+ 0.13 }_{ -0.12 }$ & 7.77 $^{+ 0.07 }_{ -0.05 }$ & 10.12 $^{+ 3.51 }_{- 1.10 }$ & -20.68 $^{+ 0.17 }_{- 0.13 }$ & -2.44 $^{+ 0.54 }_{- 0.53 }$ & 0.67 \\
JADES - ID 1 & 27143 & 53.13600 & -27.79848 & 5.94 $^{+ 0.04 }_{ -0.05 }$ & 7.76 $^{+ 0.08 }_{ -0.08 }$ & 3.01 $^{+ 3.53 }_{- 0.53 }$ & -19.16 $^{+ 0.16 }_{- 0.15 }$ & -2.20 $^{+ 0.51 }_{- 0.53 }$ & 0.65 \\
JADES - ID 1 & 36154 & 53.12488 & -27.78412 & 5.70 $^{+ 0.20 }_{ -0.05 }$ & 8.23 $^{+ 0.15 }_{ -0.16 }$ & 2.14 $^{+ 0.37 }_{- 0.28 }$ & -19.05 $^{+ 0.18 }_{- 0.16 }$ & -2.83 $^{+ 0.50 }_{- 0.52 }$ & 0.60 \\
JADES - ID 1 & 30094 & 53.12870 & -27.79826 & 5.48 $^{+ 0.03 }_{ -0.03 }$ & 7.26 $^{+ 0.23 }_{ -0.23 }$ & 0.78 $^{+ 0.12 }_{- 0.12 }$ & -17.98 $^{+ 0.18 }_{- 0.16 }$ & -3.22 $^{+ 0.52 }_{- 0.50 }$ & 0.59 \\
JADES - ID 1 & 17532 & 53.15530 & -27.80555 & 5.83 $^{+ 0.12 }_{ -0.07 }$ & 7.11 $^{+ 0.11 }_{ -0.13 }$ & 0.72 $^{+ 0.98 }_{- 0.17 }$ & -17.49 $^{+ 0.16 }_{- 0.16 }$ & -2.13 $^{+ 0.53 }_{- 0.51 }$ & 0.59 \\
JADES - ID 1 & 37094 & 53.13005 & -27.77839 & 5.60 $^{+ 0.26 }_{ -0.02 }$ & 8.22 $^{+ 0.15 }_{ -0.20 }$ & 3.94 $^{+ 0.67 }_{- 0.52 }$ & -19.69 $^{+ 0.18 }_{- 0.17 }$ & -2.60 $^{+ 0.53 }_{- 0.48 }$ & 0.57 \\
 & ... & ... & ... & ... & ... \\
 & ... & ... & ... & ... & ... \\
\hline
\end{tabular}
\end{table*}


\bsp	
\label{lastpage}
\end{document}